\documentclass[twoside,12pt]{article}
\usepackage{epsfig}

\def\NPB{{\em Nucl. Phys.} B}

\def\PLB{{\em Phys. Lett.} B}

\def\PRL{\em Phys. Rev. Lett.}

\def\PREP{\em Phys. Rep.}

\def\PRD{{\em Phys. Rev.} D}

\def\ANNP{\em Ann. Phys. (N.Y.)}

\newcommand{\be}{\begin{equation}}
\newcommand{\ee}{\end{equation}}
\newcommand{\bea}{\begin{eqnarray}}
\newcommand{\eea}{\end{eqnarray}}
\newcommand{\nn}{\nonumber}
\begin{document}

\title{
\begin{flushright}
{\small FTPI-MINN-07/34 \\
UMN-TH-2625/07 \\}
\end{flushright}
\vspace{1cm} Charmonium}
\author{M.B. Voloshin\\
William I. Fine Theoretical Physics Institute, University of
Minnesota,\\ Minneapolis, MN 55455 \\
and \\
Institute of Theoretical and Experimental Physics, Moscow, 117218
\\
{\small \it To be published in Prog. Part. Nucl. Phys. 2008. 
}}
\maketitle
\begin{abstract}
Topics in the description of the properties of charmonium states are reviewed
with an emphasis on specific theoretical ideas and methods of relating those
properties to the underlying theory of Quantum Chromodynamics.
\end{abstract}

\tableofcontents

\section{Introduction}
Time is carrying us farther away from that day of November 11, 1974, when the
news of an unusual resonance found simultaneously at BNL\cite{jbnl} and at
SLAC\cite{psislac} has swiftly spread through high-energy laboratories around
the globe. Well before the age of widespread instantaneous satellite
communications and the Internet, the new resonance became known in Moscow within
the same hour as it was publicly announced in Palo Alto. After the initial
excitement, confusion and revelations, it became clear that the new $J/\psi$
resonance was the first to have been observed state of a system containing
previously unknown (but anticipated) charmed quark and its antiquark: $c \bar
c$\,\footnote{The second charmonium resonance was found just ten days after the
$J/\psi$\,\cite{psiprim}.}. The new system, charmonium, in a close analogy with
positronium or even with a hydrogen atom, was expected to contain a spectrum of
resonances, corresponding to various excitations of the heavy quark pair.
However, unlike its analogs governed mainly by the electrostatic Coulomb force,
the properties of charmonium are determined by the strong interaction, so that
the newly found system was, in a way, the simplest object for a study of the
strong interactions. It was strongly hoped\cite{6aut} that charmonium could play
the same role for understanding hadronic dynamics as the hydrogen atom played in
understanding the atomic physics. In a way, this has indeed been the case and
the development of many methods in QCD is directly related to analyses of the
properties of charmonium and of its heavier sibling bottomonium.

Recently the physics of charmonium regained a great renewed interest due to the
massive dedicated investigation by BES and CLEO-c and the studies using decays
of $B$ mesons and the radiative return technique at the B factories with a
higher initial energy of the electron and positron beams. After a `dry spell' of
more  than two decades during which no new states of charmonium have been found
with any certainty, new observations discover charmonium and charmonium-related
resonances at a rate that outpaces the ability of the theory to fit their
properties in a consistent scheme. Furthermore, the data very strongly suggest
that among the new resonances there are exotic four-quark states, possibly
hybrid states with gluonic degrees of freedom in addition to the $c \bar c$
pair, and also loosely bound states of heavy hadrons -- charmonium molecules.
Thus it looks like that charmonium not only has provided us with a `hadronic
atomic physics' but quite possibly also with a `hadronic chemistry', and in its
mature age of 33 charmonium still offers us new intriguing puzzles.

In this review some properties of the old and new states of charmonium and
QCD-based methods of study of these properties are discussed. An emphasis is
made on selected theoretical methods rather than on presenting the whole field
and reviewing the data. An all-inclusive presentation can be found in the much
larger review\cite{qwg} and an excellent update on the most recent data and the
related theoretical developments is given in Ref.\cite{egmr}.

The topics in the spectroscopy of the traditional charmonium states  are
provided in Section 2, and the annihilation of  and the radiative transitions
between these states are discussed in respectively Sections 3 and 4. The Section
5 is devoted to hadronic transitions between charmonium levels and the related
topic of the interaction of slow charmonium with hadronic matter. Finally, in
Section 6 are discussed peculiar properties of the resonances with masses above
the open charm threshold.
\begin{figure}[tb]
\begin{center}
\begin{minipage}[t]{18 cm}
\epsfxsize=13cm
\epsfbox{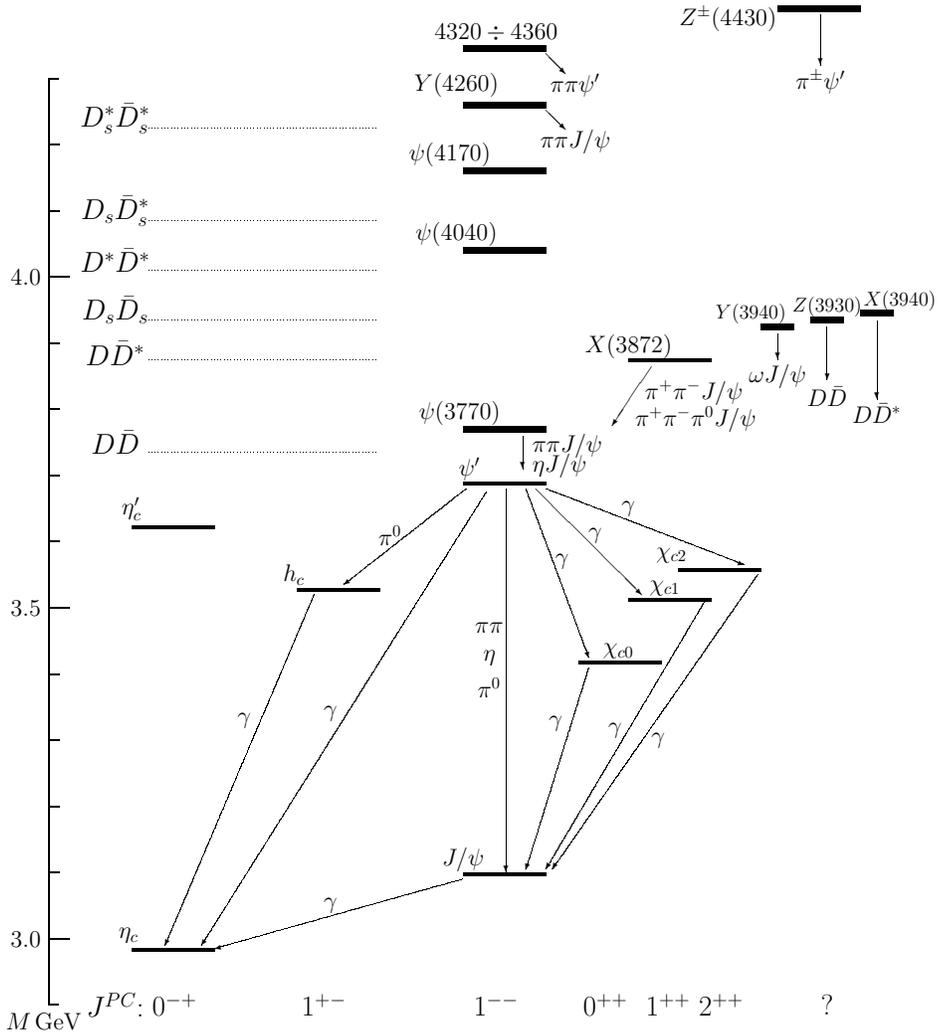}
\end{minipage}
\begin{minipage}[t]{16.5 cm}
\caption{The known charmonium and charmonium-related resonances and some
transitions between them. Also are shown (dotted lines) the thresholds for
various pairs of charmed mesons.\label{spectrum}}
\end{minipage}
\end{center}
\end{figure}

\section{The Spectrum of Charmonium States}

\subsection{\it General Considerations}
The diagram of the known charmonium and apparently charmonium-related states is
shown in Fig.\ref{spectrum}.
The quantum numbers and basic properties of most of  the  states in the
charmonium family can be described within a simple picture of a nonrelativistic
quark - antiquark pair $c {\bar c}$. In this picture the states are
characterized by the orbital angular momentum $L$, the total spin $S$ of the
quark pair, and the total angular momentum $J$, which defines the spin of the
state viewed as a particle. As usual, the total angular momentum is given by the
vector sum of the orbital and the spin momenta: ${\vec J}= {\vec L}+ {\vec S}$.
Likewise, the total spin $S$ is determined by the vector sum of the quark and
antiquark spins: ${\vec S}= {\vec s}_c+ {\vec s}_{\bar c}$. Clearly, $S$ takes
the values $0$ and $1$, thus splitting the four possible spin states of the pair
into a singlet and a triplet. Furthermore,  the excitation of the radial motion
of the $c {\bar c}$ pair results in a spectrum of levels with the same $L$, $S$
and $J$, and differing by the ``radial" quantum excitation number $n_r$ with
$n_r=0$ corresponding to the lowest state in this spectrum. It is therefore
customary to encode the values of these quantum numbers for each state of
charmonium in the form of the symbol $(n_r+1) ^{(2S+1)} L_J$. The combination
$2S+1$ conveniently indicates the spin multiplicity, while following the
tradition from atomic physics the values of $L$, $L=0,1,2,3,\ldots$ are written
as $S,P,D,F,\ldots$. In this notation the lowest state with $L=0, S=0$ and
(necessarily) $J=0$ is represented as $1^1S_0$ ($\eta_c$  resonance) while the
first excited state with the same quantum numbers is $2^1S_0$ ($\eta_c'$).

The value of $L$ determines the parity ($P$) for each of the states:
$P=(-1)^{L+1}$, while $L$ and $S$ combined also determine the charge conjugation
parity: $C=(-1)^{L+S}$. Therefore the previously mentioned $^1S_0$ states have
quantum numbers $J^{PC}=0^{-+}$, while e.g.  $^3S_1$ states have the same
quantum numbers $J^{PC}=1^{--}$ as the electromagnetic current, so that these
states ($J/\psi$, $\psi'$, \ldots) can be produced as resonances in $e^+e^-$
annihilation.

This briefly described standard nomenclature of the quark-antiquark states
originates in the strictly nonrelativistic mechanics. Relativistic effects in
the dynamics of quark and antiquark interacting with each other preserve most,
but not all of it. Indeed, the conservation of the total angular momentum
ensures that the states have definite $J$. On the contrary, the value of $L$ is
generally not preserved by the interaction. In particular, the operator of the
so-called tensor forces $S \, (S+1) - 3 ({\vec S} \cdot {\vec r}) ({\vec S}
\cdot {\vec r})/r^2$ does not commute with ${\vec L}^2$ for the states with
$S=1$, so that a mixing of states with different $L$ takes place. However the
parity conservation requires that only the states with the same parity of $L$
can get mixed, and the conservation of $J$ then implies that the values of $L$
for the mixed states can differ by at most two units. For instance a $^3S_1$
state can receive, due to relativistic effects and admixture of a $^3D_1$ state
(`$S-D$ mixing'). It can be further noticed that the mixing in $L$ is
necessarily absent for certain states. Indeed all states with $J=L$ are pure in
$L$, both the spin-singlets, $S=0$, and the spin-triplets, $S=1$, as are the
$^3P_0$ resonances.

The applicability, in a certain extent, of a nonrelativistic description to
charmonium has always been a source of interest to this system with the hope
that in it the quark dynamics can be studied being not overly complicated by the
relativistic effects. The significance of such effects in charmonium can be very
approximately estimated already from the masses of the resonances, e.g. the mass
difference $\Delta M$ between the ground $1^3S_1$ state ($J/\psi$ resonance) and
its first radial excitation  $2^3S_1$ ($\psi'$) in units of either of the masses
provides an estimate of the relativistic parameter $v^2/c^2$:
\be
{v^2 \over c^2} \sim {\Delta M \over M} \sim 0.2~.
\ee
Such moderate, but not very small magnitude of the relativistic effects in
charmonium in fact makes some of these effects visible in experiments and thus
further expands the range of dynamical details that can be studied in the
charmonium system.

\subsection{\it Potential Models}

\subsubsection{\it Leading Nonrelativistic Approximation}
One widely used approach to describing charmonium is to consider its dynamics in
analogy with atomic systems or positronium and to treat it in the
nonrelativistic limit by means of a Schr\"odinger equation with a potential
$V(r)$ depending on the distance $r$ between the quark and the antiquark. The
relativistic effects up to the order $v^2/c^2$ can then be considered as
perturbation due to relativistic terms in the potential as well as in the
kinetic energy. The shape of the potential at short distances is determined by
the perturbation theory in QCD. In the lowest order the exchange of (Coulomb)
gluons between slow quarks is fully analogous to interaction in QED, so that for
a color-singlet quark pair the interaction takes the Coulomb-like form:
\be
V_0(r)=-{4 \over 3} \, {\alpha_s \over r}~,
\label{v0}
\ee
where $\alpha_s$ is the QCD coupling. Once the scale dependence  of this
coupling is taken into account, the constant $\alpha_s$ in Eq.~(\ref{v0}) has to
be replaced by the running coupling $\alpha_s(r)$. At distances longer than the
charmed quark Compton wave length $1/m_c$, the one-loop running is described by
\be
\alpha_s(r)={2 \pi \over 9 \ln {1 \over r \,\Lambda_{QCD}} }~.
\label{asr}
\ee
In higher orders of the QCD loop expansion the precise relation of this
`Coulomb-like' coupling to the running constant defined in a specific
renormalization scheme, such as e.g. the $\overline {MS}$ scheme, is a matter of
calculations, which have been carried to the two-loop level\cite{peter,schroder}
with some partial results\cite{bpsv} in three loops.

The details of these fine calculations however are not of an immediate
significance for charmonium. The reason is that the perturbative QCD is
applicable only at short distances, which are much shorter than the typical
spatial size of the charmonium states. At the relevant intermediate and long
distances one has to resort to models for the interaction between quarks. Some
guidance in constructing such models is provided by the general idea of quark
confinement, which can be mimicked by a potential rising at long distances. The
most popular choice of the confining behavior is a linearly growing potential:
$V(r) = \sigma \, r$. Such behavior originates in the idea of the contraction of
the chromoelectric field between the quarks into a flux tube, giving a
string-like binding.

The interaction potential can also be studied as the energy of a static
infinitely heavy  quark - antiquark pair separated by the distance $r$. This
quantity can be evaluated in terms of the Wilson loop\cite{wilson} by lattice
QCD calculations. The numerical results of such analysis\cite{bali} produce a
dependence of the static energy on $r$, which is in agreement with the
Coulomb-like behavior at short distances and an approximately linearly rising
potential at larger $r$.

It should be noted however that unlike in QED, a potential approach to heavy
quarkonium in QCD is formally justified only in the limit of very high quark
mass: in tens to hundreds GeV. In this limit the low-lying bound states in the
short-distance potential (\ref{v0}) are localized at short distances, where the
perturbative potential description is applicable and the whole approach is thus
selfconsistent. Once nonperturbative effects in QCD are taken into account such
consistency becomes questionable. Already the leading
corrections\cite{mv79,mv82_1,leutwyler} to energies of the quarkonium levels in
the limit of very heavy quarks do not correspond to {\it any} potential between
the quark and the antiquark. The reason for such behavior can be readily
understood\cite{mv79}. Indeed, a potential implies an instantaneous interaction.
Any non-locality of the interaction in time would contain characteristic time
scales, that can be interpreted as the evolution time scales for additional
degrees of freedom. Once such additional degrees of freedom come into play the
system (quarkonium) can no longer be described by a potential, neither it can be
considered at all as a two-body system. In reality an interaction between the
quark and antiquark through a gluon field should necessarily invoke
nonperturbative light degrees of freedom in QCD whose typical evolution scale is
determined by the QCD infrared parameter $\Lambda_{QCD}$. The interaction
through exchange of such field can thus be viewed as instantaneous inasmuch as
the quark and antiquark are slow in this scale, i.e. the characteristic time of
evolution of the quarkonium wave function is long in this scale. For charmonium
one can estimate the time of evolution as an inverse of the typical energy
spacing between the levels, e.g. $M(\psi')-M(J/\psi) \approx 590\,$MeV, which by
any measure is certainly comparable with the QCD scale.

In other words, there in fact is no parameter that would justify a QCD-derived
description of charmonium or bottomonium as a two-body quark-antiquark system
interacting through a potential. However due to some numerical reasons, which
are yet to be understood, such simple picture works reasonably well, especially
if it is not pushed to requirements of high accuracy, or to highly excited
states of quarkonium. It is almost so that any smooth potential whose behavior
resembles Coulomb at short distances and an approximately linear rise at large
$r$ reasonably well describes the properties of the observed charmonium
resonances, after the parameters of the model are appropriately adjusted. Some
of the models for the potential discussed in the literature can be found in
Refs.~\cite{cornell,richardson,bt}.

One of  the most developed is the Cornell model\cite{cornell,elq,elq2} which
builds upon the simplest potential being just a sum of the Coulomb and linear
parts,
\be
V(r) = - {\kappa \over r} + {r \over a}~,
\label{vcorn}
\ee
and adding then finer effects, such as the relativistic terms, resulting in the
hyperfine and fine structures of charmonium levels, and also, importantly,
including the coupling of the $c {\bar c}$ system to pairs of charmed mesons,
such as $D {\bar D}$. The latter coupling effectively accounts for the fact that
above the open charm threshold the charmed quarks and antiquarks do emerge
inside the charmed hadrons, which effect is totally ignored in a pure potential
model with confining interaction.

Clearly, a leading nonrelativistic treatment can only describe gross features of
the charmonium levels, i.e. without resolving the fine splitting  between the
states with the same $L$ and $S$ and different $J$ and the hyperfine splitting
between the spin-triplet and spin-singlet states. It can be noted that even at
such approximate level of detail the resulting sequencing of the energies of the
levels with different $n_r$ and $L$ can provide useful constraints on the
properties of the potential $V(r)$\cite{martin}.

\subsubsection{\it Spin-dependent Forces}
The potential description extended to spin-dependent interactions results in
three types of interaction terms that are to be added to the discussed leading nonrelativistic interaction:
\be
V_1(r)= V_{LS}(r) \, ({\vec L} \cdot {\vec S}) + V_T(r) \, \left [ S \, (S+1) -{
3 ({\vec S} \cdot {\vec r}) ({\vec S} \cdot {\vec r}) \over r^2} \right ] +
V_{SS}(r) \, \left[ S (S+1) - {3 \over 2} \right ]~.
\label{vs}
\ee
The spin-orbit, $V_{LS}$, and the tensor, $V_T$, terms describe the fine
structure of the states, while the spin-spin term, $V_{SS}$, proportional to $2
({\vec s}_q \cdot {\vec s}_{\bar q}) = S (S+1)-3/2$ gives the spin-singlet -
triplet splittings. The interaction in Eq.(\ref{vs}) arises
among the $v^2/c^2$ effects in the nonrelativistic expansion and it generally
requires additional model-dependent assumptions about the structure of the
interquark forces. Within the phenomenological approach it is usually assumed
that parts of the static potential (similar to that in Eq.(\ref{vcorn}))
correspond to definite Lorentz structures of the relativistic interaction
between the quarks\cite{prs,schnitzer,6aut}. In other words, those structures
correspond to an ``exchange of something" with a definite spin between the quark
and the antiquark. Then the short-distance Coulomb-like part of the static
potential is naturally generalized as a limit of a vector type exchange:
$$ ({\bar u} \gamma^\mu u)({\bar v} \gamma_\mu v)\, V_V(q^2)$$
with $u$ and $v$ being the Dirac spinors for the quark and the antiquark, while
the confining part has been treated in the literature as a part of a vector
exchange\cite{prs,schnitzer}, or a scalar\cite{qwg,egmr}, or a mixture of
these\cite{efg}. With this choice of options restricted to a combination of only
vector and scalar exchange, the spin-dependent terms in Eq.(\ref{vs}) can be
written in terms of the vector, $V_V(r)$, and scalar, $V_S(r)$, parts of the
static potential by the standard Breit-Fermi expansion to order
$v^2/c^2$\cite{blp}:
\be
V_{LS}= {1 \over 2 m_c^2 r} \, \left (3 {d V_V \over dr}- {d V_S \over dr} \right
)~,
\label{vls}
\ee
where $m$ is the charmed quark mass,
\be
V_T={1 \over 6 m_c^2} \, \left ( {d^2 V_V \over dr^2} - {1 \over r} \, {d V_V
\over dr} \right )~,
\label{vt}
\ee
and
\be
V_{SS}={1 \over 3 m_c^2} \, \Delta V_V~,
\label{vss}
\ee
with $\Delta = \nabla^2$ being the three-dimensional Laplacian.

It should be emphasized that by its nature the interaction in Eq.(\ref{vs}) is a
part of the $v^2/c^2$ term in the nonrelativistic expansion. As such it can be
used only in the first order, and by no means one should iterate this potential,
or use it for an input in the Schr\"odinger equation. Therefore any accuracy of
the results found with potential of this type is intrinsically limited.
Furthermore, the formulas (\ref{vls}) - (\ref{vss}) are only approximate even in
perturbative QCD at short distances. In particular, the tree-level QCD potential
(\ref{v0}), results, according to Eq.(\ref{vss}), in a point-like spin-spin
interaction:
\be
V_{SS}={16 \, \pi \, \alpha_s \over 9 \, m^2} \, \delta^{(3)}(\vec r)~,
\label{vss0}
\ee
which is the correct tree-level expression for the corresponding interaction in
QCD. If one then improves the static potential by using the running coupling
from Eq.(\ref{asr}), the formula (\ref{vss}) would produce terms extending to
finite $r$ and behaving as $\alpha_s^2/r^3$. A real calculation\cite{bnt} of the
one-loop correction to the spin-spin interaction, however produces no such
terms, and predicts that the hyperfine splitting in perturbative QCD is still
proportional to the square of the wave function at the origin, $|\psi(0)|^2$.

Clearly, the point-like behavior of the spin-spin interaction can generally be
invalidated in higher orders in perturbative QCD and also by nonperturbative
dynamics. Moreover, the leading nonperturbative effects in the limit of
asymptotically heavy quarkonium\cite{mv82} are not reduced to a point-like
spin-spin interaction. Nevertheless, inspite of these reservations, the actual
hyperfine splitting in charmonium closely resembles that produced by a
short-distance interaction. Namely,  the proportionality of the hyperfine
splitting to $|\psi(0)|^2$ implies that the mass gap between the $^3S_1$ and
$^1S_0$ states should be proportional to the $e^+e^-$ decay width of the vector
$^3S_1$ resonance, while the hyperfine splitting in the $P$ wave should be
extremely small because of vanishing wave function at the origin. In
phenomenological terms this implies the following relations:
\be
{M(\psi')-M(\eta_c') \over M(J/\psi) - M(\eta_c)} \approx {\Gamma_{ee}(\psi')
\over \Gamma_{ee}(J/\psi)}~.
\label{dmgee}
\ee
and
\be
M(h_c) \approx {\overline M}(\chi_{cJ})~.
\label{pmsplit}
\ee
where ${\overline M}(\chi_{cJ})= [5 M(\chi_{c2})+ 3 M(\chi_{c1})+M(\chi_{c2})]/9
$ is the `center of gravity' of the $^3P_J$ states which is not shifted by
either the spin-orbital or the tensor interactions from Eq.(\ref{vs}).
According to the Tables\cite{pdg} the ratio of the $e^+e^-$ decay rates in the
r.h.s. of Eq.(\ref{dmgee}) is $0.45 \pm 0.02$, while the ratio of the mass
splittings in the l.h.s. is $0.44 \pm 0.04$. Furthermore, the center of gravity
of the spin-triplet $\chi_{cJ}$ states is at ${\overline M}(\chi_{cJ})= 3525.36
\pm 0.06 \,$MeV. The spin-singlet $^1P_1$ state, the $h_c$, has been
sighted\cite{e760} as a resonance in the $p {\bar p}$ annihilation with the mass
$3526.28 \pm 0.18 \pm 0.19\,$MeV, then in the same process\cite{e835} at the
mass $3525.8 \pm 0.2 \pm 0.2\,$MeV, and eventually\cite{cleohc} in the decays
$\psi' \to \pi^0 \, h_c$ at $3524 \pm 0.6 \pm 0.4\,$MeV, with the most recent
improvement in the precision yielding\cite{seth_talk} the $h_c$ mass of 
$3525.35 \pm 0.19 \pm 0.15\,$MeV. The data with the smallest claimed
experimental
errors point at an extremely small violation, if any, of the relation
(\ref{pmsplit}): $M(h_c) - {\overline M}(\chi_{cJ}) = -0.05 \pm 0.19 \pm
0.16\,$MeV\cite{seth_talk}.
Thus the simple relations (\ref{dmgee}) and (\ref{pmsplit}) both hold amazingly
well.
In fact it is still a challenge for  experiment to measure the violation of
these relations, and even a greater challenge to correctly predict such
violation theoretically.

The spin-orbit and tensor terms in Eq.(\ref{vs}) produce the fine structure of
the charmonium levels, and the tensor term also gives rise to mixing of states
with $L$ differing by two units, such as $^3S_1- {^3D_1}$ mixing. The
phenomenological effects of the mixing are somewhat more subtle and will be
discussed further in Sec.3.1.2. Here we concentrate on the fine structure of the
$^3P_J$ states $\chi_{cJ}$. The shifts of the masses of the states with
different $J$ with respect to ${\overline M}(\chi_{cJ})$ are given in terms of the averages $<V_{LS}>$ and $<V_T>$ over the
$P$ wave coordinate wave function as
\be
\delta M(^3P_0)=-2 \langle V_{LS} \rangle + 2 \langle V_T \rangle~,~~~\delta
M(^3P_1)=- \langle V_{LS} \rangle - \langle V_T \rangle~,~~~\delta M(^3P_2)=
\langle V_{LS} \rangle + {1 \over 5} \,  \langle V_T \rangle~.
\label{pshifts}
\ee
Using the measured differences between the masses of the $\chi_{cJ}$ charmonium
resonances\cite{pdg} one can find the average values over the $1P$ charmonium:
\bea
 \langle V_{LS} \rangle &=& {1 \over 12} \, \left [ 5\, M(\chi_{c2})-3\,
M(\chi_{c1})-2 \, M(\chi_{c0}) \right] \approx 35\,{\rm MeV}~, \nonumber \\
 \langle V_{T} \rangle &=& {5 \over 36} \, \left [  M(\chi_{c2})-3 \,
M(\chi_{c1})+ 2 \, M(\chi_{c0}) \right] \approx -20\,{\rm MeV}~.
\label{vlst}
\eea
This estimate illustrates that in order to describe the observed mass splitting
between the $\chi_{cJ}$ states both the LS and tensor interactions are required
with comparable strength. Furthermore, it can be noted that these forces,
generated by a pure tree-level gluon exchange, as can be found from using the
Coulomb-like potential $V_V$ in the formulas (\ref{vls}) and (\ref{vt}), have
correct signs, in agreement with the estimate (\ref{vlst}), but with a relative
strength of the LS interaction enhanced in comparison with this estimate:
$V_{LS}/V_T|_{Coul}=-3$. Thus a certain reduction in the spin-orbit term due to
a contribution of the Lorentz scalar potential $V_S$ in Eq.(\ref{vls}) is indeed
helpful from this point of view.

\subsubsection{\it Potential Models and Predictions for New States}

Once the parameters of a specific potential model are fixed from the data on the
known states of charmonium, it is natural to use the same approach for
predicting the masses of yet unobserved resonances corresponding to higher
energy levels in the system.  A large number of such predictions can be found in
the literature spanning last three decades. Some of the results can be found in
the papers \cite{cornell,elq,elq2,efg,gi,fulcher,zvor} to name a few.  The
predictions for masses of the excited states considerably differ between models,
which is not surprising in view of the nature of the approach. Moreover, given
that the considered models do not have a controllable accuracy, it would be
troublesome to assess what deviation of a prediction from the actual value of
the mass should be regarded as successful.

Nevertheless, such application of the models appears far from being an entirely
empty exercise, and in a way provides some gross features of the expected
spectrum of higher excitations. Namely all the `reasonable' models predict the
same sequencing of levels: $M(1D) < M(2P) < M(3S)$, which can possibly be traced
to the general properties\cite{martin}, and the specific masses being generally
in the following ranges: $M(1D) \approx 3.8 - 3.9\,$GeV, $M(2P) \approx 3.9 -
4.0\,$GeV, and $M(3S) > 4.0\,$GeV. The fine and hyperfine splittings of the
levels are smaller than the uncertainty in the overall positions of the levels.
All these states are above the $D {\bar D}$ threshold, so that most of them are
expected to be broad due to decay into pairs of $D$ mesons. The exception from
this behavior can be found in the $^3D_2$ ($2^{--}$) and $^1D_2$ ($2^{-+}$)
resonances if they are below the $D {\bar D}^*$ threshold at $3872\,$MeV. Indeed
the unnatural spin-parity of these resonances forbids them to decay in the $D
{\bar D}$ pairs, the only kinematically allowed states with open charm at such
mass. The $2P_1$ and $3^1S_0$ states are expected to be well above the $D {\bar
D}^*$ threshold and for them such argument does not work.

The issue of higher excitation of charmonium has recently gained a great
attention due to observation of a whole `zoo' of new charmonium-like states in
experiment. In particular those, which seem to relatively well fit the expected
pattern of the levels are the resonances $Z(3930)$ and $Y(3940)$. The former
state is found by Belle\cite{z3930} in $\gamma \gamma$ production with the mass
and width $M(Z)=3929 \pm 5 \pm 2\,$MeV and $\Gamma(Z)=29 \pm 10 \pm 2\,$MeV
decaying mostly to $D {\bar D}$, and which reasonably fits\cite{elq2} the slot
for the $2^3P_2$ ($\chi_{c2}'$) state of charmonium. The latter resonance,
$Y(3940)$, observed by Belle\cite{y3940} in the $\omega J/\psi$ channel in the
decays $B \to \omega J/\psi K$ has the parameters $M(Y)=3943 \pm 11 \pm 13\,$MeV
and $\Gamma(Y)=87 \pm 22 \pm 26\,$MeV. It appears to not decay into the pairs of
pseudoscalar mesons $D {\bar D}$ and can be considered\cite{barnes} as a
candidate for the $2^3P_1$ ($\chi_{c1}'$) state of charmonium. If this
interpretation proves to be correct an interesting spectroscopic question would
arise related to the apparently inverted or small mass splitting between the
$2^3P_1$ and $^3P_2$ states. Most recently the peak in the invariant mass of the
system $\omega J/\psi$ possibly consistent with $Y(3940)$ was observed by
BaBar\cite{y3940babar} in the decays $B^+ \to \omega J/\psi K^+$ and $B^0 \to
\omega J/\psi K_S$. The values for the mass and width of the observed peak are
somewhat off compared to the initial observation: $M(Y)=3914^{+3.8}_{-3.4} \pm
1.9\,$MeV and $\Gamma(Y)=33^{+12}_{-8} \pm 5\,$MeV. Thus the status of this
resonance is still not clear. The suggested interpretations include an excited
$P$ wave quarkonium\cite{egmr}, hybrid $c\bar c g$ state\cite{close07}, and a
four-quark molecular state.

Another recently found resonance $X(3940)$ is observed\cite{x3940} as recoiling
against $J/\psi$ in $e^+ e^- \to J/\psi + X$, which implies that its $C$ parity
is positive. Furthermore it appears to decay into $D {\bar D}^*$ but not into $D
{\bar D}$, so that it likely has unnatural spin-parity. These properties invite
an interpretation\cite{elq2,barnes} of the resonance as a $0^{-+}$ charmonium
state which would then be $\eta_c(3S)$. A possible problem of such
interpretation is that most of the models expect the $3S$ level in charmonium to
be somewhat above $4.0\,$GeV, so that if further study of $X(3940)$ indeed
identifies it as a $0^{-+}$ resonance, this may bring some new interesting
understanding of hadron dynamics.

\subsection{\it Spectral Methods}

The potential models of heavy quarkonia and of charmonium in particular, are
intuitively appealing, versatile and very convenient for estimates of various
characteristics of the heavy resonances, but they can not be entirely
satisfactory due to their model-dependent relation to the underlying theory of
QCD. More directly related to the first principles of QCD are the methods based
on the spectral relations for correlators in QCD. Such approach can be
illustrated in its most basic form by considering the correlation function of
the type $F(x)= \langle 0 | T \, \{O^{\dagger}(x),O(0)\}|0 \rangle$, where
$O(x)$ is a local operator and  $|0 \rangle$ is the vacuum state in QCD. Of
relevance to charmonium is the choice of the operator $O(x)$ where it contains a
factor ${\bar c} \Gamma c$ (with some structure $\Gamma$) and thus produces
states of charmonium. The correlation function can be written in terms of the
spectral sum over the physical states $|n \rangle$ containing a $c {\bar c}$
pair:
\be
F(x)= \sum_{n} \, |\langle n | O | 0 \rangle|^2 D_n(x)~,
\label{srep}
\ee
where $D_n(x)$ is the propagator of the state $n$. The lowest mass states
contributing to the sum are the one-particle states i.e. the charmonium
resonances, while at higher mass the sum is also contributed by the continuum of
states containing the charm - anticharm quark pair. By an appropriate choice of
the operator $O(x)$ one projects out the states with particular quantum numbers,
while a suitable choice of $x$ allows to make the sum being dominated by the
charmonium resonances of interest. The direct relation to `the first principles'
arises when the correlator $F$ can be also calculated in the interesting range
of $x$ by methods of the underlying QCD theory, thereby relating the
phenomenological properties of hadrons to the results of a QCD calculation.

The two approaches to calculating the correlator $F$ are the numerical
calculations in lattice QCD and the short-distance QCD analytical treatment. The
lattice approach in principle allows to evaluate the correlator at large
Euclidean separation $x$ where the spectral sum is given by only the lowest mass
state, so that e.g. the mass of this state can be fully determined. On the other
hand, the short-distance QCD methods are restricted to relatively small values
of the interval $x$, so that the spectral sum still contains some contribution
from higher states as well as the lowest one. For this reason the relations
resulting from calculations of this type are known as the QCD sum rules. The
usefulness of the sum rules for description of the lowest state in a given
channel depends on the existence of an intermediate range of a parameter
analogous to $x$, where both the theoretical uncertainty in a short-distance
calculation and the phenomenological uncertainty of the contribution of the
higher mass states to the spectral sum can be reasonably controlled.

\subsubsection{\it QCD Sum Rules}

It is due to the $e^+e^-$ annihilation data that the most well studied channel
with hidden charm is the vector one, i.e. corresponding to the charm - anticharm
production by the electromagnetic current of the charmed quarks $j_\mu = ({\bar
c} \gamma_\mu c)$, and this is the channel for which the original QCD sum rules
were developed\cite{6sr}. The relevant correlator is then the vacuum
polarization $P(q^2)$, considered in the momentum, rather than the position
space:
\be
P(q^2) \, \left ( - q^2 \, g_{\mu \nu} + q_\mu q_\nu \right )= i \, \int \, d^4
x \, e^{iqx} \, \left \langle 0 \left | T \left \{ j_\mu(x), j_\nu(0) \right \}
\right | 0 \right \rangle~.
\label{pdef}
\ee
The spectral sum then takes the form of the dispersion relation for $P(q^2)$,
\be
P(q^2)={q^2 \over \pi} \, \int {{\rm Im} P(s) \over s \, (s-q^2-i \epsilon)}
ds~,
\label{disp}
\ee
and the imaginary part Im$P(s)$ is related to the contribution of the states
with hidden charm $R_c(s)$ to the measured cross section ratio
$R(s)=\sigma(e^+e^- \to hadrons)/(4 \pi \alpha^2/3s)$:
\be
{\rm Im} P(s) = {R_c(s) \over 12 \pi}~.
\label{imrp}
\ee

At values of $q^2$ sufficiently below the (perturbative) threshold at $4 m_c^2$
the vacuum polarization $P(q^2)$ is determined by the QCD dynamics at short
distances and can be calculated by using the Operator Product Expansion (OPE)
for the $T$ product in Eq.(\ref{pdef}),
\be
T \left \{ j(x), j(0) \right \} = \sum_d c_d(x) \, {\cal O}_d(0)
\label{ope}
\ee
in terms of local operators  ${\cal O}_d(0)$ with increasing dimension $d$, and
$c_d(x)$ being the coefficient functions calculable in QCD. The leading operator
of lowest dimension in this expansion is the unit operator ${\cal I}$, and the
corresponding coefficient $c_0(x)$ includes all the QCD perturbation theory
result for $P(q^2)$. The first nontrivial operator in the series is the next one
with dimension $d=4$ and is quadratic in the gluon field strength tensor $G_{\mu
\nu}^a$, so that its contribution to the vacuum polarization is proportional to
the gluon vacuum condensate\cite{svvz,svz0,6aut,svz}. The relation between the
theoretical expression for $P(q^2)$ and the phenomenological integral over the
observed cross section can be studied as a function of $q^2$ far below the
threshold. Alternatively, one can compare the expressions for the derivatives of
the vacuum polarization with respect to $q^2$ at $q^2=0$, for which the
`phenomenological' side of the sum rules is given by the moments of the ratio
$R_c$:
\be
{\cal M}_n = \int \, {R_c(s) \over s^{n+1}} \, ds~,~~~~~n=1,2,\ldots~,
\label{moms}
\ee
${\cal M}_n =(12 \pi^2/n!)\, (d^n P(z)/dz^n)|_{z \equiv q^2=0}$. Theoretically,
the Taylor expansion of the vacuum polarization including the leading
nonperturbative term can be written as
\be
P(q^2)= \sum_{n=1} \, \left ( {\cal C}_n + {\cal D}_n {\langle 0| G^2 |0 \rangle
\over m_c^4} \right ) \, \left ( {q^2 \over 4 m_c^2} \right )^n~,
\label{ptay}
\ee
where ${\cal C}_n$ and ${\cal D}_n$ are dimensionless coefficients that are
calculated as series in powers of $\alpha_s$. The coefficients ${\cal C}_n$ are
known\cite{cks} at arbitrary $n$ up to $\alpha_s^2$, and ${\cal C}_1$ has been
recently evaluated\cite{cks0,bcs} to order $\alpha_s^3$. The coefficients ${\cal
D}_n$ are known in the lowest\cite{svvz,6aut} and the next to lowest\cite{istc}
orders in $\alpha_s$.

The contribution of a narrow resonance to $R_c$ can be approximated by a
$\delta$ function:
\be
R_c(s)={9 \pi \over \alpha^2} \, \delta(s-M^2) \, \Gamma_{ee} \, M~,
\label{rres}
\ee
where $M$ is the mass of the resonance and $\Gamma_{ee}$ is the width of its
decay into $e^+e^-$. The lowest charmonium state contributing to the dispersion
integral in Eq.(\ref{disp}) is the $J/\psi$ resonance, so that the $n-$th moment
can be written as
\be
{\cal M}_n = {9 \pi \over \alpha^2} \, \left [ {\Gamma_{ee}(J/\psi) \over
M^{2n+1}(J/\psi)}+ {\Gamma_{ee}(\psi') \over M^{2n+1}(\psi')} \right]+ \int_{s >
M^2(\psi')} \, {R_c(s) \over s^{n+1}} \, ds~,
\label{momph}
\ee
where the latter integral runs over c.m. energies above the $\psi'$ resonance.
One can readily see that the relative weight of the lowest resonance grows with
the number of the moment $n$. (E.g. the $J/\psi$ contribution essentially
dominates the moments of charm cross section already at $n=3 \div 4$.) However
the
correction terms, both the perturbative and nonperturbative, in the theoretical
calculation of the moments also grow with $n$, which is a reflection of the fact
that higher derivatives are sensitive to the threshold singularity in the
correlator where the essential distances are no longer short. In particular the
parameter for the perturbative expansion for the moments is effectively
$\alpha_s\, \sqrt{n}$ (corresponding to the $\alpha_s/v$ behavior of effects of
the Coulomb-like interaction near the threshold), and for the leading
nonperturbative term due to the gluon condensate behaves as $n^3 \, \langle G^2
\rangle/m_c^4$. It was found nevertheless\cite{6sr,6aut} that the perturbative
one-loop expression for the moments can be relied on up to $n \approx 4$, and
somewhat higher moments (up to $n=7 \div 8$) then can be used in order to
determine\cite{svvz,svz0} the value of the gluon vacuum condensate $\langle G^2
\rangle$. Once other relevant QCD parameters ($\alpha_s$, the charmed quark mass
$m_c$) were determined from the data and the sum rules for the vector channel,
they could be used for other channels as well. In particular, the sum rules for
the spectral density of the pseudoscalar operator $({\bar c}\gamma_5 c)$
correctly predicted\cite{svvz} the mass of the lowest $0^{-+}$ charmonium
resonance, the $\eta_c$.

The best theoretically known is the first moment of $R_c(s)$, due to the
accuracy ($\alpha_s^3$) of the available perturbative calculation and due to
very small nonperturbative term proportional to $\langle G^2 \rangle$. On the
experimental side this moment is rather sensitive to the details of the charm
production cross section in $e^+e^-$ annihilation at around $4.0\,$GeV where the
open charm production sets in, and the energy behavior of the cross section is
quite complicated. Recent data in this region \cite{bes2,bes6} have allowed to
put to use the attained theoretical accuracy in ${\cal M}_1$. In recent analyses
the sum rule for the first charm moment was used for precision determination of
the short-distance charmed quark mass parameter, ${\overline m}_c({\overline
m}_c)$: $1295 \pm 15\,$MeV\cite{bcs} and $1286 \pm 13\,$MeV\cite{kss}.

\subsubsection{\it Lattice Methods and Limitations of the Spectral Approach}
The QCD sum rule approach, based on the OPE and analytical calculations is
certainly limited by the applicability of the perturbative expansion in
$\alpha_s$. This requires a careful choice of parameters, e.g. the number of the
moment $n$,  to ensure such applicability and still get a phenomenologically
useful relation. It is widely believed that a ticket to calculations beyond the
perturbation theory in $\alpha_s$ that are not limited to short-distance QCD is
the numerical lattice approach. In particular, a natural application of this
approach would be a calculation of correlators at large separation between the
points in the Euclidean space, so that the contribution of the lowest states in
each channel could be completely determined. However practical implementations
of lattice calculations to heavy quarkonium including charmonium still run into
difficulties of their own. A detailed discussion of recent developments of the
lattice methods and of the associated difficulties can be found in the
review\cite{qwg}. As of this writing the accuracy of the lattice results even
for the masses of the lowest states of charmonium in each $J^{PC}$ channel
leaves an ample room for further improvement.

The spectral approach in general, based on analytical or numerical calculations,
has `built in' a certain deficiency with regards to excited states. Namely the
spectral sum for a correlator at a Euclidean separation necessarily receives the
largest contribution from the lowest physical state. By increasing the
separation one can at best enhance the sensitivity to the properties of the
lowest state and thereby evaluate those properties. There is however no simple
way of `focusing' spectral relations on excited states, so that a study of those
states by spectral methods runs into additional difficulties.

Furthermore, even at the level of sum rules, or of a study of the lowest
charmonium states in each channel by lattice methods, the spectral approach
eventually runs into limits of its accuracy arising due to annihilation of
charmonium into light hadrons. This limitation can also be viewed as a version
of the problem of excited states: the lowest state of charmonium in a channel
with given quantum numbers $J^{PC}$ is certainly not the lowest hadronic state
in that channel. In other words, at some level an operator $({\bar c} \Gamma c)$
produces states containing only light quarks and gluons and no charmed quarks.
The old phenomenological rule, according to which a mixing of hidden quark
flavors, although not forbidden by any conservation laws, is dynamically
suppressed, is traditionally referred to as the Okubo\cite{okubo}-Zweig\cite{zweig}-Iizuka\cite{iizuka} (OZI) rule,
so that the discussed effect is precisely the violation of this rule. The
significance of this effect generally depends on the channel considered, and as
also will be mentioned in the Section 3.2.3, it can be essential in the
properties of the $\eta_c$ resonance. In the considered above case of the vector
channel this effect is likely to be quite small, fully in line with the known
strong suppression of OZI violation in vector mesons. Indeed, the partial width
of the $J/\psi$ resonance associated with strong annihilation of the hidden
charm into light hadrons is only about $70\,$keV, which can be viewed as the
imaginary part of the shift of the mass of the resonance due to the OZI
violation. Generally the real part of the shift is of the same order as the
imaginary, which for the $J/\psi$ meson would be much smaller than any current
accuracy of a theoretical calculation of its mass.

\begin{figure}[tb]
\begin{center}
\begin{minipage}[t]{18 cm}
\epsfxsize=17cm
\epsfbox{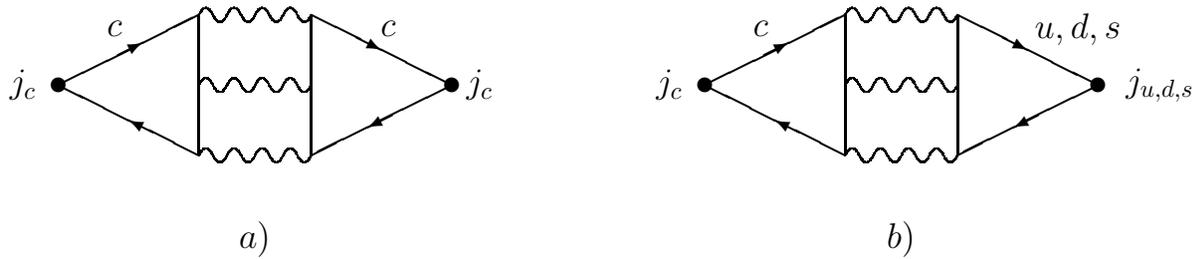}
\end{minipage}
\begin{minipage}[t]{16.5 cm}
\caption{Representative graphs for OZI rule violation in the vacuum polarization
by the electromagnetic current $j_c$ of charmed quarks ($a$) and for the
interference between the charmed quark current and that of the light quarks
($b$). The wavy lines show the gluons.\label{cmix}}
\end{minipage}
\end{center}
\end{figure}

It can however be noted that the already achieved accuracy in the first moment
of $R_c(s)$ is actually on the verge of being sensitive to the effects of the
OZI rule violation. Indeed, the separation of the hidden charm from light
degrees of freedom in the electromagnetic production is broken in order
$\alpha_s^3$ due to the mechanisms illustrated in Fig.\ref{cmix}.
The  theoretical calculation of the first moment ${\cal M}_1$ happens to be
insensitive\cite{gp1,gp2} to the contribution of the three-gluon intermediate
states shown in Fig.\ref{cmix}a, while the cross term between the currents of
the light and charmed quarks, shown in Fig.\ref{cmix}b does not enter the
correlator in Eq.(\ref{pdef}) by definition. The absence of the contribution
from the graphs of the type shown in  Fig.\ref{cmix}a in ${\cal M}_1$ (as well
as in ${\cal M}_2$ and ${\cal M}_3$) can be assured by the following reasoning.
Consider this mechanism at small $q^2$, so that $q^2 \ll m_c^2$. In this region
the charmed quark loop can be contracted into a point thus generating an
effective point-like Lagrangian for coupling of the current to three gluons.
Similarly to the well-known Heisenberg-Euler Lagrangian for the photons, the
current conservation and the QCD gauge invariance ensure that the mass of the
fermion in the loop enters as $1/m_c^4$. The graph of Fig.\ref{cmix}a contains
two such heavy quark loops, so that the low-energy expansion of the contribution
of the discussed mechanism to the correlator of the currents starts as
$\alpha_s^3 \, (q^2/m_c^2)^4 \, \ln (q^2/m_c^2)$, and therefore no contribution
to the first three moments ${\cal M}$ results from this mechanism.

The cross-talk between the electromagnetic currents of the light and charmed
quarks illustrated in Fig.\ref{cmix}b in fact relates to the general problem of
phenomenological separation of the hidden-charm and light-quark contribution
both to the measured cross section and to the sources of the observed final
states. The former separation of the hidden-charm part from the total cross
section is usually done by subtracting from the data the smooth background
associated with the light states, while the latter separation of the sources
corresponds to evaluating how much of light states are produced by the current
of charmed quarks  and how much of the hidden charm is produced by the
electromagnetic current of the light quarks. The cross-talk effect, however is
suppressed by the flavor $SU(3)$ symmetry. Indeed, the electromagnetic current
of the light quarks is a pure component of $SU(3)_{fl}$ octet, so that in the
limit of exact symmetry this current does not produce the heavy quarks, which
are $SU(3)_{fl}$ singlets. This effect is further suppressed in the first moment
${\cal M}_1$ (in this case only in the first moment). The reasoning is similar
to the one regarding Fig.\ref{cmix}a. In this case there is only one heavy quark
loop proportional to $1/m_c^4$. Thus the low-energy expansion of the total sum
of the graphs of the type shown in Fig.\ref{cmix}b necessarily starts with
$(q^2/m_c^2)^2 \, \ln (q^2/m_c^2)$, giving no contribution to ${\cal M}_1$.

It can be also mentioned that in the vector channel the mixing of light hadrons
and the hidden charm starts in the order $\alpha_s^3$ due to the $C$ parity. In
the $C$-even channels such mixing starts in the order $\alpha_s^2$ (except for
the $J=1$ channels, where it is forbidden by the Landau-Yang
theorem\cite{landau,yang,6aut}), so that the OZI violating effects should be
larger in the spectroscopy of these channels. Furthermore, the $J=0$ channels
$0^{-+}$ and $0^{++}$ generally suffer from large quarks-glue mixing effects,
related to direct instantons\cite{alike}, which may hold the clues to
understanding some properties of the $\eta_c$ and $\chi_0$ resonances.

\section{Charmonium Annihilation}

The same OZI violating effects that  are complicating an improvement in the
precision of calculation of the masses of charmonium resonances are also the
sole reason for the eventual disappearance of the hidden charm, i.e. for
annihilation of the $c {\bar c}$ quark pair into light states. The two
interactions contributing to the annihilation decays are the strong interaction
responsible for most processes of this type, and the electromagnetic interaction
which is relevant to annihilation decays of the vector $1^{--}$ states and the
$2\gamma$ decay widths of the $C$ even states with $J \neq 1$. We first consider
in this section the electromagnetic processes.

\subsection{\it Electromagnetic Annihilation}
\subsubsection{\it Annihilation of $^3S_1$ States through Virtual Photon}
The $^3S_1$ states have quantum numbers of a virtual photon, $J^{PC}=1^{--}$ and
can annihilate into lepton pairs or light hadrons through one photon. This is
also the process, which when reversed, gives rise to the formation of the
$^3S_1$ states as resonances in $e^+e^-$ annihilation. The rate of the decay can
be estimated in the extreme-nonrelativistic picture, where the system is
described by the wave function $\psi({\vec r})$ for the quark-antiquark pair and
depending on their relative position ${\vec r}={\vec r}_c-{\vec r}_{\bar c}$.
The annihilation takes place at the characteristic distances of order $1/m_c$
which are to be viewed as $r \to 0$ for a nonrelativistic pair, so that the
annihilation amplitude is proportional to the wave function at the origin. For
an n-th $S$-wave state the wave function can be written entirely in terms of its
radial part $R_{nS}$: $\psi_{nS}({\vec r})= R_{nS}/\sqrt{4\pi}$, and the
expression for the rate of annihilation into $e^+e^-$ takes the form
\be
\Gamma_{ee}(n ^3S_1)= {4 \, \alpha^2 \, e_c^2 \over M^2} \, |R_{nS}(0)|^2 \,
\left ( 1 - {16  \alpha_s \over 3 \pi} \right )~,
\label{geer}
\ee
where $e_c=2/3$ is the electric charge of the charmed quark in units of the
fundamental charge, $M$ is the mass of charmonium, and finally, the term with
$\alpha_s$ in the parenthesis gives the first QCD correction. The coupling
constants $\alpha_s$ and $\alpha$ should be taken at the scale $m_c$ (or $M$) as
appropriate for a process proceeding at distances $\sim 1/m_c$. The correction
(sometimes forgotten) due to the running of the QED coupling is reasonably
tractable and results in approximately 7\% enhancement of the rate, while the
first QCD correction is quite discouraging. Indeed its numerical value, $-1.7
\alpha_s$,  amounts to $-(0.35 \div 0.5)$ for $\alpha_s(m_c)$ in the realistic
range $0.2 \div 0.3$, which tells us that higher QCD corrections can be quite
essential\cite{py}.

There are however other uncertainties of a conceptual nature associated with the
formula (\ref{geer}). One source of such uncertainties arises from the use of
the nonrelativistic (essentially static) approximation. One of the indications
of this approximation is the parameter $M$: ``the mass of charmonium" entering
Eq.(\ref{geer}). It would be impossible to specify at this level of
approximation whether $M$ should be set at twice the quark mass $2 m_c$, or the
mass of the specific decaying state, or some combination of those, since the
difference between these values for $M$ is formally of order $v^2/c^2$. Clearly,
the first relativistic correction can amount to tens percent. It should be noted
that the specific form of this correction does not come from  purely kinematical
effects in the annihilation of moving, as opposed to static, quarks, but is also
sensitive to the interaction, since the average kinetic energy is of the order
of the average interaction potential as follows from the virial theorem. Another
fundamental uncertainty of Eq.(\ref{geer}) is related to the assumption that the
quarkonium can be described by a quark-antiquark two-body wave function, which
assumption is formally not valid in nonperturbative QCD, as was discussed in
Sec.2.2.1. The nonperturbative correction to the width $\Gamma_{ee}$ of the
$^3S_1$ states has been calculated\cite{mv82_1} in the limit of very heavy
quarkonium, where the correction is parametrically small. An extrapolation down
to charmonium would produce an unreasonably large result\cite{py}, neither it
would be justified. Thus the issue of the applicability of a description of
charmonium as a two-body system arises again as previously in the general
discussion of potential models. If in spite of this issue potential models are
used in conjunction with Eq.(\ref{geer}) and similar formulas for other
annihilation rates that will be discussed in this section, the results are in a
qualitative agreement with the observed pattern of the decay rates, although no
controllable accuracy can be assigned to such results.

The electromagnetic annihilation contributes a sizable fraction, about a
quarter, of the total decay width of the $J/\psi$ resonance. The latest precise
data\cite{cleo_ee} have moved the world average\cite{pdg} for the $e^+e^-$
branching ratio to ${\cal B}_{ee}(J/\psi)=(5.94 \pm 0.06)\%$. The decay into
$\mu^+\mu^-$ goes with essentially the same rate, and the rate of
electromagnetic annihilation into light hadrons is given by the $e^+e^-$ rate
scaled by the ratio $R$ measured in $e^+e^-$ annihilation just below the
$J/\psi$ resonance\cite{bes2}: $R(3.0\,{\rm GeV})=2.21 \pm 0.05 \pm 0.11$, thus
giving the total branching fraction for the decays of $J/\psi$ through a virtual
photon as
\be
{\cal B}(J/\psi \to \gamma^* \to X)=(2+R) \, {\cal B}_{ee}(J/\psi)=(25.0 \pm
0.8)\%~.
\label{bgstar}
\ee

For the $\psi'$ resonance the contribution of the electromagnetic annihilation
is by far less prominent (partly due to larger total decay width), given that
${\cal B}_{ee}(\psi')=(7.35 \pm 0.18) \times 10^{-3}$ and using\cite{bes2}
$R(3.7\,{\rm GeV})=2.23 \pm 0.08 \pm 0.08$, one can estimate
\be
{\cal B}(\psi' \to \gamma^* \to X)=(2+0.39+R) \, {\cal B}_{ee}(\psi') = (3.39
\pm 0.11)\%~,
\label{bgstarp}
\ee
where the term 0.39 in the parenthesis accounts for the decay $\psi' \to \tau^+
\tau^-$. The theoretical ratio of the rate of this decay to $\Gamma_{ee}(\psi')$
is in agreement with the direct measurement\cite{pdg}, but has about ten times
smaller error. In terms of potential models the ratio of the $e^+e^-$ decay
rates for $\psi'$ and $J/\psi$ can serve according to Eq.(\ref{geer}) as a
measure of the ratio of the wave functions at the origin, and this estimate was
used above in connection with the relation in Eq.(\ref{dmgee}).

The $e^+e^-$ decay rates of the vector resonances appear directly in the
spectral approach (Eq.(\ref{rres})), so that the relativistic effects and
perturbative and nonperturbative QCD terms are all included inasmuch as these
effects are taken into account in calculation of the correlation function.
However, as mentioned previously, the spectral relations are mostly sensitive to
the contribution of the lowest resonance, i.e. the $J/\psi$ in charmonium. It is
known since long ago\cite{6sr,6aut} that the  QCD sum rules are in an excellent
agreement with the observed value of $\Gamma_{ee}(J/\psi)$ of about $5\,$keV.

\subsubsection{\it Annihilation of mixed $^3D_1- {^3S_1}$ States through Virtual
Photon}
The quantum numbers of a $^3D_1$ state are $J^{PC}=1^{--}$ and allow such state
to decay through one virtual photon. However in a pure $D$ wave state the wave
function at the origin is vanishing, so that in the leading nonrelativistic
approximation the annihilation amplitude is zero. A non-vanishing contribution
to this amplitude arises in the order $v^2/c^2$ due to two mechanisms. The first
one is the $^3D_1 - ^3S_1$ mixing due to the tensor force, while the second
arises from the expansion of the $D$ wave coordinate wave function at the
annihilation distances $1/m_c$ and is proportional to the second derivative of
the radial function at the origin: $R''_{nD}(0)/m_c^2$, which is also of order
$p^2/m^2 \sim v^2/c^2$ in comparison with the $S$-wave wave function at the
origin. The latter mechanism alone would result in the expression for the
width\cite{6aut}
\be
\Gamma_{ee}(n^3D_1)={200 \, \alpha^2 \, e_c^2 \over M^6} \, |R''_{nD}(0)|^2~,
\label{geed}
\ee
where no short-distance QCD correction is included. The direct annihilation
amplitude and the one due to the mixing fully interfere with each other and are
of the same order in $v^2/c^2$. Thus it would generally be unjustified to
consider one mechanism but not the other.

Phenomenologically, the resonance $\psi(3770)$ is considered to be such
dominantly $1^3D_1$ state with an admixture of $^3S_1$ wave function. The
simplest model for the latter admixture uses the proximity in mass of the
$\psi'$ resonance and considers only the two state $\psi(3770) - \psi'$ mixture.
Then, using the data\cite{pdg} for
$\Gamma_{ee}[\psi(3770)]=0.242^{+0.027}_{-0.024}\,$keV and ignoring the direct
annihilation amplitude one can estimate\cite{rosner,rosner2} the mixing angle
$\theta$ as being about $0.2$. Such estimate certainly agrees with the
expectation for the size of the $v^2/c^2$ effects in charmonium, although the
particular numerical value should likely be taken with certain reservations
given the very simplistic nature of the model.

\subsubsection{\it Two-photon Annihilation of $C$-even States}
The $c {\bar c}$ quark pair in $C$ even states with $J \neq 1$ can annihilate
into two photons\cite{bgk,6aut}. For the $n^1S_0$ states the amplitude is
proportional, in a potential model approach, to the wave function at the origin,
$R_{nS}(0)$, so that it makes sense\cite{kmrr} to consider the ratio of the
$^1S_0 \to 2 \gamma$ and $J/\psi \to e^+e^-$ decay rates where the value of the
wave function at the origin cancels. Including also the first short-distance QCD
correction\cite{becr} for the $2 \gamma$ decay, which can be traced back to the
positronium result\cite{hb}, one can write
\be
{\Gamma(n^1S_0 \to \gamma \gamma) \over \Gamma_{ee}(n^3S_1)}=3 \, e_c^2 \, \left
[1+ { \alpha_s \over 3 \pi} \, (\pi^2-4) \right ] = {4 \over 3} \, \left ( 1 +
1.96 \, {\alpha_s \over \pi} \right )~.
\label{ejp}
\ee
Experimentally the $2 \gamma$ decay rate is measured for the $\eta_c$, albeit
with a large uncertainty\cite{pdg}, so that for the $1S$ charmonium the ratio of
the rates in Eq.(\ref{ejp}) is experimentally $1.3 \pm 0.4$. Although this value
can be considered as being in agreement with the theoretical expectation,
measurements with smaller experimental uncertainty are clearly needed for a more
meaningful comparison. It would also be of a great interest to test
Eq.(\ref{ejp}) for the $2S$ charmonium, i.e. for the corresponding decay rates
of the $\psi'$ and $\eta_c'$ resonances. So far the process $2 \gamma \to
\eta_c'$ has been seen\cite{cleogg,babargg} as a source of $\eta_c'$ in $\gamma
\gamma$ collisions, however quantitative data on decay widths of the $\eta_c'$
are still in flux.

It can be mentioned that the rate of the decay $\eta_c \to 2 \gamma$ can be
analysed by considering the QCD sum rules for the moments of the $\gamma \gamma$
scattering cross section due to the electromagnetic current of the charmed
quarks\cite{6ce,6aut}. The $\eta_c$ resonance is the lowest state contributing
to this cross section for perpendicular linear polarizations of the photons. The
result of such estimate, not including QCD corrections and nonperturbative terms
is\cite{6aut} $\Gamma(\eta_c \to 2 \gamma) = (6.1 \div 6.5)\,$keV, which is also
in agreement with the existing data\cite{pdg}.

The amplitudes of the two-photon annihilation of the $P$-wave states, $^3P_0$
and $^3P_2$ are proportional to the first derivative of the radial wave function
at the origin: $R'_P(0)/m_c$. The specific expressions\cite{bgk} with the first
short-distance QCD correction\cite{bcgr} are given by
\bea
&&\Gamma(^3P_0 \to \gamma \gamma)= {2^4 \,3^3 \, e_c^4 \, \alpha^2 \over M^4}\,
|R'_P(0)|^2 \, \left [ 1+{\alpha_s \over 3 \pi} \left ( \pi^2 - {28 \over 3}
\right ) \right ]~,\nonumber \\
&&\Gamma(^3P_2 \to \gamma \gamma)= {2^6 \,3^2 \, e_c^4 \, \alpha^2 \over 5 \,
M^4}\, |R'_P(0)|^2 \, \left(1-{16 \, \alpha_s \over 3 \pi} \right)~.
\label{pggw}
\eea
Both of these rates are of order $v^2/c^2$ in comparison with the similar rate
for the $^1S_0$ state. Thus one should expect a certain suppression of these
decays of the $\chi_{cJ}$ resonances as compared to the rate $\Gamma(\eta_c \to
\gamma \gamma)$, although an absolute prediction of the rates would be quite
model dependent due to uncertainty in the value of the wave function (as well as
due to other uncertain factors discussed in connection with Eq.(\ref{geer})).
The experimental values\cite{pdg} for these decay rates are:  $\Gamma(\chi_{c0}
\to \gamma \gamma) = 2.9 \pm 0.4\,$keV and  $\Gamma(\chi_{c2} \to \gamma \gamma)
= 0.534 \pm 0.050\,$keV, which can indeed be considered as somewhat suppressed
in comparison with $\Gamma(\eta_c \to \gamma \gamma)$. As usual, some of the
theoretical uncertainty goes away if one considers an appropriate ratio, in this
case the ratio of the two decay rates in Eq.(\ref{pggw}):
\be
{\Gamma(^3P_2 \to \gamma \gamma) \over \Gamma(^3P_0 \to \gamma \gamma)}= {4
\over 15} \, \left [ 1- {\alpha_s \over 3 \pi} \left ( \pi^2+ {20 \over 3}
\right ) \right ] = {4 \over 15} \left ( 1- 5.51 {\alpha_s \over \pi} \right )~.
\label{pggwr}
\ee
The experimental value of the ratio calculated from the world averages\cite{pdg}
for the decay rates, $0.185 \pm 0.025$ is indeed
smaller than the uncorrected value 4/15 = 0.267 in Eq.(\ref{pggwr}), which
agrees with the negative value of the QCD correction. However, the latest data
with the best precision in a single experiment\cite{mahlke_talk} give $0.235 \pm
0.042 \pm 0.005 \pm 0.030$. In either case it would still be
troublesome to draw a more quantitative conclusion due to the large coefficient
of the first radiative term.

The same considerations regarding the annihilation to $\gamma \gamma$ can be
applied to the recently observed state $Z(3930)$ interpreted as the radial
excitation $\chi'_{c2}$. The observation of this resonance\cite{z3930} is in
fact due to its discussed coupling to $2 \gamma$. However no quantitative data
on the electromagnetic decay width are available so far.

The subject of the two-photon annihilation may also become of relevance for the
yet unobserved $1^1D_2$ ($2^{-+}$) state of charmonium. If the mass of this
resonance is below the $D {\bar D}^*$ threshold it should be quite narrow since
its unnatural spin-parity forbids decay into $D {\bar D}$ pairs. No mixing due
to the tensor forces is possible for this state, therefore its two-photon
annihilation would provide an access to the amplitude of direct annihilation
from $D$-wave. A similar amplitude enters the previously discussed annihilation
of a $^3D_1$ state into $e^+e^-$, where however it gets tangled with the $S-D$
mixing effects. Thus a measurement of the decay $1^1D_2 \to \gamma \gamma$ can
be helpful in separating the direct annihilation from mixing in the properties
of $\psi(3770)$. The rate of the decay is given\cite{6aut} as
\be
\Gamma(n^1D_2 \to \gamma \gamma)= {2^6 \, 3 \, e_c^4 \, \alpha^2 \over M^6} \,
|R''_{nD}(0)|^2
\label{gdgg}
\ee
in terms of the second derivative at the origin of the same radial function as
in Eq.(\ref{geed}).

\subsubsection{\it $J/\psi \to \gamma \gamma \gamma$}
The decays $^3S_1 \to 3 \gamma$ have very small rates proportional to
$\alpha^3$. However a measurement of such decay for the $J/\psi$ resonance does
not appear to be unrealistic. The rate of the decay in the potential approach is
given by
\be
\Gamma(J/\psi \to 3 \gamma)={16\, (\pi^2-9) \, e_c^6 \, \alpha^3 \over 3 \pi
M^2} \, |R_{1S}(0)|^2 \, \left ( 1 - 12.6 {\alpha_s \over \pi} \right )~,
\label{g3g}
\ee
where the lowest-order result is an adaptation of the orthopositronium decay
formula\cite{op} and the first QCD correction is also an adaptation\cite{kmrr}
of the numerical result\cite{cls} for the one-loop QED correction to the
orthopositronium decay rate. As usual, model-dependence can be reduced by
considering the ratio of this decay rate to $\Gamma_{ee}$ (Eq.(\ref{geer})):
\be
{\Gamma(J/\psi \to 3 \gamma) \over \Gamma_{ee}(J/\psi)} = {64 \, (\pi^2-9) \over
243 \pi} \, \alpha \, \left ( 1 - 7.3{\alpha_s \over \pi} \right ) \approx 5.3
\times 10^{-4} \,  \left ( 1 - 7.3{\alpha_s \over \pi} \right )~.
\label{3gee}
\ee
The QCD radiative correction in this estimate is large, and it is not clear what
numerical value it should be assigned. The uncorrected number puts the
three-photon decay rate in the ballpark of $3\,$eV corresponding to the
branching fraction ${\cal B}(J/\psi \to 3 \gamma) \sim 3 \times 10^{-5}$, which
can be compared with the current upper limit (at 90\% CL): $5.5 \times 10^{-5}$.

The three-photon decay, inspite of its small rate, presents a very interesting
object for a study of dynamics of charmonium. Indeed, the ratio of the rates in
Eq.(\ref{3gee}) is sensitive to only the QCD corrections, so that a measurement
of this ratio can possibly shed some light on understanding of the behavior of
the QCD radiative effects in a situation where the coefficients in the loop
expansion are large. Since such behavior of the expansion coefficients is
typical for various effects in heavy quarkonium a better understanding of the
QCD expansion can be of help in considering those other effects as well.
Furthermore, it would be of great interest if one-photon energy spectrum in the
$3 \gamma$ decay could be studied, since the photons provide a very clean ``CT
scan" of the internal structure of $J/\psi$. Namely, the photon spectrum at
energy $\omega$ provides sensitivity to distances $\sim 1/\sqrt{m_c \omega}$, so
that dynamics of the charmed quarks can be probed by this spectrum at distances
ranging from the short ones, $\sim 1/m_c$, up to the typical charmonium size.

\subsection{\it Strong Annihilation into Light Hadrons}
Charmonium decay into light hadrons through the strong interaction is viewed in
QCD as a two-stage factorized process. First the $c {\bar c}$ pair annihilates
into gluons at `short' distances of order $1/m_c$ and then the gluons fragment
into specific hadronic final states. The total inclusive probability of the
latter fragmentation is considered to be equal to one, so that the total decay
rate is determined by the short-distance annihilation to on-shell gluons.
Clearly, this approach based on a perfect `gluon - hadron duality', involves an
uncertainty of its own in addition to the previously mentioned uncertainties
involved in calculation of charmonium annihilation. Inspite of this reservation,
such treatment first considered in Ref.~\cite{ap}, is extremely successful in
explaining, at least  semi-quantitatively, the pattern of OZI violating narrow
widths of the charmonium resonances below the open charm threshold, in
particular in understanding the very narrow width of the $J/\psi$ resonance,
which has so profoundly awed particle physicists in November 1974.

\subsubsection{\it Three-gluon Annihilation of $^3S_1$ Charmonium.}
The minimal number of gluons into which a $^3S_1$ state of a heavy quark pair
can annihilate is three, since the process through one virtual gluon is
forbidden by color and a two-gluon final state is excluded by the negative $C$
parity of the initial state. The decay rate in the lowest order in QCD\cite{ap}
can be found by decorating the orthopositronium decay formula\cite{op} with the
appropriate color factor, while the first QCD radiative correction is known only
numerically\cite{ml}, and being expressed in terms of $\alpha_s$ normalized at
the scale $m_c$ in the ${\overline MS}$ scheme reads as
\be
\Gamma(n^3S_1 \to 3 g \to {\rm light~hadrons})= {40 \over 81} \, {\pi^2 - 9
\over \pi} \, {\alpha_s^3(m_c) \over M^2} \, |R_{nS}(0)|^2 \, \left ( 1- 3.7 \,
{\alpha_s \over \pi} \right )~.
\label{g3gl}
\ee
The ratio of this rate to $\Gamma_{ee}$ is then sensitive, in this approach, to
only the coupling constants:
\be
{\Gamma(n^3S_1 \to 3 g \to {\rm light~hadrons}) \over \Gamma_{ee}(n^3S_1)}= {5
\over 18} \, {\pi^2 -9 \over \pi} \, {\alpha_s^3(m_c) \over \alpha^2} \, \left (
1+ 1.6 \, {\alpha_s \over \pi} \right )~.
\label{g3geer}
\ee
One is naturally tempted to use this formula for an estimate of the QCD coupling
$\alpha_s$. In order to make such estimate one has to evaluate the branching
fraction for the direct strong annihilation of $J/\psi$, which amounts to
subtraction from the total sum the electromagnetic annihilation contribution,
estimated in Eq.(\ref{bgstar}), the small contribution of the decay $J/\psi \to
\gamma \eta_c$ with the branching fraction\cite{pdg} $(1.3 \pm 0.4)$\%, and the
contribution of the direct photon emission, originating in the process $J/\psi
\to \gamma g g$, which will be discussed in some detail few lines below. The
branching fraction for the latter process is somewhat uncertain experimentally
due to a large background of secondary photons in the low-energy part of the
photon spectrum. Namely, the direct photon emission has been observed and
reliably measured\cite{mkII} at $x > 0.6$, where $x=2 \omega_\gamma/M_{J/\psi}$
is the ratio of the photon energy to the maximally allowed by kinematics. The
integral over the observed spectrum in this region corresponds to the branching
fraction ${\cal B}(J/\psi \to \gamma gg)|_{x > 0.6} = (4.1 \pm 0.8)\%$. The
experimental shape of the spectrum reasonably suggests that the observed part
makes
about one half of the total decay rate with another half being hidden under the
background at $x < 0.6$. Thus one can rather conservatively estimate the total
branching fraction as ${\cal B}(J/\psi \to \gamma gg) \approx (8 \pm 3)\% $. As
a result the fraction of the direct strong annihilation of $J/\psi$ can be
estimated as ${\cal B} (J/\psi \to 3g) \approx (66 \pm 3)\%$, and the ratio in
Eq.(\ref{g3geer}) as\footnote{This estimate also agrees with the result of a
recent similar evaluation\cite{egmr} done in a slightly different way.}
\be
{{\cal B} (J/\psi \to 3g) \over {\cal B}_{ee}(J/\psi)} = 11.1 \pm 0.5~,
\label{g3eerpsi}
\ee
thus providing one with the estimate $\alpha_s(m_c) \approx 0.19$. It would
however be difficult to assign a reliable error bar to this number, given the
reservations mentioned previously regarding the assumptions and approximations
made in connection with the described calculations of the annihilation rates.

A marginal suitability of the direct hadronic decay rate of charmonium for
precision determination of QCD parameters can also be illustrated by comparing
the data on the annihilation decays of $J/\psi$ and $\psi'$. The proportionality
of the annihilation rates to $|R_{nS}(0)|^2$ implies a similarity between the
decays of $J/\psi$ and $\psi'$. Namely, the ratios between these two resonances
of their similar decay rates  should be the all equal to each other. In
particular the ratia of the branching fractions for similar decays should all be
equal to ${\cal B}_{ee}(\psi')/{\cal B}(J/\psi)=(12.4 \pm 0.3)\%$ (the so-called
``12\% rule"). A recent CLEO-c dedicated study\cite{cleop2p} of the decays of
$\psi'$ ending up in $J/\psi$ in the final state allowes to separate the
branching fraction for the direct decays of $\psi'$ into light hadrons, which
includes the electromagnetic decays of this type and the $\gamma gg$ decays with
direct photon: ${\cal B}(\psi' \to {\rm light hadrons})= (16.9 \pm 2.6)\% $. The
same branching fraction for $J/\psi$ is listed in the Tables\cite{pdg}: ${\cal
B}(J/\psi \to {\rm light hadrons})= (87.7 \pm 0.5)\% $. The ratio of these
numbers gives ${\cal B}(\psi' \to {\rm light hadrons})/{\cal B}(J/\psi \to {\rm
light hadrons})=(19.3 \pm 3.0)\%$, which substantially differs from the ``12\%
rule". One would also arrive at a similar contradiction with the simple picture
of charmonium annihilation if first subtracted from ${\cal B}(\psi' \to {\rm
light hadrons})$ the small contribution of the electromagnetic processes $\psi'
\to \gamma^* \to {\rm light hadrons}$ and the (estimated) fraction of $\psi' \to
\gamma gg$, leaving ${\cal B} (\psi' \to 3g) \approx (14.3 \pm 2.9)\%$, which
gives the estimate for the ratio of the $ggg$ and $e^+e^-$ annihilation rates of
$\psi'$ equal to $19.5 \pm 4.0$, i.e. different from the same ratio for $J/\psi$
in Eq.(\ref{g3eerpsi}).

That there is more dynamics to the hadronic annihilation of the $^3S_1$
charmonium than the simple factorized approach suggests, is indicated by that
the 12\% rule is also quite conspicuously broken in exclusive decay modes, the
most famous in this respect being the decay into $\rho \pi$ which is strongly
suppressed for $\psi'$, well below the ``12\% rule". A number of other exclusive
modes with a (seemingly random) deviation from this rule have been measured
experimentally\cite{pdg,cleo12p}. Although deviations from the simple picture
are expected on general grounds, the specific mechanisms for such deviations are
currently unknown even at the level of inclusive decay rates. We will discuss
some considerations on this subject at the end of the current section on
charmonium annihilation.

\subsubsection{\it The Decay into $\gamma g g$}
The mixed electromagnetic and strong annihilation into a photon and two gluons,
briefly discussed above in connection with the estimate in Eq.(\ref{g3eerpsi}),
is actually quite interesting on its own\cite{chanowitz}. The ratio of the total
decay rate to that of the decay into $e^+e^-$ in the simple annihilation picture
depends only on the QED and QCD coupling constants:
\be
{\Gamma(n^3S_1 \to \gamma g g) \over \Gamma_{ee}(n^3S_1)}= {8 \over 9} \,{\pi^2
-9 \over \pi} \, {\alpha_s^2(m_c) \over \alpha} \, \left ( 1- 1.3 {\alpha_s
\over \pi} \right )~,
\label{gamggee}
\ee
where the QCD radiative correction is from the result of numerical calculation
of Ref.~\cite{ml}. Using the known value of ${\cal B}_{ee}(J/\psi)$ and
$\alpha_s(m_c) \approx 0.19$ as estimated from Eq.(\ref{g3eerpsi}), one can
estimate ${\cal B}(J/\psi \to \gamma gg) \approx 6.7\%$, which is in a
reasonable agreement with the available data\cite{mkII}.

Besides the total rate of this radiative hadronic decay, of a great interest is
the spectrum of the direct photons\cite{6aut}, since this spectrum provides some
insight into how the gluon - hadron duality sets in. Indeed the invariant mass
squared, $q^2$, of the hadronic final state into which the two gluons fragment
is related to the photon energy $\omega$ as $q^2=M_{J/\psi}^2 \, (1-x)$, where
$x=2 \omega_\gamma/M_{J/\psi}$. The `parton' spectrum, corresponding to on-shell
gluons extends all the way to the kinematical boundary at $x=1$, corresponding
to $q^2=0$. In reality, one certainly does not expect a gluon-hadron duality at
low $q^2$, so that the spectrum of direct photons at high energy should be
suppressed by hadronic effects, as is observed in experiment\cite{mkII}. At
larger $q^2$, corresponding to lower $x$, the actual spectrum should approach
the parton curve due to the onset of the gluon-hadron duality. It should be
noted however that, as previously mentioned, the direct photon spectrum at lower
$x$ (in practice at $x \approx 0.5$ for charmonium) goes under an overwhelming
background of secondary photons which are present in hadronic final states. Thus
in the decays of charmonium it is unlikely that the fragmentation can be studied
at $q^2$ beyond 4 or 5 GeV$^2$, and a much better testing ground for such study
is provided by the decays of the bottomonium $\Upsilon$
resonances\cite{cleo-ups}.

\subsubsection{\it Two-gluon Annihilation of $C$-even States}
The $C$-even states of quarkonium with $J \neq 1$ can annihilate into two
gluons, much in the same way as they decay into two photons. In fact in the
lowest order the $gg$ and $\gamma \gamma$ decay rates are simply
related\cite{bgk,6aut} as
$\Gamma_{gg}/\Gamma_{\gamma \gamma}=2 \, \alpha_s^2 /(9 \, e_c^4 \alpha^2 ) =
(9/8) \, (\alpha_s/\alpha)^2$. The coefficients of the first QCD correction are,
naturally, different for individual states\cite{becr,bcgr}:
\be
{\Gamma(^1S_0 \to gg) \over \Gamma(^1S_0 \to \gamma \gamma)}= {9 \over 8} \,
\left [ {\alpha_s^2(m_c) \over \alpha} \right ]^2 \, \left ( 1+8.2 {\alpha_s
\over \pi} \right )~,
\label{sggam}
\ee
\be
{\Gamma(^3P_0 \to gg) \over \Gamma(^3P_0 \to \gamma \gamma)}= {9 \over 8} \,
\left [ {\alpha_s^2(m_c) \over \alpha} \right ]^2 \, \left ( 1+9.3 {\alpha_s
\over \pi} \right )~,~~~~{\Gamma(^3P_2 \to gg) \over \Gamma(^3P_2 \to \gamma
\gamma)}= {9 \over 8} \, \left [ {\alpha_s^2(m_c) \over \alpha} \right ]^2 \,
\left ( 1+3.1 {\alpha_s \over \pi} \right )~.
\label{p0ggam}
\ee
The correction for the decay rate of the $^1D_2$ state is presently unknown. One
can see that the coefficients of the first QCD correction for  $^1S_0$ and
$^3P_0$ states are unusually large, which makes difficult a reliable
quantitative comparison with the data. The uncorrected formulas with $\alpha_s
\approx 0.2$ would give for the ratio of the width the value of about 850, which
is not even close to the observed values\cite{pdg}:
$\Gamma_{gg}(\eta_c)/\Gamma_{\gamma \gamma} (\eta_c) = (3.57 \pm 1.15) \times
10^3$, $\Gamma_{gg}(\chi_{c0})/\Gamma_{\gamma \gamma} (\chi_{c0}) = (3.62 \pm
0.43) \times 10^3$, and $\Gamma_{gg}(\chi_{c2})/\Gamma_{\gamma \gamma}
(\chi_{c2}) = (3.09 \pm 0.23) \times 10^3$. The experimental data can be
somewhat understood using a significantly larger QCD coupling $\alpha_s(m_c)
\approx 0.3$ than is inferred from the three-gluon decays of $J/\psi$. Such
larger value of $\alpha_s(m_c)$ better complies with the
determination\cite{menke} of the QCD coupling from $\tau$ decays and $e^+e^-$
annihilation data, but then it entirely misses the observed rate of the
three-gluon annihilation. It is not clear at present how to avoid this crunch.
One could possibly argue for the $\eta_c$ and $\chi_{c0}$ resonances that their
hadronic decay rates are enhanced by the instanton-type quarkonium glue mixing
in the $0^{-+}$ and $0^{++}$ channels\cite{alike}. However such argument would
still leave the observed $2g/2\gamma$ ratio for the $\chi_{c2}$ unexplained.

\subsubsection{\it Strong Annihilation of $^1P_1$ and $^3P_1$ States}
The so far discussed amplitudes of the decays into on-shell gluons are not
sensitive to the infrared behavior of quarks and gluons, at least at the
considered here level of perturbative calculation at the leading and the
one-loop level. In other words, these amplitudes are all expressed in terms of
the wave function of the quark pair at the typical annihilation distances of
order $1/m_c$. In the two-gluon decays this behavior is guaranteed by the fixed
and large energy of each of the gluons in the c.m. system, while for the
three-gluon decays of the $^3S_1$ states the kinematical region, where one of
the gluons is soft, is not enhanced due to considering the heavy quarks being
essentially static and thus not radiating soft gluons. Clearly, such ``good"
infrared behavior generally becomes invalid in higher orders of the perturbative
expansion, as well as beyond the leading nonrelativistic approximation. For the
$^1P_1$ and $^3P_1$ states the infrared behavior of the amplitudes describing
their annihilation shows up already in the leading approximation\cite{bgr}. The
reason for the infrared sensitivity of these amplitudes is that a $P$ wave
necessarily involves a motion of the quarks, so that a $P$-wave state can go
into an $S$-wave by an E1 emission of a soft gluon. The spin of the quarks is
not involved in the emission. Therefore the colorless $^1P_1$ state goes into a
color-octet $^1S_0$ state of the quark pair, while the $^3P_1$ state goes into a
color-octet $^3S_1$. The intermediate $^1S_0$ state decays into two gluons,
while the colored $^3S_1$ can decay either into a light quark-antiquark pair, $q
{\bar q}$ through a virtual gluon, or, generally, into a gluon pair. It turns
out however\cite{bgr,6aut} that for a static colored $^3S_1$ pair of heavy
quarks the amplitude of annihilation into two gluons is identically zero in the
leading order, thus the resulting decay processes are $^1P_1 \to ggg$ and $^3P_1
\to g q {\bar q}$. In calculating the total probability one has to integrate
over the energy $\omega$ of the soft gluon, which integration, as is standard
for a soft gauge quantum emission, is infrared-divergent as $\int
d\omega/\omega$. With a logarithmic accuracy the upper limit in this integral is
set by the quark mass, and the lower limit is a typical QCD mass scale.

The leading-logarithm expressions for the hadronic widths of the $^1P_1$ and
$^3P_1$ states are given as\cite{bgr,6aut}
\be
\Gamma(^1P_1 \to 3g \to {\rm hadrons})={20 \over 9} \, {\alpha_s^3 \over m_c^4}
\, |R'_P(0)|^2 \, \ln {m_c \over \Lambda_{QCD}}~,
\label{g1p1}
\ee
and
\be
\Gamma(^3P_1  \to {\rm hadrons})=\sum_{q=u,d,s} \Gamma(^3P_1 \to g \, q {\bar q}
\to {\rm hadrons})= {8 \over 3 \pi} \, {\alpha_s^3 \over m_c^4} \, |R'_P(0)|^2
\, \ln {m_c \over \Lambda_{QCD}}~.
\label{g3p1}
\ee
One can see that the numerical coefficients in these formulas correspond to the
hadronic width of the $^1P_1$  being larger than that of the $^3P_1$ by the
factor $5 \pi/6 \approx 2.6$. The experimental value of the hadronic width of
the $1^3P_1$ charmonium resonance, $\chi_{c1}$, is approximately 0.6\,MeV, so
that based on these formulas one would expect the hadronic width of the $1^1P_1$
resonance, $h_c$, to be about 1.5\,MeV. The latter value does not contradict
the CLEO-c data\cite{cleohc}, but it might be in a contradiction with the
Fermilab E835 results\cite{e835} claiming the width of the observed $h_c$
resonance to be $\Gamma \le 1\,$MeV. Apriori one would expect the applicability
of these logarithmic estimates of the annihilation rates to be marginal at best.

\subsection{\it Non-perturbative Effects in Hadronic Annihilation}
The logarithmic formulas become formally applicable in the limit of very heavy
quarks, where the resulting logarithm can be considered as a large parameter,
and one can develop a formalism\cite{bbl} of a logarithmic theory of the
processes with soft-gluon transitions between color-octet and color-singlet
heavy quarkonium states, or consider purely phenomenologically the mixing
between pure colorless $c {\bar c}$ states and states where a color-octet quark
pair is accompanied by a soft gluon field. An extensive overview of this
approach, called in the literature Non-Relativistic QCD (NRQCD), and a list of
references can be found in the review \cite{qwg}.

The effects of soft gluons can be addressed not only in the decays of the
$^3P_1$ and $^1P_1$ states, where such effects are dominant, but also for `well
behaved' processes, where these effects give rise to small or moderate
corrections. In particular the corrections arising from nonperturbative soft
gluon field in the annihilation of the $^3S_1$ states may be responsible for the
previously discussed deviations from predictions based on the simple
perturbative picture, such as the ``12\% rule".  A consistent analysis of the
leading nonperturbative corrections to the annihilation rates is however
possible only in the limit of very heavy quarkonium, for which such corrections
are expressed\cite{mv84} in terms of the gluon vacuum condensate.

\begin{figure}[tb]
\begin{center}
\begin{minipage}[t]{18 cm}
\epsfxsize=17cm
\epsfbox{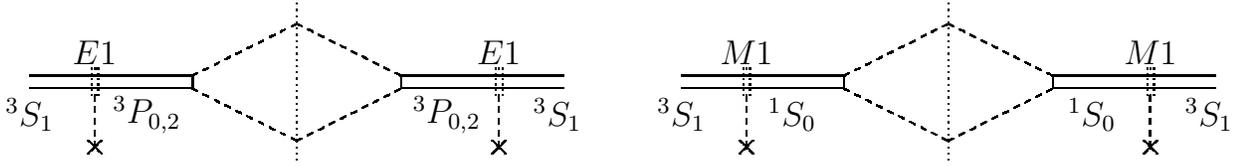}
\end{minipage}
\begin{minipage}[t]{16.5 cm}
\caption{The mechanism describing the leading nonperturbative correction to the
annihilation rate of a heavy $^3S_1$ quarkonium shown as a unitary cut across
two hard gluons. The soft gluons (ending in a ``$\times$") are described by the
field in the vacuum condensate $\langle 0 | G^2 | 0 \rangle$.\label{npann}}
\end{minipage}
\end{center}
\end{figure}

The leading nonperturbative corrections to the three-gluon annihilation rate of
a heavy $^3S_1$ quarkonium are described by the mechanism shown in
Fig.\ref{npann}. The
soft gluon field converts the color-singlet $^3S_1$ state into a color-octet
$^3P_{0,2}$ state through the $E1$ chromoelectric interaction, or into a
color-octet $^1S_0$ state due to the chromomagnetic $M1$ interaction. The final
state in either of these transitions then decays into two hard gluons. The
contribution of this mechanism can be calculated as the imaginary part of the
graphs shown in Fig.\ref{npann} given by the unitarity cut across the hard
gluons, while
the averaging of the quadratic expression in the soft gluon field makes the
effect proportional to the gluon condensate $\langle 0 | G^2 | 0 \rangle$.

It is quite essential that both the contributions of the $^3P_J$ and $^1S_0$ are
comparable even for a nonrelativistic quarkonium.   Indeed, the chromomagnetic
$M1$ transition to the $^1S_0$ state is suppressed by the inverse of the heavy
quark mass, $1/m_Q$, in the amplitude, while the amplitude of annihilation of
the $^3P_{0,2}$ states contains an extra factor of $1/m_Q$ in comparison with
that for the $^1S_0$. Furthermore, the contribution of the $^3P_J$ intermediate
states is proportional to the vacuum condensate of the chromoelectric field,
$\langle 0 |{\vec E}^2|0 \rangle$, while that of the $^1S_0$ intermediate state
is given by the chromomagnetic condensate, $\langle 0 |{\vec B}^2|0 \rangle$.
The Lorentz invariance of the vacuum however requires that these two condensates
have opposite sign\cite{mv79}:
\be
\langle 0 |{\vec B}^2|0 \rangle = - \langle 0 |{\vec E}^2|0 \rangle= {1 \over 4}
\, \langle 0 |G^2|0 \rangle~.
\label{licon}
\ee

The important feature of the discussed mechanism is that its dependence on the
wave function of the $^3S_1$ state is not reduced to $\psi(0)$, but rather is
determined by the corresponding overlap integrals. Therefore such mechanism
generally breaks the similarity between the decays of different $^3S_1$ states.
The same behavior however implies that the overlap integrals can be reliably
evaluated only for an asymptotically heavy quarkonium, whose dynamics is
dominated by the Coulomb-like potential due to the gluon exchange between the
quark and the antiquark. In this limit one finds for the relative correction to
the decay rate of an $n^3S_1$ state the expression\cite{mv84}
\be
{\delta \Gamma  \over \Gamma} (n^3S_1)= - { \pi  \langle 0 |G^2|0 \rangle \over
2^{11}\, 9 \, (\pi^2-9) \, \alpha_s (m_Q) \, k_n^4 } \, {a_n \over n^4}~,
\label{npcor3g}
\ee
where $a_n$ are numerical coefficients given by
\be
a_n={ 2^4 \, 7 \,  n^4 \over 9} \left [ {64 \, (n^2-64) \over (81 \, n^2-256) \,
(81 \, n^2 -64) } \right ]^2 -  2^{14} \, 9 \, {n^2 \over (81 \, n^2 - 64)^2}~,
\label{ann}
\ee
and $k^n$ is the `Bohr momentum' for the $n$-th state determined from the
relation
\be
k_n={4 \over 3 \, n} \, m_Q \, \alpha_s(k_n)~.
\label{kn}
\ee
It should be mentioned that the normalization gluon field $G_{\mu \nu}$ used
here corresponds to the gluon term in the QCD Lagrangian of the form $L = -(1/4
g^2) \, G^2$, so that numerically the gluon condensate is $\langle 0 |G^2|0
\rangle \approx 0.7\,$GeV$^4$.

As previously discussed the chromoelectric and the chromomagnetic contributions
have opposite sign, so that the coefficients $a_n$ in Eq.(\ref{ann}), have the
form of a difference: $a_n=a_n(E)-a_n(M)$. Furthermore, the coefficients
$a_n(E)$ at $n \ge 3$ and $a_n(M)$ at $n \ge 2$ are very small. This is a
consequence of a relatively weak Coulomb-like interaction in the octet state. If
the latter interaction were neglected altogether (the $N_c \to \infty$ limit)
those coefficients would also vanish, and the only remaining nonzero would be
$a_1(E)=a_2(E)=28$ and $a_1(M)=27$. The real world values given by
Eq.(\ref{ann}) are $a_1=22.858-18.897 = 3.961$ and $a_2=9.393-0.323=9.070$.
These values illustrate that for the ground $^3S_1$ state the discussed
correction is very sensitive to the details of the quarkonium dynamics, so that
any extrapolation of the asymptotic formulas down to the realistic bottomonium
and charmonium necessarily is quite uncertain.

If such extrapolation is done by using $k_1 \approx 1\,$GeV for the $\Upsilon$
resonance and $k_1 \approx 0.5\,$GeV for $J/\psi$, one would estimate the
relative effect in the hadronic annihilation of $\Upsilon$ as about 0.5\,\%, and
about 5\,\% for the $J/\psi$ resonance. If however the cancellation between the
chromoelectric and chromomagnetic terms does not occur for these realistic
quarkonia, the effect can be few times larger.

A similar mechanism, with one of the hard gluons replaced by photon, describes
the nonperturbative correction to the rate of the decay $^3S_1 \to \gamma+2g$.
The relative correction in this case is three times smaller than in
Eq.(\ref{npcor3g}).

\section{Radiative Transitions}
The treatment of charmonium as a nonrelativistic system suggests that one can
apply the standard multipole expansion in electrodynamics to calculation of the
transitions between charmonium levels with emission of a photon. The leading
terms in this expansion are the $E1$ and $M1$ terms, which can be described for
the discussed here transitions by the corresponding terms in the Hamiltonian:
\be
H_{E1}=-e_c \, e \,({\vec r} \cdot {\vec E})~~~~{\rm and}~~~~H_{M1}=-\mu_c \,
({\vec \Delta} \cdot {\vec B})~,
\label{e1m1e}
\ee
where $e_c=2/3$ is the charge of the charmed quark,  $\mu_c=e_c \, e/(2m_c)$ is
its magnetic moment, $\vec E$ and $\vec B$ are standing for the electric and the
magnetic field, and ${\vec \Delta}$ being a spin operator defined as $\vec
\Delta= \vec \sigma_1 - \vec \sigma_2$ with $\vec \sigma_1$ and $\vec \sigma_2$
acting respectively on the quark and the antiquark. The electric dipole term is
responsible for the transitions between the $S$ and $P$ states with the same
spin $S$ of the quark pair, while the $M1$ term describes the transitions
between
$S=1$ and $S=0$ states with the same orbital momentum $L$.

\subsection{\it $E1$ Transitions}
\subsubsection{$\psi' \to \gamma \, \chi_{cJ}$}
The rate for the transitions from a $^3S_1$ state to $^3P_J$ induced by the $E1$
term in Eq.(\ref{e1m1e}) is given by
\be
\Gamma(^3S_1 \to \gamma \, ^3P_J)=(2J+1) \, {4 \over 27} \, e_c^2 \, \alpha \,
\omega_\gamma^3 \, |I_{PS}|^2~,
\label{e1g}
\ee
where $\omega_\gamma$ is the energy of the emitted photon, and $I_{PS}$ is the
radial overlap integral:
\be
I_{PS}=\langle P |r |S \rangle = \int_0^\infty r^3 \, R_P(r) \, R_S(r) \, dr~,
\label{ips}
\ee
with $R_{S,P}(r)$ being the normalized radial wave functions for the
corresponding states.

Clearly, the overlap integral has dimension of length, and its particular value
is somewhat model dependent.(With certain restrictions following from general
Quantum Mechanical considerations\cite{jackson,6aut}.) One can however assess
the overall validity of the nonrelativistic formula (\ref{e1g}) for charmonium
by
comparing with each other the experimentally known rates of the transitions from
$\psi'$ to the $\chi_{cJ}$ levels with different $J$. This can be done in terms
of the value of the overlap integral, $|I_{PS}|$, extracted from the rate of
each of the transition. Proceeding in this way one finds
\be
|\langle 1P|r|2S\rangle| \approx \left \{ \begin{array}{ll}
                                 0.37\, \mbox{fm from} & \psi' \to \chi_{c0} \,
\gamma \\
                                  0.39\, \mbox{fm from} & \psi' \to \chi_{c1} \,
\gamma \\
                                  0.45\, \mbox{fm from} & \psi' \to \chi_{c2} \,
\gamma ~,
                                     \end{array} \right .
\ee
so that the extracted values of the overlap integral agree with each other
within the expected accuracy of the expression (\ref{e1g}), and the absolute
value of the radial integral also agrees well with a general understanding of
the typical size of charmonium.
\subsubsection{$\chi_{cJ} \to \gamma \, J/\psi$}
The transitions from  $^3P_J$ levels to a $^3S_1$ state are described by the
expression for the rate
\be
\Gamma(^3P_J \to \gamma \, ^3S_1)= {4 \over 9} \, e_c^2 \, \alpha \,
\omega_\gamma^3 \, |I_{SP}|^2~.
\label{e1g2}
\ee
Performing a similar to previous extraction of the overlap integral from the
experimental data on the rates of the $\chi_{cJ} \to J/\psi$ decays, one finds
the overlap integral for the $1P \to 1S$ transitions as
\be
|\langle 1S|r|1P\rangle|= \left \{ \begin{array}{ll}
                                 0.36\, \mbox{fm from} & \chi_{c0} \to J/\psi \,
\gamma \\
                                  0.38\, \mbox{fm from} & \chi_{c1} \to J/\psi
\, \gamma \\
                                  0.37\, \mbox{fm from} & \chi_{c2} \to J/\psi
\, \gamma~.
                                     \end{array} \right .
\ee
Thus one can conclude that the observed $E1$ transitions in charmonium present
us with no unusual behavior.

\subsubsection{$h_c \to \gamma \, \eta_c$}
The decay $h_c \to \gamma \, \eta_c$ is crucial for identifying the charmonium
$^1P_1$ resonance in the experiments\cite{e835,cleohc}.
A straightforward application of the $E1$ transition formula results in the
expression for the rate of this decay
\be
\Gamma(^1P_1 \to \gamma \, ^1S_0)= {4 \over 9} \, e_c^2 \, \alpha \,
\omega_\gamma^3 \, |I_{SP}|^2~,
\ee
where the overlap integral is the same as is just estimated from the $\chi_{cJ}
\to \gamma \, J/\psi$ decays, up to the relativistic corrections $O(v^2/c^2)$.
Thus using the extracted value of $|I_{SP}|$ one can readily predict the width
of the decay in absolute terms:
\be
\Gamma(h_c \to \gamma \, \eta_c) \approx 0.65\,{\rm MeV}~.
\ee
As discussed previously, the hadronic width of the $h_c$ is quite uncertain. The
logarithmic formulas and the data on the hadronic width of $\chi_{c1}$ indicate
that hadronic decay rate of $h_c$ can be as large as $1.5\,$MeV, in which case
the radiative decay branching ratio should be about 30\%. It is believed that
this branching ratio, with all the uncertainties involved, should be in the
range
between 30\% and 50\%.

\subsubsection{\it Relativistic Effects}

In the next order of the relativistic expansion the leading amplitudes of the
considered (dominantly) $E1$ transitions, receive $O(v^2/c^2)$ corrections from
the magnetic quadrupole $M2$ and the electric octupole $E3$ terms\cite{gos,sgr}.
These corrections are expected to be suppressed relative to the leading term by
a factor $v^2/c^2 \approx 0.2$. In the total decay rate these terms do not
interfere with the leading one, so that their contribution is only of order few
percent and hardly can be measured. However these amplitudes generally interfere
with the $E1$ term in the angular distribution. In particular for the $1P \to
1S$ transitions the $M2$ and $E3$ contributions can be studied by measurements
of the angular distributions and helicity amplitudes in the processes
$p \bar p \to \chi_{c1,2} \to \gamma \, J/\psi$. The most precise available
measurements were performed by the E835 experiment\cite{e835m2}. The current
averages\cite{pdg} for the relative values of the amplitudes are: $M2/|E1|=
-0.13 \pm 0.05$ and $E3/|E1|=0.011^{+0.041}_{-0.033}$ from the $\chi_{c2} \to
\gamma \, J/\psi$ decays, and $M2/|E1| = -0.002^{+0.008}_{-0.017}$ from
$\chi_{c1} \to \gamma \, J/\psi$.

Thus the data on the radiative decay of the $\chi_{c2}$ generally support the
expectations: the suppression of the $M2$ amplitude is indeed of order
$v^2/c^2$, while the $E3$ amplitude was predicted to be small\cite{sgr}. On the
other hand the result of the extraction of $M2$ from the $\chi_{c1}$ decay is
neither compatible with that for the $\chi_{c2}$, nor does it comply with
theoretical expectations. Clearly, this situation strongly suggests that further
studies of the relativistic terms are needed.

\subsection{\it M1 Transitions and the Puzzle of $\eta_c$}
The radiative transitions induced by the $M1$ term (Eq.(\ref{e1m1e})) appear to
be simpler than the $E1$ ones, since the $M1$ interaction does not contain any
coordinate dependence, while the spin matrix elements are trivial inasmuch as
the spin and coordinate degrees of freedom are factorized. Phenomenologically
the most interesting are the $M1$ transitions $J/\psi \to \gamma \, \eta_c$ and
$\psi'  \to \gamma \, \eta_c$, whose rates are given by
\be
\Gamma(n^3S_1 \to \gamma \, m^1S_0) = {4 \over 3} \, e_c^2 \, \alpha \,
{\omega_\gamma^3 \over m_c^2} \, |I_{mn}|^2~,
\label{m1g}
\ee
where $I_{mn}$ is the overlap integral for unit operator between the coordinate
wave functions of the initial and the final charmonium states. In the leading
nonrelativistic order, where there is no effect of the spin-spin interaction on
the coordinate wave function, the overlap integrals are determined by the
orthonormality condition for the coordinate wave functions:
$I_{mn}=\delta_{mn}$. Using for an estimate $m_c = 1.4\,$GeV one thus calculates
the rate of the transition $J/\psi \to \gamma \, \eta_c$ as $\Gamma(J/\psi \to
\gamma \, \eta_c) \approx 3.3\,$keV,  while the overlap matrix element vanishes
for the $2S \to 1S$ transition $\psi'  \to \gamma \, \eta_c$. The latter decay
becomes possible due to the relativistic effects, but at present any estimate of
the actual rate would be quite unreliable.

In reality both decays have very similar rates: $\Gamma(J/\psi \to \gamma \,
\eta_c) \approx 1.2\,$keV and $\Gamma(\psi' \to \gamma \, \eta_c) \approx
0.9\,$keV. Although for the latter transition the small value of the rate is not
surprising and can be readily attributed to the suppression of the relativistic
effects, the failure of the seemingly robust theoretical expectation for the
former transition is quite paradoxical.

One can certainly attempt to explain the discrepancy of almost a factor of 3
between the experimental value and the straightforward theoretical estimate for
the rate of the transition from the $J/\psi$ in terms of a larger charmed quark
mass, modified magnetic moment of the quark, and enhanced relativistic effects,
or a combination of these factors. However such a `post diction' would
inevitably
have to be quite contrived. Furthermore, Shifman has presented a
phenomenological argument\cite{mash}, which eliminates the uncertainty related
to the charmed quark mass and magnetic moment, and still leaves a considerable
gap between the theoretical and the experimental values of the rate. The
argument is based on applying the standard dispersion relation to the amplitude
of the decay $\eta_c \to \gamma \gamma$. Namely, this amplitude is described by
just one form factor $F$:
\be
A(\eta_c \to \gamma \gamma) = F \, \varepsilon^{\mu \nu \lambda \sigma} \,
F^{(1)}_{\mu \nu} \, F^{(2)}_{\lambda \sigma}
\ee
with $F^{(1,2)}_{\mu \nu}$ being the field tensors for the two photons. The form
factor is a function of the squares of the three 4-momenta, involved in the
process: $F(m_{\eta_c}^2, k_1^2,k_2^2)$, and the width is determined by its
value for on-shell photons: $\Gamma(\eta_c \to \gamma
\gamma)=|F(m_{\eta_c}^2,0,0)|^2 m_{\eta_c}^3/4 \pi$. One can further use the
dispersion relation for the form factor in one of the photon momenta:
\be
F(m_{\eta_c}^2,0,k^2)={1 \over \pi} \, \int {ds \over s-k^2} \, {\rm Im}
F(m_{\eta_c}^2, 0,s)~,
\label{fdisp}
\ee
and express the imaginary part in the integrand in terms of contribution of real
intermediate states. These states with quantum numbers $J^{PC}=1^{--}$ include
the $J/\psi$ resonance, higher charmonium resonances: $\psi'$, $\psi(3770)$ and
so on, and the continuum of the states with charmed $D(D^*)$ mesons.

The contribution of the $J/\psi$ is expressed in terms of the product of the
amplitudes $A(\eta_c \to \gamma \, J/\psi) \, A(J/\psi \to \gamma^*(s))$, where
the coupling of $J/\psi$ to a virtual photon is obviously related to its decay
width into $e^+e^-$. The contribution of the higher charmonium resonances is
vanishing, as discussed, in the leading nonrelativistic approximation and is
indeed very small phenomenologically. Namely, the effective overlap integral
$I_{21}$, that would fit the observed rate of the decay $\psi' \to \gamma \,
\eta_c$, is numerically only about 0.04.  Similarly the contribution of the
continuum to the integral in Eq.(\ref{fdisp}) can also be argued\cite{mash} to
be quite small compared to the contribution of the $J/\psi$. Therefore the
dispersion relation (\ref{fdisp}) should be saturated to a good accuracy by the
$J/\psi$ pole thus relating the amplitudes of the decays $\eta_c \to \gamma
\gamma$, $J/\psi \to e^+e^-$ and $J/\psi \to \gamma \, \eta_c$ and the masses of
$\eta_c$ and $J/\psi$, i.e. without the need to invoke the mass of the charmed
quark. The resulting relation between the rates reads as
\be
\Gamma(J/\psi \to \gamma \, \eta_c)= {2 \over 9} \, {\Gamma(\eta_c \to \gamma
\gamma) \over \Gamma(J/\psi \to e^+e^-)} \, \alpha \, {m_{J/\psi}^4 \over
m_{\eta_c}^3} \, \left ( 1-  {m_{\eta_c}^2 \over m_{J/\psi}^2} \right )^3 \,
\left [ 1 + O(\alpha_s) \right ]~.
\label{mashres}
\ee
Using the experimental data in the r.h.s. of this relation, one finds the
central value of the discussed rate (modulo the QCD corrections) as
$\Gamma(J/\psi \to \gamma \, \eta_c) = 2.9\,$keV, which is pretty close to the
previous simple nonrelativistic estimate, and is still very far away from the
experimental value of this rate.

It may well be that the puzzles of the $\eta_c$: its large total width, small
branching fraction of decay into $\gamma \gamma$ and the suppression of the
decay $J/\psi \to \gamma \, \eta_c$ are all tied together and possibly can be
resolved by a strong mixing in the $0^{-+}$ channel\cite{alike}, so that the
$\eta_c$ has a sizable admixture of light quark and gluon states. Qualitatively
such admixture would enhance the hadronic decay rate and suppress the radiative
transition from a pure charmonium state $J/\psi$. In terms of the Shifman's
argument, there would arise an additional contribution to the imaginary part of
the form factor $F$ from $1^{--}$ states lighter than $J/\psi$, which would
generally violate the relation (\ref{mashres}). However a quantitative
description of these phenomena cannot be offered at present.

\section{Hadronic Transitions Between Charmonium Resonances}
Similarly to the radiative transitions between heavy quarkonium levels, caused
by the interaction of quarks with photons, the hadronic transitions  arise
through the interaction of the heavy quarks with gluons and the gluons
materializing as light mesons, the pions and $\eta$. Also in complete analogy
with the radiative transitions, the interaction of a nonrelativistic quarkonium
with the gluon field in hadronic transitions can be described within the
multipole expansion in QCD\cite{gottfried}. The leading terms, that will be
important in our further discussion of the realistic processes are the
chromoelectric and the chromomagnetic dipoles, $E1$ and $M1$ and the
chromomagnetic quadrupole $M2$. The corresponding terms in the Hamiltonian can
be written as
\be
H_{E1}=-{1 \over 2} \xi^a \, {\vec r} \cdot {\vec E}^a (0)~,
~~~H_{M1}= - {1 \over 2
\, M}\, \xi^a \, ({\vec \Delta} \cdot {\vec B}^a)~,~~~{\rm and}~~~H_{M2}=- (4 \,
m_Q)^{-1} \, \xi^a \, S_j \, r_i \, \left ( D_i  B_j(0)
\right )^a ~,
\label{cme}
\ee
where $\xi^a=t_1^a-t_2^a$ is the difference of the color generators
acting on the quark and antiquark (e.g. $t_1^a = \lambda^a/2$ with
$\lambda^a$ being the Gell-Mann matrices),  ${\vec r}$ is the vector
for relative position of the quark and the antiquark, ${\vec S}=({\vec \sigma}_Q
+{\vec \sigma}_{\bar Q})/2$ is the
operator of the total spin of the quark-antiquark pair, and $\vec D$ is the QCD
covariant derivative. Finally ${\vec E}$ and $\vec B$ are the chromoelectric and
chromomagnetic components of the gluon field strength tensor.

An important difference of the hadronic transitions from the radiative ones is
that the physical amplitudes arise in at least the second order in the
interaction with the gluon field due to the color requirements. Considering
quarkonium as a compact object interacting with soft gluonic fields, one can
further approximate the quarkonium transition in the second order in the
interaction Hamiltonian in terms of a local colorless glounic operator, which
operator produces the light mesons in the actual hadronic transition. Thus the
full transition amplitude factorizes into the heavy quarkonium part determined
by the terms in the multipole expansion (Eq.(\ref{cme})) and the production
amplitudes of light mesons by the gluonic operators. The heavy quarkonium part
is to a great extent model dependent, and here we discuss this part only by
relying on the general dynamical properties, while the production of soft light
mesons by local gluonic operators can be described using chiral algebra and
certain low-energy theorems in QCD. In what follows we first concentrate on the
latter description, and then apply the results to specific hadronic transitions.

\subsection{\it Production of Light Mesons by Gluonic Operators}
The structures arising in the second order in the interaction terms in
Eq.(\ref{cme}) are quadratic in the gluon field strength tensor. Therefore the
interesting for the present consideration amplitudes are those for the
production of one or two light pseudoscalar mesons by the operator of the
general form $G^a_{\mu \nu} G^a_{\lambda \sigma}$. As will be described, in the
low-energy limit the two-meson production by this operator is
determined\cite{vz,ns}, up to a small constant, from the chiral algebra and the
QCD anomaly in the trace of the energy-momentum
tensor\cite{crewther,chel,fmw,cdjo}, while the one-meson production amplitude is
fully fixed by the anomaly in the light quark axial current\cite{aanom1,aanom2}.
Furthermore, the matrix element for one-meson production by the operator
containing one extra covariant derivative, $G_{\mu \nu} D G_{\lambda \sigma}$ is
entirely determined\cite{vz,mv03} (in the same low-energy limit) by the fixed
matrix element without the derivative, and is therefore relatively well
understood. The knowledge of the production amplitudes for the light mesons in
fact results in relations between the observed transitions which appear to hold
quite well when compared with the data.
\subsubsection{\it Two-pion Production Amplitude}
Let us consider first the two-pion production amplitude  $\langle \pi^+(p_1)
\pi^-(p_2)| G_{\mu \nu}^a G_{\lambda \sigma}^a|0 \rangle$\,\footnote{The charged
pions are considered for definiteness. The amplitude for the neutral pion pair
production is trivially related by the isospin symmetry.}. In the leading chiral
limit the momenta $p_1$ and $p_2$ of the pions as well as the pion mass $m_\pi$
are
to be considered as small parameters, and the expression for the amplitude,
quadratic in these parameters, can be written in the following general form
\be
- \langle \pi^+(p_1) \pi^-(p_2)| G_{\mu \nu}^a G_{\lambda \sigma}^a|0 \rangle =
\left [ X\, (p_1 \cdot p_2) + Y \,(p_1^2+p_2^2- m_\pi^2) \right ] \, (g_{\mu
\lambda} g_{\nu \sigma} - g_{\mu \sigma} g_{\nu \lambda})+ Z \, t_{\mu \nu
\lambda \sigma}~,
\label{genf}
\ee
where the structure
\begin{eqnarray}
t_{\mu \nu \lambda \sigma}&=&(p_{1 \mu} p_{2\lambda}+p_{1 \lambda} p_{2 \mu}) \,
g_{\nu \sigma} +(p_{1 \nu} p_{2\sigma}+p_{1 \sigma} p_{2 \nu}) \, g_{\mu
\lambda} \nonumber \\ &-& (p_{1 \mu} p_{2\sigma}+p_{1 \sigma} p_{2 \mu}) \,
g_{\nu \lambda}  -
(p_{1 \nu} p_{2\lambda}+p_{1 \lambda} p_{2 \nu}) \, g_{\mu \sigma} - (p_1 \cdot
p_2) \,(g_{\mu \lambda} g_{\nu \sigma} - g_{\mu \sigma} g_{\nu \lambda})
\label{gent}
\end{eqnarray}
has zero overall trace: ${t_{\mu \nu}}^{\mu \nu}=0$, and $X$, $Y$, and $Z$ are
yet undetermined coefficients. The form of the amplitude in Eq.(\ref{genf}) is
uniquely determined by the symmetry (with respect to the indices) of the
operator $G_{\mu \nu} G_{\lambda \sigma}$ and by the Adler zero condition, which
requires that the amplitude goes to zero when either one of the two pion momenta
is set to zero and the other one is on the mass shell. One can also notice that
the proper index symmetry and the Adler zero condition also automatically ensure
that the amplitude is C even, i.e. symmetric under the permutation of the pion
momenta: $p_1 \leftrightarrow p_2$.

The coefficients $X$ and $Y$ in Eq.(\ref{genf}) are in fact determined~\cite{vz}
by the anomaly in the trace of the energy-momentum tensor $\theta_{\mu \nu}$ in
QCD. Indeed, in the low-energy limit in QCD, i.e. in QCD with three light
quarks, one finds
\be
\theta_\mu^\mu=-{b \over 32 \pi^2} \, G^a_{\mu \nu} G^{a \, \mu
\nu}+\sum_{q=u,d,s} m_q (\bar q q)~,
\label{anom}
\ee
where $b=9$ is the first coefficient in the beta function for QCD with three
quark flavors. The first term in Eq.(\ref{anom}) represents the conformal
anomaly, while the quark mass term arises due to the explicit breaking of the
scale invariance by the quark masses. On the other hand, the matrix element of
the
energy-momentum tensor $\theta_{\mu \nu}$ over the pions: $\theta_{\mu
\nu}(p_1,p_2) \equiv\langle \pi^+(p_1) \pi^-(p_2)|\theta_{\mu \nu}|0 \rangle$ is
fully determined~\cite{vz,ns,dv} in the quadratic in $p_1, p_2$ and $m_\pi$
order by
the conditions of symmetry in $\mu$ and $\nu$, conservation on the mass shell
($(p_1+p_2)^\mu \, \theta_{\mu \nu}(p_1,p_2) =0$ at $p_1^2=p_2^2=m_\pi^2$),
normalization  ($\theta_{\mu \nu}(p,-p)=2\, p_\mu p_\nu$ at $p^2=m_\pi^2$), and
the
Adler zero condition ($\theta_{\mu \nu}(p,0)|_{p^2=m_\pi^2}=0$):
\be
\theta_{\mu \nu}(p_1,p_2)=\left [(p_1 \cdot p_2) +p_1^2 +p_2^2 -m_\pi^2 \right
]\,
g_{\mu \nu} -p_{1 \mu} p_{2\nu}- p_{1 \nu} p_{2 \mu}~.
\label{st}
\ee
The equations (\ref{genf}) and (\ref{st}) allow for the pion momenta to be
off-shell in order to demonstrate the Adler zero. However
in what follows only the amplitudes with pions on the mass shell will be
considered, so that it will be henceforth implied that $p_1^2=p_2^2=m_\pi^2$.
In particular one finds for the trace of the expression in Eq.(\ref{st})
\be
\theta_\mu^\mu(p_1,p_2)=2\,(p_1 \cdot p_2) + 4\,m_\pi^2~.
\label{tst}
\ee
Furthermore, the quark mass term in Eq.(\ref{anom}), corresponding to the
explicit breaking of the chiral symmetry in QCD corresponds to the same symmetry
breaking by the pion mass term in the pion theory. Thus one finds to the
quadratic order in $m_\pi^2$:
\be
\langle \pi^+ \pi^- |\sum_{q=u,d} m_q (\bar q q)| 0 \rangle=m_\pi^2~,
\label{mt}
\ee
while the term with the strange quark makes no contribution to the discussed
amplitude.

Combining the formula in Eq.(\ref{anom}) for $\theta_\mu^\mu$ with the
expressions (\ref{tst}) and (\ref{mt}) one finds the matrix element of the
gluonic operator over the pions in the form\footnote{It can be mentioned that
this relation, taking into account the pion mass, was used in Ref.~\cite{mv86},
and was also derived in a particular chiral model in Refs.~\cite{ccgm,ccggm}.}
\be
- \langle \pi^+(p_1) \pi^-(p_2)| G_{\mu \nu}^a G^{a\, \mu \nu}|0 \rangle =
{32 \pi^2 \over b} \, \left [ 2\,(p_1 \cdot p_2) + 3 \,m_\pi^2 \right ]~
\label{gt}
\ee
which thus allows to determine the coefficients $X$ and $Y$ in Eq.(\ref{genf}):
$X=16 \pi^2/(3 b)$ and $Y=3 X/2= 8 \pi^2/b$.

The coefficient $Z$ of the traceless part in Eq.(\ref{genf}) cannot be found
from the trace relation (\ref{anom}). Novikov and Shifman~\cite{ns} estimated
this coefficient by relating this part to the matrix element of the traceless
(twist-two) energy-momentum tensor of the gluons only: $\theta_{\mu \nu}^G = 4
\pi \alpha_s \, (- G_{\mu \lambda}^a {G_{\nu}^a}^\lambda + {1 \over 4} \, g_{\mu
\nu} \, G_{\lambda \sigma}^a G^{a\, \lambda \sigma} )$,
\be
Z\, {t_{\mu \lambda \nu}}^\lambda = 4 \pi \alpha_s \,  \langle \pi^+(p_1)
\pi^-(p_2)|\theta_{\mu \nu}^G |0 \rangle~.
\label{nsr}
\ee
They then assume that the matrix element of the twist-two operator is
proportional to the traceless part of the phenomenological energy momentum
tensor of the pions,
\be
\langle \pi^+(p_1) \pi^-(p_2)|\theta_{\mu \nu}^G |0 \rangle= \rho_G \, \left [
{1 \over 2} \, (p_1 \cdot p_2) \, g_{\mu \nu} - p_{1 \mu} p_{2\nu}- p_{1 \nu}
p_{2 \mu} \right ]
\label{rhog}
\ee
with the proportionality coefficient  interpreted, similarly to the deep
inelastic scattering, as ``the fraction of the pion momentum carried by gluons".
They further introduce a related parameter $\kappa = b \alpha_s \rho_G/(6 \pi)$
and conjecture that numerically $\kappa \approx 0.15 - 0.20$. However, for the
purpose of considering the two-pion transitions it is not necessary to pursue
the interpretation of $\kappa$ as being related to the gluon structure function
of pion, but rather it can be treated as a phenomenological parameter which can
be determined from the data.

Summarizing the results so far in this section one can write the expression for
the general matrix element (\ref{genf}) for on-shell pions as
\be
- \langle \pi^+(p_1) \pi^-(p_2)| F_{\mu \nu}^a F_{\lambda \sigma}^a|0 \rangle =
{8 \pi^2 \over 3 b} \,\left [ (q^2+m_\pi^2) \, (g_{\mu \lambda} g_{\nu \sigma} -
g_{\mu \sigma} g_{\nu \lambda})- {9 \over 2} \, \kappa \, t_{\mu \nu \lambda
\sigma} \right ] ~,
\label{kapf}
\ee
where $q=p_1+p_2$ is the total four-momentum of the dipion.

Few remarks are due regarding effects of higher order in $\alpha_s$. The trace
term in Eq.(\ref{kapf}) receives no renormalization, provided that the
coefficient $b$ is replaced by $\beta(\alpha_s)/\alpha_s^2$ with
$\beta(\alpha_s)=b \, \alpha_s^2 + O(\alpha_s^3)$ being the full beta function
in QCD. This modification however only affects the overall normalization of the
trace part, and can in fact be absorbed into the definition of the heavy
quarkonium amplitudes. On the contrary, the relative coefficient of the
traceless term in Eq.(\ref{kapf}), i.e. the parameter $\kappa$, does depend on
the normalization scale, which scale is appropriate to be chosen as the
characteristic size of the heavy quarkonium~\cite{ns}. However, given other
uncertainties in the analysis of the two-pion transitions, the slow logarithmic
variation of $\kappa$ is a small effect.

\subsubsection{\it Production of $\eta$ by Gluonic Operators}
The amplitude of production of the $\eta$ meson by the quadratic gluonic
operator is described by only one form factor:
$\langle \eta | G_{\mu \nu}  G_{\lambda \sigma}| 0
\rangle = A\,\epsilon_{\mu \nu \lambda \sigma}$ and is therefore entirely
determined by the expression\cite{aanom1,aanom2} following from the chiral
algebra and the anomaly in the divergence of the singlet axial current in QCD,
\be
\epsilon^{\mu \nu \lambda \sigma} \, \langle \eta | G_{\mu \nu}  G_{\lambda
\sigma}| 0
\rangle = 16 \pi^2 \, \sqrt{2 \over 3}
\, F_\eta \, m_\eta^2~,
\label{eanom}
\ee
where $F_\eta$ is the $\eta$ `decay
constant', equal to the pion decay constant $F_\pi \approx 130 \, MeV$
in the limit of exact flavor SU(3) symmetry, and $F_\eta$ is likely to
be larger due to effects of the SU(3) violation.

A more interesting case is presented by the matrix elements relevant to
discussion of transitions in quarkonium involving the $M2$ interaction from
Eq.(\ref{cme}) and containing an extra covariant derivative:
$\langle \eta(p) | G_{\mu \nu} D_\rho G_{\lambda \sigma}| 0
\rangle$ and $\langle \eta |(D_\rho G_{\mu \nu})^a G_{\lambda \sigma}^a|0
\rangle$. It turns out that these matrix elements are also entirely determined
by the relation (\ref{eanom}). The
possibility of the reduction of the structures with extra derivative to the
expression in Eq.(\ref{eanom}) follows from the general
theory\cite{sv}. The reasoning in this specific case\cite{mv03} makes use of the
following identity, valid for arbitrary four-vector $p$:
\be
p_\rho \epsilon_{\mu \nu \lambda \sigma}=p_\lambda \epsilon_{\mu \nu
\rho \sigma}-p_\sigma \epsilon_{\mu \nu \rho \lambda}-p_\mu
\epsilon_{\nu \rho \lambda \sigma}+p_\nu \epsilon_{\mu \rho\lambda
\sigma}~,
\label{eid}
\ee
where the convention $\epsilon_{0123}=1$ is assumed. The antisymmetry of
the field tensor $G_{\mu \nu}$ then allows one to write the general form
of the first of the discussed matrix elements in the linear order in the
$\eta$ momentum $p$ in terms of two scalars $X$ and $Y$:
\be
i \,\langle \eta(p) |G_{\mu \nu}^a (D_\rho G_{\lambda \sigma})^a|0
\rangle=X \, p_\rho \epsilon_{\mu \nu \lambda \sigma}+ Y \, \left (
p_\lambda \epsilon_{\mu \nu \rho \sigma}-p_\sigma \epsilon_{\mu \nu \rho
\lambda} \right )~.
\label{gst1}
\ee
The third structure, allowed by the symmetry and proportional to $(p_\mu
\epsilon_{\nu \rho \lambda \sigma}-p_\nu \epsilon_{\mu \rho\lambda
\sigma})$, is reduced to the first two due to the identity (\ref{eid}).
Furthermore, applying in Eq.(\ref{gst1}) the equations of motion (the
Jacobi identity): $D_\rho G_{\lambda \sigma}+D_\sigma G_{\rho \lambda
}+D_\lambda G_{\sigma \rho }=0$, one arrives at the relation $X=2Y$.

Likewise, writing the second of the discussed matrix elements in terms
of two scalars ${\tilde X}$ and ${\tilde Y}$ as
\be
i \, \langle \eta(p) |(D_\rho G_{\mu \nu})^a G_{\lambda \sigma}^a|0
\rangle={\tilde X} \, p_\rho \epsilon_{\mu \nu \lambda \sigma}+ {\tilde
Y} \, \left ( p_\mu \epsilon_{\lambda \sigma \rho \nu}-p_\nu
\epsilon_{\lambda \sigma \rho \mu} \right )~,
\label{gst2}
\ee
and applying the Jacobi identity, one finds the relation ${\tilde X}=2
{\tilde Y}$.

The sum of the expressions (\ref{gst1}) and (\ref{gst2}) should combine
into a total derivative, i.e. the sum should be proportional to
$p_\rho$. This
is possible due to the identity (\ref{eid}) under the condition that
${\tilde Y}=Y$, so that all the considered scalar form factors are
expressed in terms of one of them, e.g. in terms of $X$:\,\footnote{An
alternative derivation of two of these relations, namely ${\tilde X}=X$
and ${\tilde Y}=Y$, would be by arguing that in the particular
amplitudes
(\ref{gst1}) and (\ref{gst2}) the operators $G^a$ and $(DG)^a$ can be
considered as commuting with each other, so that the expressions
(\ref{gst1}) and (\ref{gst2}) differ only by re-labeling the indices.}
\be
{\tilde X}=X,~~~~{\tilde Y}=Y={1 \over 2}\,X~.
\label{xy}
\ee
Using this relation and contracting the sum of the expressions
(\ref{gst1}) and (\ref{gst2}) with ${1 \over 2}\epsilon^{\mu \nu \lambda
\sigma}$ the form factor $X$ is identified from the equation
(\ref{eanom}) as
\be
X=-{1 \over 60} \, \epsilon^{\mu \nu \lambda \sigma} \, \langle \eta | G_{\mu
\nu}  G_{\lambda \sigma}| 0
\rangle  =- {4
\pi^2 \over 15} \, \sqrt{2 \over 3} \, F_\eta \, m_\eta^2~.
\label{xeta}
\ee

\subsubsection{\it Production of Single $\pi^0$ by Gluonic Operators}
The production of a single neutral pion by gluonic operators obviously requires
a
breaking of the isotopic symmetry. Within the chiral approach the isospin
breaking originates in the different masses of the up and down quarks, $m_u \neq
m_d$. Allowing for this mass difference, the gluonic matrix elements for a
single $\pi^0$ can be considered in essentially the same way as those for the
$\eta$ meson. In particular the `master equation' for the anomalous amplitude in
the case of pion takes the form\cite{aanom1}
\be
\epsilon^{\mu \nu \lambda \sigma} \, \langle \eta | G_{\mu \nu}  G_{\lambda
\sigma}| 0
\rangle = 16 \pi^2 \, \sqrt{2 }
\, {m_d-m_u \over m_d+m_u} \,  F_\pi \, m_\pi^2~.
\label{pianom}
\ee
Thus the simple rule of conversion from the amplitudes of the processes with
$\eta$ emission to similar processes with single $\pi^0$ can be done by
replacing the coefficient $F_\eta \, m_\eta^2 \to \sqrt{3} \, F_\pi \, m_\pi^2
\, (m_d-m_u)/(m_d+m_u)$ and replacing the $\eta$ momentum with that of the pion,
$p_\eta \to p_\pi$. Naturally, these replacement rules can be also viewed in
terms of a fixed $\eta-\pi$ mixing.

\subsection{\it Two-pion Transitions}
\subsubsection{\it The Transition $\psi' \to \pi \pi J/\psi$}
The general soft-pion relations of the chiral algebra require that the amplitude
of a two-pion transition is bilinear in the components of the four-momenta of
the pions\cite{bc,mv75} in the chiral limit. For the most studied transition
$\psi' \to \pi \pi J/\psi$ this implies that in the limit of soft pions the
amplitude can be generally parametrized as\cite{bc}
\bea
A(\psi' \to \pi^+ \pi^- J/\psi) &=&
\left [ A \, (q^2-2 m_\pi^2) + \lambda \, m_\pi^2 \right ] \, \left ( \epsilon_1
\cdot \epsilon_2 \right ) + B \, E_1 E_2 \, \left ( \epsilon_1 \cdot \epsilon_2
\right ) \nn \\ &+& C \, \left [ (p_1 \cdot \epsilon_1) (p_2 \cdot \epsilon_2)+
(p_2 \cdot \epsilon_1) (p_1 \cdot \epsilon_2) \right ]~,
\label{abc}
\eea
where $\epsilon_1^{\mu}$ and $\epsilon_2^\mu$ are the polarization amplitudes of
the initial and the final vector resonances, $p_1$ and $p_2$ are the 4-momenta
of the two pions, $E_1$ and $E_2$ are their energies in the rest frame of the
initial state, and $q=p_1+p_2$ is the total 4-momentum of the dipion, so that
$q^2=m_{\pi \pi}^2$. Finally, $A$, $B$, $C$, and $\lambda$ are the form factors,
which
should be considered constant in the soft pion limit, but are generally
functions of kinematic variables beyond this limit. The spin-dependent term,
proportional to $C$ should be suppressed inasmuch as the charmed quark can be
considered as heavy, since the spin-dependent interaction is proportional to the
inverse of the heavy quark mass. The remaining constants in Eq.(\ref{abc}) can
be related to the parameters in the leading order of the multipole expansion in
QCD.

The two-pion transition between $^3S_1$ states is generated in the second order
in the leading $E1$ term (Eq.(\ref{cme}) of the multipole expansion, so that the
amplitude of the process can be written as
\be
A(\psi' \to \pi^+ \pi^- J/\psi) = {1 \over 2} \langle \pi^+ \pi^- | E^a_i E^a_j
| 0 \rangle \, \alpha^{(12)}_{i j}~,
\label{e1ampg}
\ee
where $\alpha^{(12)}_{i j}$ can be termed, in complete analogy with the atomic
properties in electric field, as the transitional chromo-polarizability of the
quarkonium. In other words, the $\psi_2 \to \psi_1$ transition in the
chromo-electric field is described by the effective Hamiltonian
\be
H_{eff}=-{1 \over 2}\, \alpha^{(12)}_{i j} \, E^a_i E^a_j~,
\label{heff}
\ee
with the chromo-polarizability given by
\be
\alpha^{(12)}_{i j}={1 \over 16} \, \langle 1S | \xi^a \, r_i \, {\cal G} \, r_j
\, \xi^a | 2S \rangle~,
\label{aij}
\ee
where ${\cal G}$ is the Green's function of the heavy quark pair in a color
octet state. The latter function is not well understood presently, so that an
{\it ab initio} calculation of the chromo-polarizability would be at least
highly model dependent. Generally $\alpha_{ij}$ is a symmetric tensor. In the
leading nonrelativistic order the coordinate and spin degrees of freedom
factorize, so that the discussed chromo-polarizability for a transition between
$S$ wave states is actually reduced to a scalar:
\be
\alpha^{(12)}_{i j}=\alpha^{(12)} \, \delta_{ij} \, (\vec \epsilon_1 \cdot \vec
\epsilon_2)~.
\label{a12}
\ee

The matrix element for production of the pion pair by the quadratic
chromoelectric operator is then found from the general expression (\ref{kapf}),
and the result for the amplitude of the decay $\psi' \to \pi^+ \pi^- J/\psi$
is\cite{vz,ns,mv06}
\bea
\label{apipi0}
&&A(\psi' \to \pi^+ \pi^- J/\psi)= \\ \nonumber
&&-{4 \pi^2 \over b} \, \alpha^{(12)} \, \left [ (q^2+m_\pi^2)-  \kappa \, \left
(1+{2 m_\pi^2 \over q^2} \right)\,  \left ( {(q\cdot P)^2 \over P^2} - {1 \over
4} \, q^2 \right )+ {3 \kappa \over 2} \, {\ell_{\mu \nu} P^\mu P^\nu \over P^2}
\right ] \,(\epsilon_1 \cdot \epsilon_2)~,
\eea
where $P$ is the 4-momentum of the initial $\psi'$ resonance, and the tensor
$\ell_{\mu \nu}$ corresponds to a $D$ wave motion in the c.m. frame of the
dipion and is written in terms of the relative momentum $r=p_1-p_2$ of the two
pions as
\be
\ell_{\mu \nu}=r_\mu r_\nu  + {1 \over 3}\, \left ( 1- {4 m_\pi^2 \over q^2}
\right ) (q^2 \, g_{\mu \nu} - q_\mu q _\nu)~.
\label{lmn}
\ee

The constants in the parametrization (\ref{abc}) can be thus found as
\be
A=-{4 \pi^2 \over b} \, \alpha^{(12)} \, \left ( 1+ {3 \over 4} \, \kappa \right
)~, ~~~~~ B = 6 \, \kappa \, {4 \pi^2 \over b} \, \alpha^{(12)}~,~~~~~ \lambda=
{12 \kappa \over 4 + 3 \, \kappa}~,
\label{abckap}
\ee
while the parameter $C$ is clearly equal to zero, since the $E1$ interaction
carries no spin dependence. It can be also noted that Eq.(\ref{apipi0}) differs
from (\ref{abc}) at $C=0$ in that it contains only two parameters, while the
third, $\lambda$ is fully fixed as shown in Eq.(\ref{abckap}). This is due to
the chiral constraints and the conformal anomaly (Eq.(\ref{anom})) fully
determining the terms of order $m_\pi^2$ relative to those quadratic in the pion
momenta.

\begin{figure}[tb]
\begin{center}
\begin{minipage}[t]{18 cm}
   \epsfxsize=9cm
    \epsfbox{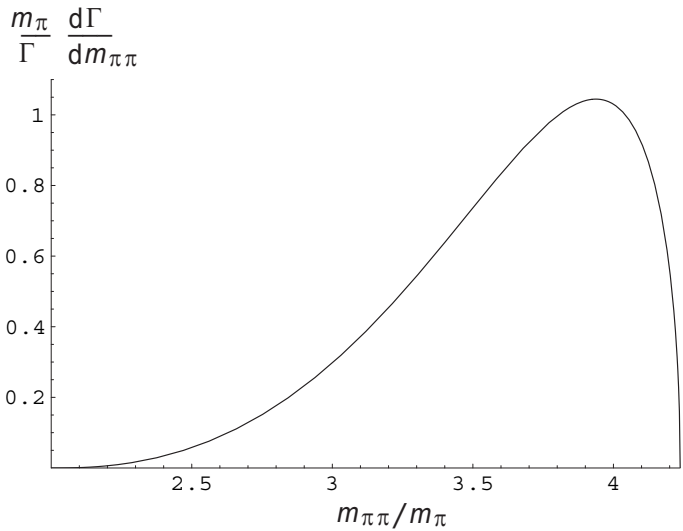}
\end{minipage}
\begin{minipage}[t]{16.5 cm}
\caption{The spectrum of the dipion invariant masses $m_{\pi \pi}$ in the decay
$\psi' \to \pi^+ \pi^- \, J/\psi$ as given by the amplitude (\ref{apipi0}) with
$\kappa=0.2$. \label{pipispec}}
\end{minipage}
\end{center}
\end{figure}

One can readily see from Eq.(\ref{apipi0}) that the overall rate of the decay is
determined by the chromo-polarizability $\alpha^{(12)}$, while the shape of the
spectrum depends on the parameter $\kappa$. Furthermore, the last term in the
braces in Eq.(\ref{apipi0}) is proportional to $\kappa$ and describes the $D$
wave motion in the c.m. frame of the two pions, which $D$ wave correlates with
the motion of the dipion as a whole in the rest frame of the decaying $\psi'$.
Thus the formula (\ref{apipi0}) based on the leading $E1$ term of the multipole
expansion predicts\cite{ns} a relation between the amplitude of the $D$ wave and
the parameter in the shape of the spectrum in the decay. The most detailed to
date experimental study of the decay  $\psi' \to \pi^+ \pi^- J/\psi$ using this
formula was done by BES\cite{bes}. The fit to the spectrum of the invariant mass
of the produced dipion resulted in the value $\kappa=0.186 \pm 0.003 \pm 0.006$,
while the fit to the ratio of the $D$ and $S$ wave amplitudes from the angular
distribution gave $\kappa=0.210 \pm 0.027 \pm 0.042$. Clearly, the consistency
of these two values implies that the discussed approach correctly predicts the
ratio of the $D$ wave in terms of the sub-dominant term proportional to $\kappa$
in the $S$ wave amplitude. It can be also noted that at $\kappa \approx 0.2$ the
contribution of the $D$ wave to the total rate is quite small -- only about
2\%\,\cite{bes}. The familiar characteristic shape of the spectrum of the
two-pion invariant masses is shown at $\kappa=0.2$ in Fig.\ref{pipispec}.
Comparing the
integral over this spectrum with the experimental total decay rate one can
evaluate the transitional chromo-polarizability as $\alpha^{(12)} \approx
2\,$GeV$^{-3}$.

It can be also noted that in the discussed picture the pions are produced in
pure isoscalar $I=0$ state, which implies fixed relation between the
differential rates of emission of pairs of neutral and of charged pions
$d\Gamma(\psi' \to \pi^0 \pi^0 J/\psi)= 0.5 d\Gamma(\psi' \to \pi^+ \pi^-
J/\psi)$. The most precise test of this relation has been done by the CLEO-c
experiment\cite{cleopsipsi}, which resulted in the measured ratio of the total
decay rates $0.4924 \pm 0.0047 \pm 0.0086$.

\subsubsection{\it Effects of the Final State Interaction Between Pions}
So far the amplitude of the two-pion transition was considered in the chiral
limit. The formula in Eq.(\ref{kapf}) is exact in the leading chiral order, i.e.
as far as only the quadratic terms in the pion momenta and mass are concerned.
In particular this expression receives no corrections due to the final state
interaction (FSI) between the pions. The latter interaction however can give
rise to the terms whose expansion starts with the fourth power of momenta and
the pion mass, and generally can modify the amplitude at momenta of the pions
relevant for actual transitions in quarkonium. The effects of FSI in chiral
treatment of the two-pion transitions in quarkonium were a matter of concern
ever since the earlier theoretical analyses~\cite{bc}. The effect in the phases
of the amplitudes is well known: these phases for the production amplitudes are
equal to the two-pion scattering phases in the corresponding partial waves:
$S=|S| \, \exp(i \delta_0)$, $D=|D| \, \exp(i \delta_2)$, where the $I=0$ phases
for the $S$ wave, $\delta_0$, and for the $D$ wave, $\delta_2$ are quite well
studied.\footnote{It can be mentioned that the analysis~\cite{bes} of the data
on the $\psi' \to \pi^+ \pi^- J/\psi$ decay does not take into account the
relative phase between the $S$ and $D$ wave pion production amplitudes. Thus it
would be interesting to know whether including the phase factor in the angular
analysis, produces a significant impact on the results.} It is also generally
estimated both on theoretical and phenomenological grounds that the FSI
corrections are not big (at most 20 - 25\%) in the transitions $\psi' \to \pi
\pi J/\psi$ and $\Upsilon' \to \pi \pi \Upsilon$. (For a discussion see the
review~\cite{vzai}.) Some phenomenological arguments in favor of such estimate
will also be discussed here after presenting  a theoretical estimate of the
onset of the higher term in the chiral expansion.

The interaction of pions at low energy in the $D$ wave is quite weak, so that
any modification by FSI of the $D$ wave production amplitude  can be safely
neglected, and only the modification of the $S$ wave amplitude is of interest
for present phenomenology. The imaginary part of the correction at $q^2 > 4m^2$
is found from the unitarity relation in terms of the isospin I=0 $\pi \pi \to
\pi \pi$ scattering amplitude $T(q^2)$ in the $S$ wave as
\be
{\rm Im}\, ( \delta S) = \sqrt{1-{4m_\pi^2 \over q^2}}\, {T(q^2) \over 16 \pi}
\, S ~.
\label{unit}
\ee
The amplitude $T(q^2)$ is well known in the chiral limit, i.e. in the quadratic
in $q$ and $m_\pi$ approximation, since the work of Weinberg~\cite{weinberg}. In
the normalization used here this amplitude has the form
\be
T(q^2)={2 \, q^2 - m_\pi^2 \over F_\pi^2}~,
\label{tpipi}
\ee
where $F_\pi \approx 130\,$MeV is the $\pi^+ \to \mu^+ \nu$ decay constant.
Clearly, the expression in Eq.(\ref{unit}) is of the fourth power in $q$ and
$m$.

The real part of the correction to the $S$ wave production amplitude $\delta S$
can then be estimated  from Eq.(\ref{unit}) using the dispersion relation in
$q^2$ for the amplitude $S$. In doing so one should set the condition for the
subtraction constants that this real part does not contain quadratic (and
certainly also constant) terms in $q$ and $m$, since these are given by
Eq.(\ref{kapf}). After these subtractions the remaining dispersion integral is
still logarithmically divergent and contains the well known `chiral logarithm',
depending on the ultraviolet cutoff $\Lambda$, which is usually set at $\Lambda
\sim 1\,$GeV, i.e. the scale where any chiral expansion certainly breaks down.
Using this approach one can estimate\cite{mv06} the corrections in the $S$ wave
part of the amplitude given by Eq.(\ref{apipi0}) as
\bea
\label{apipi1}
&&A(\psi' \to \pi^+ \pi^- J/\psi)= \\ \nonumber
&&-{4 \pi^2 \over b} \, \alpha^{(12)} \, \left [ (q^2+m_\pi^2)(1+\xi_1)-  \kappa
\, \left (1+{2 m_\pi^2 \over q^2} \right)\, (1+\xi_2) \, \left ( {(q\cdot P)^2
\over P^2} - {1 \over 4} \, q^2 \right )+ {3 \kappa \over 2} \, {\ell_{\mu \nu}
P^\mu P^\nu \over P^2} \right ] \,(\epsilon_1 \cdot \epsilon_2)~,
\eea
where the correction terms $\xi_1$ and $\xi_2$  are given by
\be
\xi_1 = {2 \, (q^2)^2- 7 \, q^2 \, m_\pi^2 +3 \, m_\pi^4 \over 16 \pi^2 \,
F_\pi^2 \, (q^2+m_\pi^2)}\, \ln {\Lambda^2 \over m_\pi^2}+ {\rm i} \, {2 \, q^2
- m_\pi^2 \over 16 \pi \, F_\pi^2}\, \sqrt{1-{4m_\pi^2 \over q^2}} ~,
\label{xi1}
\ee
and
\be
\xi_2 = {2 \, (q^2)^2- 9 \, q^2 \, m_\pi^2 +8 \, m_\pi^4 \over 16 \pi^2 \,
F_\pi^2 \, (q^2+2 m_\pi^2)}\, \ln {\Lambda^2 \over m_\pi^2}+{\rm i} \, {2 \, q^2
-
m_\pi^2 \over 16 \pi \, F_\pi^2} \, \sqrt{1-{4m_\pi^2 \over q^2}}~,
\label{xi2}
\ee
where the non-logarithmic imaginary part is retained for reference regarding the
normalization. The lower limit under the logarithm is generally a function of
both $q^2$ and $m_\pi^2$, however any difference of this function from the value
$m^2$ used in Eqs.(\ref{xi1}) and (\ref{xi2}) is a non-logarithmic term, i.e.
beyond the accuracy of these equations. Since $m_\pi^2$ is the smallest of the
two parameters in the physical region of pion production, it can be expected
that using this parameter provides a conservative estimate of the effect of FSI.

Estimating the corrections in Eq.(\ref{xi1}) and Eq.(\ref{xi2}), one finds that
at the lower end of the physical phase space, i.e. near $q^2=4 \, m_\pi^2$,
these terms do not exceed few percent. Thus the corrections only weakly modify
the normalization of the pion production amplitude near the threshold. A
theoretical extrapolation to higher values of $q^2$ is problematic, and, most
likely, one would have to resort to using actual data on the dipion spectra in
order to judge upon the significance of deviation from the essentially linear in
$q^2$ behavior of the amplitude described by Eq.(\ref{apipi0}). A quantitative
estimate of the deviation from this behavior has been attempted~\cite{vzai}
using the data~\cite{argus} on $\Upsilon' \to \pi^+\pi^- \Upsilon$ by
parametrizing the deviation as a factor $(1+q^2/M^2)$ in the amplitude with $M$
being a parameter. The thus obtained lower limit on $M$ is $1\,$GeV at 90\% C.L.

Another phenomenological argument in favor of a relatively moderate FSI effect
in the absolute value of the dominant $S$ wave in the decay $\psi' \to \pi^+
\pi^- J/\psi$ stems from the previously mentioned agreement of the
observed~\cite{bes} value of the ratio $D/S$ with the parameter $\kappa$
entering the expression for the $S$ wave and extracted from the two-pion
invariant mass spectrum. Clearly such an agreement would be ruined if there was
a significant enhancement of the $S$ wave by FSI.

The treatment of the FSI effects in the two-pion transition amplitude in fact
reveals a certain inadequacy of the parametrization in Eq.(\ref{abc}). Indeed
the factors $B$ and $C$ are each contributed by both the $S$ and $D$ wave motion
in the c.m. frame of the two pions, and those contributions are differently
modified by the FSI effects. In this situation it is more reasonable to use the
parametrization\cite{mv06,dv07} in terms of separate partial waves:
\be
A(\psi' \to \pi^+ \pi^- J/\psi)=S \, (\epsilon_1 \cdot \epsilon_2) + D_1 \,
\ell_{\mu \nu} \, {P^\mu P
^\nu \over P^2} \, (\epsilon_1 \cdot \epsilon_2) + D_2\, q_\mu \, q_\nu \,
\epsilon^{\mu \nu}+ D_3 \, \ell_{\mu \nu} \, \epsilon^{\mu \nu}~,
\label{sddd}
\ee
where  $\epsilon^{\mu \nu}$ stands for the spin-2 tensor made from the
polarization amplitudes of the $^3S_1$ resonances
\be
\epsilon^{\mu \nu}=\epsilon_1^\mu \epsilon_2^\nu +  \epsilon_1^\nu
\epsilon_2^\mu +{2 \over 3} \, (\epsilon_1 \cdot \epsilon_2) \, \left ( {P^\mu P
^\nu \over P^2} - g_{\mu \nu} \right )~.
\label{emn}
\ee
The terms in the expression (\ref{sddd}) describe an $S$ wave and three possible
types of $D$-wave motion: the term with $D_1$ corresponds to a $D$ wave in the
c.m. system of the two pions correlated with the overall motion of the dipion in
the rest frame of the initial state, the $D_2$ term describes the $D$-wave
motion of the dipion as a whole, correlated with the spins of the quarkonium
resonances, and finally, the $D_3$ term corresponds to the correlation between
the spins and the $D$-wave motion in the c.m. frame of the dipion. Clearly, the
factors $S$ and $D_2$ are modified by the FSI in the $S$-wave dipion, while the
factors $D_1$ and $D_3$ receive a modification from the two-pion $D$ wave FSI.
The partial-wave form factors are expressed in terms of the parameters in
Eq.(\ref{abc}) as
\bea
S &=& \left ( A+ {1 \over 3} \, C \right ) \, (q^2-2 \, m_\pi^2) + \lambda \,
m_\pi^2 + {1 \over 12} \, \left ( B-  {2 \over 3} \, C \right ) \, \left [ 3 \,
q_0^2 - (q_0^2-q^2) \, \left ( 1- {4 m_\pi^2 \over q^2} \right ) \right ] \nn \\
D_1 &=& - {1 \over 4 } \, \left ( B-  {2 \over 3} \, C \right ) ~,  ~~~
D_2 = {1 \over 6} \, C \,  \left ( 1+ {2 m_\pi^2 \over q^2} \right ) ~,~~~
D_3 = - {1 \over 4} \, C~.
\label{sdab}
\eea
Naturally, in the presented treatment, based on the multipole expansion in the
leading nonrelativistic order, only the spin-independent amplitudes $S$ and
$D_1$ are present, as described by Eq.(\ref{apipi0}) and Eq.(\ref{apipi1}).

\subsection{\it Single $\eta$ and Single Pion Transitions}
\subsubsection{$\psi' \to \eta \, J/\psi$}
The decay $\psi' \to \eta J/\psi$ arises due to the interference of the $E1$ and
$M2$ terms in the multipole expansion (Eq.(\ref{cme})).  The amplitude of the
transition can thus be written as
\be
A(\psi' \to \eta \, J/\psi)= m_Q^{-1} \, \left
\langle \eta \left |  E^a_i \, (D_j B_k)^a +  (D_j B_k)^a \, E^a_i
\right | 0 \right \rangle \, A_{ijk}~,
\label{aeta}
\ee
where
\be
A_{ijk}={1 \over 64} \,  \langle 2^3S_1 | \xi^a r_i {\cal G} r_j
\xi^a S_k | 1^3S_1 \rangle
\label{aijk}
\ee
is the quarkonium transition amplitude. In the approximation of factorized
coordinate and spin degrees of freedom the matrix element of the spin operator
is expressed in terms of the polarization amplitudes $\vec \epsilon_1$ and $\vec
\epsilon_2$ of the initial and final quarkonium resonances, while the remaining
coordinate overlap is the same as in the expression (\ref{aij}) for the
chromo-polarizability, so that the amplitude is expressed in terms of
$\alpha^{(12)}$ as
\be
A(\psi' \to \eta \, J/\psi)= {\rm i}\, {\alpha^{(12)}\over 4 \, m_Q} \, \left
\langle \eta \left |  E^a_i \, (D_i B_k)^a +  (D_i B_k)^a \, E^a_i
\right | 0 \right \rangle \, \epsilon_{klm} \, \epsilon_{1l} \, \epsilon_{2m}~,
\label{aeta1}
\ee
The amplitude of the $\eta$ production  by the gluonic operator in this
expression is readily found from the relations (\ref{gst1}) - (\ref{xeta}), and
is given by
\be
{\rm i}\, \left
\langle \eta \left |  E^a_i \, (D_i B_k)^a +  (D_i B_k)^a \, E^a_i
\right | 0 \right \rangle = {16 \, \pi^2 \over 15} \, \sqrt{2 \over 3} \, F_\eta
\, m_\eta^2 \, p_k
\label{geta}
\ee
with $\vec p$ being the momentum of the $\eta$ meson. As a result one finds the
following simple formula for the amplitude of the $\eta$ transition
\be
A(\psi' \to \eta \, J/\psi)= {4 \, \pi^2 \over 15} \, \sqrt{2 \over 3}
\,{\alpha^{(12)}\over  m_Q} \,  F_\eta \, m_\eta^2 \, \epsilon_{klm} \, p_k \,
\epsilon_{1l} \, \epsilon_{2m}~.
\label{aeta2}
\ee

Comparing the latter formula with Eq.(\ref{apipi0}) one finds that the
quarkonium part of the amplitude, $\alpha^{(12)}$, cancels in the ratio of the
amplitudes for the $\eta$ and the two-pion emission and the ratio is essentially
determined by the two anomalies in QCD: the conformal anomaly and the one in the
divergence of the singlet axial current\cite{vz}. This remarkable relation can
be written in terms of the decay rates as
\be
{\Gamma(\psi' \to  \eta \, J/\psi) \over d\Gamma(\psi' \to  \pi^+ \pi^-
J/\psi)/dq^2}={64\pi^2 \over 25} \, {F_\eta^2 \over m_c^2} \,  {p_\eta^3 \over
|{\vec q}|}
 \, \left ( {m_\eta^2 \over q^2 } \right )^2 \, \left (1- {4
m_\pi^2 \over q^2} \right )^{-1/2} \, {1 \over {\cal F}}~,
\label{gamrat}
\ee
where the factor ${\cal F}$ describes the deviation of the two-pion transition
amplitude from the limit of dominance of the anomaly contribution ($\kappa \to
0$) and of massless pion:
\be
{\cal F}= \left | 1+ {m_\pi^2 \over q^2}-{\kappa \over q^2} \, \left (1+{2
m_\pi^2 \over q^2} \right)\,  \left [ {(q\cdot P)^2 \over P^2} - {1 \over 4} \,
q^2 \right ] \right |^2 + {\kappa^2 \over 5} \, \left [ {(q\cdot P)^2 \over q^2
\, P^2} -1 \right ]^2 \, \left ( 1- {4 \, m_\pi^2 \over q^2} \right )^2~,
\label{calf}
\ee
where the last term describes the small contribution of the $D$ wave.
Using $\kappa=0.2$ and integrating over the phase space of the two-pion
transition one can arrive at the estimate of the ratio of the total rates:
\be
{\Gamma(\psi' \to  \eta \, J/\psi) \over \Gamma(\psi' \to  \pi^+ \pi^- J/\psi)}
=0.09 \, \left ( {F_\eta \over 130\,{\rm MeV}} \right )^2 \, \left ( {1.4 \,
{\rm GeV} \over m_c} \right )^2~.
\label{gamratn}
\ee
Given all the uncertainties, this estimate agrees very well with experimental
value $0.097 \pm 0.003$ for this ratio.

It can be mentioned as a sidenote, that being applied to the transitions in
bottomonium between $\Upsilon'$ and $\Upsilon$, Eq.(\ref{gamrat}), using the
appropriate for these transitions value $\kappa \approx 0.15$, predicts the
ratio of the rates
\be
{\Gamma(\Upsilon' \to  \eta \, \Upsilon) \over \Gamma(\Upsilon' \to  \pi^+ \pi^-
\Upsilon)}
=\left ( 2.2 \times 10^{-3} \right ) \, \left ( {F_\eta \over 130\,{\rm MeV}}
\right )^2 \, \left ( {4.8 \,
{\rm GeV} \over m_b} \right )^2~,
\label{gamratups}
\ee
corresponding to ${\cal B}(\Upsilon' \to  \eta \, \Upsilon) \approx 4.3 \times
10^{-4}$. This prediction can be compared with the recent preliminary data from
CLEO\cite{cleoupseta}: ${\cal B}(\Upsilon' \to  \eta \, \Upsilon) = (2.5 \pm 0.7
\pm 0.5) \times 10^{-4}$.

Another sidenote regarding the decay $\psi' \to  \eta \, J/\psi$ is that
recently this process has been used not for its own sake, but rather as a
precision source of the $\eta$ mesons for high accuracy measurements of the
$\eta$ mass\cite{cleoetam} and its decays\cite{cleoetab}.

\subsubsection{$\psi' \to \pi^0 \, J/\psi$}
The decay $\psi' \to \pi^0 \, J/\psi$ requires breaking of the isotopic
symmetry. If the relevant to this process breaking is due to the mass difference
between the $u$ and $d$ quarks the amplitude of this decay can be found from the
amplitude of the $\eta$ transition by applying the previously discussed
conversion factor, so that the ratio of the rates for these two transitions can
be estimated\cite{is} as
\be
{\Gamma(\psi' \to \pi^0 \, J/\psi) \over \Gamma(\psi' \to \eta \, J/\psi)}= 3 \,
\left ( {m_d - m_u \over m_d + m_u} \right )^2 \, {F_\pi^2 \over F_\eta^2}  \,
{m_\pi^4 \over m_\eta^4 } \, {p_\pi^3 \over  p_\eta^3}~.
\label{petar}
\ee
In the
limit of the flavor SU(3) symmetry one has $F_\pi=F_\eta$. In reality it is
known from
comparison of $F_\pi$ and $F_K$ that the presence of heavier strange quarks
increases the constant $F$, so that $F_\eta$ is expected to be larger than
$F_\pi$. Therefore the limit $F_\pi=F_\eta$ can be used as an upper bound on the
ratio of the rates in Eq.(\ref{petar}). The ratio of the masses of the $u$ and
$d$ quarks, describing the explicit breaking of the chiral symmetry and the
isospin violation in this breaking was studied years ago in great detail by
Gasser and Leutwyler\cite{gl}. The largest value for the ratio $(m_d-m_u)/(m_d +
m_u)$ allowed by that study is approximately 0.35.
Thus the theoretical upper bound for the ratio of the decay rates in
Eq.(\ref{petar}) is approximately $2.3 \%$, which is still by more than
$4\sigma$ below the experimental result\cite{cleopsipsi}: $(4.1 \pm 0.4 \pm
0.1)$\%.
It can be also mentioned in connection with the light quark mass ratio that the
well known Weinberg's formula\cite{weinberg2} gives
\be
{m_d-m_u \over m_d + m_u }={m_{K^0}^2-m_{K^+}^2+m_{\pi^+}^2-m_{\pi^0}^2 \over
m_{\pi^0}^2} = 0.285~,
\label{weinf}
\ee
and results in a still smaller ratio of the decay rates if used in
Eq.(\ref{petar}).
It is certainly understood\cite{is} that the formula (\ref{petar}) may receive
unexpectedly large corrections from the effects of the SU(3) violation, however
such corrections can also significantly affect the analysis of the chiral
phenomenology in Ref.\cite{gl}, and the whole subject then would have to be
revisited anew. It should be mentioned that the largest theoretical estimate of
the discussed ratio of the transition rates found in the literature\cite{kty}
corresponds to $3.4\%$,
which is also significantly lower than the experimental number. However,
the latter estimate does not fully take into account the proper QCD structure of
the relevant amplitude for the meson production by the gluonic operator.

It therefore looks like the isospin violation by the light quark masses is not
sufficient to describe the data\cite{cleopsipsi} and one has to assume that at
least one of the charmonium states in the transition is in fact not a pure
isoscalar, but contains a small admixture of an isovector $I=1$ four-quark
state. We shall further discuss such possibility in the Section 6.1.2.

\subsubsection{\it $\psi' \to \pi^0 \, h_c$ and $h_c \to \pi^0 \, J/\psi$}
The transitions between the $^3S_1$ and $^1P_1$ states with emission of $\pi^0$
arise in the discussed picture through the interference of the $E1$ and $M1$
interaction terms in Eq.(\ref{cme}). The pion is emitted in the $S$ wave, and
the amplitude for the transition can be written as
\be
A(^3S_1 \to \pi^0 \,  ^3P_1) = \langle \pi^0 | E_i^a B_k^a | 0 \rangle \, I \,
\epsilon_{1k} \epsilon_{2i}~,
\label{apsih}
\ee
where $I$ is the quarkonium radial overlap integral
\be
I= -{1 \over 96 \, m_Q} \, \langle P |\, \xi^a \, (r \, {\cal G}_S + {\cal G}_P
\, r) \, \xi^a |S \rangle
\label{ipsih}
\ee
with ${\cal G}_S$ and ${\cal G}_S$ being the octet-state Green's function in the
corresponding partial waves.

The gluonic matrix element for the single pion production is readily found from
Eq.(\ref{pianom}), and one finds the following expression for the decay
rate\cite{mv86:2}
\be
\Gamma(^3S_1 \to \pi^0 \,  ^3P_1) = \left ( {2  \pi^2 \over 3} \, {m_d-m_u \over
m_d+m_u} \, F_\pi \, m_\pi^2 \right )^2 \, |I|^2 \, {p_\pi \over \pi}~.
\label{gpsih}
\ee

A numerical evaluation of the decay rate for the transition $\psi' \to \pi^0 \,
h_c$ greatly suffers from a poor knowledge of the overlap integral in
Eq.(\ref{ipsih}).  If, for an order-of-magnitude estimate, one uses typical $r$
as in the radiative transitions, $r \sim 0.4\,$fm, and also approximates ${\cal
G} \sim 1\,$GeV$^{-1}$, then one very approximately estimates $|I| \sim
0.15\,$GeV$^{-3}$, and also estimates the rate from Eq.(\ref{gpsih}) as
$\Gamma(\psi' \to \pi^0 \, h_c) \sim 15\,$eV, which corresponds to the branching
fraction of only about $5 \times 10^{-5}$. The latter number is by an order of
magnitude smaller than the measured\cite{cleohc} combined branching fraction
${\cal B}(\psi' \to \pi^0 \, h_c) \times {\cal B}(h_c \to \gamma \, \eta_c)=
(4.0 \pm 0.8 \pm 0.7) \times 10^{-4}$. It is not clear whether the estimate of
the overlap $I$ is totally off the mark, or the apparently enhanced decay $\psi'
\to \pi^0 \, h_c$ also proceeds due to a small four-quark admixture in the
$\psi'$.

A similar estimate of at most few tens eV is applicable to the transition from
the $h_c$ resonance, $h_c \to \pi^0 \, J/\psi$. The experimental status of this
decay is still not clear. An early observation of this decay in the E760
experiment\cite{e760} has not been confirmed by the E835 data\cite{e835} a
decade later. Clearly, an additional experimental input on this decay would be
of a great interest.

\subsection{\it Chromo-polarizability and Slow Charmonium in Matter}
\subsubsection{\it Diagonal Chromo-polarizability of $J/\psi$ and $\psi'$}
As discussed, the hadronic transitions from $\psi'$ to $J/\psi$ are determined
by the transitional chromo-polarizability $\alpha^{(12)}$, defined in
Eq.(\ref{aij}). Similar diagonal quantities $\alpha^{(11)} = \alpha_\psi$ and
$\alpha^{(22)} = \alpha_{\psi'}$ determine the interaction of the $J/\psi$ and
$\psi'$ resonances with soft gluonic fields and are of a great importance for
description of a number of processes including the interaction of the charmonium
resonances with nuclear matter.

At present these characteristics of the charmonium resonances are not well
known. An early calculation\cite{bpes} of the chromo-polarizability of
quarkonium ground state relied on the treatment of the system as being purely
Coulomb-like, which is a valid limit for very heavy quarkonium, but which hardly
can be applied to charmonium. One guidance for the value of the discussed
parameters is set by the transitional chromo-polarizability $\alpha^{(12)}
\approx 2\,$GeV$^{-3}$. Namely the diagonal terms $\alpha_\psi$ and
$\alpha_{\psi'}$ should satisfy the Schwartz inequality
\be
\alpha_\psi \, \alpha_{\psi'} \ge \left ( \alpha^{(12)} \right )^2~,
\label{aineq}
\ee
and they both should be real and positive since there are no intermediate states
that would enter the second-order correlator in the $E1$ interaction over either
of the vector resonances. Thus it is reasonable to consider the value of
$\alpha^{(12)}$ as a `reference' benchmark for either of the diagonal terms.

\subsubsection{\it The Decay $J/\psi \to \pi \pi \ell^+ \ell^-$}
The chromo-polarizability of the $J/\psi$ resonance can in fact be measured
experimentally in the decay $J/\psi \to \pi^+ \pi^- \, \ell^+ \ell^-$ with a
soft pion pair\cite{mv04}. Using the same description based on the QCD multipole
expansion as for the two-pion transitions, one can write the amplitude of the
decay in the form
\be
A(J/\psi \to \pi^+ \pi^- \, \ell^+ \ell^-)=  {1 \over 2} \, \langle \pi^+
\pi^- | \,  ({\vec E}^a)^2 | 0 \rangle \, \sum_{n=1} \, {\alpha^{(1n)}  \over
m(nS)-m_{J/\psi}+q_0 } \, A(n^3S_1 \to \ell^+ \ell^-)~,
\label{adec}
\ee
where the sum goes over the discrete states as well as the continuum, and
$q_0=(q \cdot P)/m_{J/\psi}$. In
writing this expression it is taken into account that the soft pion
approximation is only valid at $q_0 \ll m_{J/\psi}$, so that any recoil of
the heavy quarkonium upon emission of the pion pair can be and is
neglected. In this limit the relation between $q_0$ and the total
momentum $l$ of the lepton pair can also be written as $m_{J/\psi}^2-l^2=2 \,
q_0 \, m_{J/\psi}$.

In the chiral limit the first term (with $n=1$) in the sum in
Eq.(\ref{adec}) dominates for soft pions, due to its singular behavior
as $1/q_0$. It can be noticed however that the decay amplitude itself is
not singular due to the (even faster) vanishing of the pion production
amplitude (\ref{kapf}) in the limit of soft pions. In the `real life'
the minimal practical energy $q_0$ is not much less than the spacing of
the quarkonium levels, and the contribution of higher states mixes into
the amplitude. In what follows we first consider the contribution of
only the first term of the sum in Eq.(\ref{adec}) and then discuss the
effect of the higher terms. Keeping only the contribution of the first
term in the sum in Eq.(\ref{adec}), one can write the differential rate
of the discussed decay in terms of the chromo-polarizability
$\alpha_{\psi}$ and the leptonic width $\Gamma_{ee}(J/\psi)$ in the form
\be
d \Gamma(J/\psi \to \pi^+ \pi^- \ell^+ \ell^-) = {(q^2)^2 \, {\cal F} \over 4
b^2 \, q_0^2} \, |\alpha_\psi|^2  \, \sqrt{1-{4 m_\pi^2 \over q^2}} \,
\sqrt{q_0^2-q^2} \, \Gamma_{ee}(J/\psi) \, d q^2 \, d q_0 ~,
\label{dgam}
\ee
where the factor ${\cal F}$ is the same as defined in  Eq.(\ref{calf}).

In order to assess the feasibility of observing the discussed decays it
can be noted that at a given constraint on the maximal value of $q_0$:
$q_0 < \Delta$ (or equivalently at a lower cutoff on the invariant mass
of the lepton pair) the probability described by Eq.(\ref{dgam})
strongly peaks near the highest values of both $q^2$ and $q_0$, i.e $q^2
\sim \Delta^2$ and $q_0 \sim \Delta$, and the total probability in the
kinematical region constrained as $q_0 < \Delta$ scales approximately as
$\Delta^6$. However at higher $q^2$ both the dominance of the diagonal
$1S - 1S$ transition in the sum in Eq.(\ref{adec}) becomes weaker and
the behavior of the amplitude in Eq.(\ref{kapf}) derived
for soft pions becomes questionable. It is still quite likely that with
these limitations the discussed here approach can be used up to somewhat
higher values of $\Delta$: $\Delta \approx 0.8 - 0.9 \, GeV$, than those
observed in the pionic transitions from $\psi^\prime$ and
$\Upsilon^\prime$ and that the $F_0(980)$ resonance places
a natural upper bound on the region of applicability of Eq.(\ref{kapf}).
As to the contribution of higher quarkonium states in the sum in
Eq.(\ref{adec}), for each of these states the magnitude of this
contribution relative to that of the diagonal transition is given  by
\be
r_n = \left | {\alpha^{(1n)} \over \alpha_{\psi}} \, {q_0 \over
m_n-m_{J/\psi}+q_0} \right | \, \left [ {\Gamma_{ee}(n^3S_1) \over
\Gamma_{ee}(J/\psi)} \right ]^{1/2}~.
\label{rn}
\ee
The transition polarizability $\alpha^{(1n)}$ should considerably
decrease with $n$. This is supported by the very small experimental rate
of the decay $\Upsilon(3S) \to \pi \pi \, \Upsilon$ (a discussion of this decay
in terms of the transitional chromo-polarizability can be found in
Ref.\cite{mv06}). Thus, most likely, the only real effect of higher
states up to $\Delta \sim 0.9 \, GeV$ reduces to that of the $\psi'$
resonance. This effect however can be accounted for in the data analysis,
since for this resonances all the parameters (except for the overall
relative sign of its contribution) in Eq.(\ref{rn}) are known.
Furthermore an observation and analysis of the discussed decays at
higher values of $q_0$, where the expression (\ref{kapf}) is no longer
valid, would be of a great interest for studies of the pion-pion
scattering beyond the soft-pion region.

Numerically one can estimate from Eq.(\ref{dgam}) the total rate in the
kinematical region constrained by $\Delta = 0.9 \, GeV$ as
\be
\Gamma(J/\psi \to \pi^+ \pi^- \ell^+ \ell^-)|_{q_0 < 0.9 \, GeV} \approx
10^{-4} \left | {\alpha_{\psi} \over 2 \, GeV^{-3}} \right |^2 \,
\Gamma_{ee}(J/\psi)~.
\label{numres}
\ee
Given that the diagonal polarizability is likely to be larger than the
transition one, it can be expected that for the $J/\psi$ resonance the
branching ratio of the discussed decay in a useable kinematical range
should be at the level of $10^{-5}$ which looks to be well within the
reach with the expected CLEO-c data sample.

\subsubsection{\it Slow $J/\psi$ in  Nuclear Medium}
Understanding the charmonium interaction with nuclear matter is
important for description of the photo- and hadro- production of
charmonium and charmed hadrons on the nuclear targets as
well as for diagnostics of the hadronic final states in heavy-ion
collisions and search for Quark Gluon Plasma. Such interaction has
been a subject of numerous studies with a broad range of theoretical
predictions.

First perturbative QCD calculations~\cite{peskin,bpes} predicted  very small
$J/\psi$  dissociation cross section by  hadrons, on the order of few $\mu$barn.
With all the great interest to the problem of charmonium interaction
with nucleons and nuclear matter and its practical importance, the discussion of
this interaction is still wide open. In particular,  the estimates of the
strength of the interaction of $J/\psi$ and $\psi'$ with the nucleon range, in
terms
of the scattering cross section at low energy, from a fraction of
millibarn~\cite{Kharzeev1,Kaidalov} up to 10~mb or
more~\cite{Brodsky2,Huefner,Sibirtsev2}. Recent reviews of the subject and
further references can be found in the Refs.~\cite{Vogt,Barnes2,Song}.

In many of these applications the most interesting energy region is
usually well above the threshold, where the complexity of the
problem becomes more confounding due to the multitude of possible
inelastic processes contributing to charmonium scattering on nuclear
matter. However the strength of the interaction at energy close to
the threshold is also measurable~\cite{Brodsky2}
and its reliable estimate can serve as a useful reference point
for analyses of the behavior of the interaction at higher energies.
Furthermore, the $J/\psi$ and $\psi'$ interactions at low energies are
of explicit importance for high energy heavy ion collisions since
the relative motion between the co-moving charmonium and nuclear
matter is rather slow. Moreover the forward elastic scattering amplitude
can be related to the $J/\psi$ and $\psi'$ mass shift in  matter predicted by
a number of models~\cite{Klingl,Hayashigaki,Kim,Tsushima1}.

The interaction of a slow charmonium with nucleons can be considered within the
multipole expansion in QCD in terms of the
chromo-polarizability\cite{Kaidalov,sibmv}, which parametrizes the interaction
of charmonium with soft gluon fields inside the nucleons. For the $J/\psi$
resonance the amplitude of elastic scattering on a nucleon is then given by
\be
A(J/\psi \, N \to J\psi \, N)={\alpha_\psi \over 2} \, \langle N | ({\vec E}^a
\cdot {\vec E}^a) |N \rangle~.
\label{apsin}
\ee
It can be mentioned that at low energies, below the threshold for the process
$J/\psi + N \to \Lambda_c + D$, the only kinematically allowed inelastic
reaction is $J/\psi + N \to \eta_c + N$. However the latter reaction involves
the heavy quark spin-flip and should be suppressed in comparison with the
elastic scattering. In order to evaluate the matrix element of the gluonic
operator over the nucleon in Eq.(\ref{apsin}) one can make use of the conformal
anomaly relation (\ref{anom}). Namely one can write
\be
\langle N | ({\vec E}^a \cdot {\vec E}^a) - ({\vec B}^a \cdot {\vec B}^a) |N
\rangle = -{1 \over 2} \langle N | (G_{\mu \nu}^a)^2 | N \rangle = {16 \pi^2
\over 9} \langle N | \theta^\mu_\mu | N \rangle = {16 \pi^2 \over 9} \, 2
m_N^2~.
\label{e2n}
\ee
Given that the average of the operator $({\vec B}^a \cdot {\vec B}^a)$ over
nucleon is non-negative, one arrives at the inequality\cite{sibmv}
\be
\langle N | ({\vec E}^a \cdot {\vec E}^a) |N \rangle \ge {16 \pi^2 \over 9} \, 2
m_N^2~.
\label{e2ninec}
\ee
It is expected however\cite{sibmv} that the chromomagnetic contribution to the
nucleon mass is substantially smaller than the chromo-electric one, so that the
actual value of the average in Eq.(\ref{e2ninec}) should be close to the lower
bound.

The elastic $J/\psi N$ scattering amplitude in the low energy limit is thus
estimated as
\be
A(J/\psi \, N \to J\psi \, N) \ge {16 \pi^2 \over 9} \, \alpha_\psi \, m_N^2
\label{apsin2}
\ee
and the corresponding scattering length is then given by
\be
a_{J/\psi N} = {1 \over 4 \pi}\, A(J/\psi \, N \to J\psi \, N) {m_{J/\psi} \over
m_{J/\psi} + m_N} \ge 0.37\,{\rm fm} \, \left ( {\alpha_\psi \over 2\,{\rm
GeV}^{-3} } \right )~.
\label{slpsin}
\ee
Accordingly, the elastic cross section at the threshold is found as
\be
\sigma(J/\psi \, N \to J\psi \, N)=4 \pi \, a_{J/\psi N}^2 \ge 17\, {\rm mb} \,
\left ( {\alpha_\psi \over 2\,{\rm GeV}^{-3} } \right )^2~.
\label{xspsin}
\ee

The positive sign of the scattering length implies attraction of the $J/\psi$ to
nucleons, and the strength of this attraction can be expressed in terms of the
binding potential for $J/\psi$ in nuclear matter with the number density of
nucleons $\rho_N = 0.16\,{\rm fm}^{-3}$:
\be
V_{J\psi}=- A(J/\psi \, N \to J\psi \, N) \, {\rho_N \over 2 m_N} \le -21\,{\rm
MeV} \, \left ( {\alpha_\psi \over 2\,{\rm GeV}^{-3} } \right )~.
\label{vpsin}
\ee
Such rather strong attraction indicates a possibility of existence of bound
states
of $J/\psi$ in light nuclei. Indeed, the condition for existence of a
bound state in the approximation, where a nucleus is considered as
being of a uniform density $\rho_N$ up to the sharp boundary at the
radius $R_A$ reads as
\be
R_A^2 > {\pi^2 \over 8 m_{J/\psi} \, (- V_{J\psi})}~.
\label{ramin}
\ee
With the minimal estimate of the binding potential in Eq.(\ref{vpsin})
this condition is satisfied already at $R_A{>}0.9\,{\rm fm}$, which points to a
relevance of the problem of bound states to light nuclei.
Although the criterion in Eq.(\ref{ramin}) is not directly applicable for light
nuclei, the resulting estimate gives credibility to the claims
\cite{Brodsky2,Wasson} that bound states of the $J/\psi$ resonance
in nuclei do exist starting from light nuclei.
With regards to existence of a near-threshold bound or resonant
state of the $J/\psi$ and a single nucleon, the present understanding is
generally insufficient for arriving at a
definite conclusion.

\subsubsection{\it Interaction of slow $\psi'$ with Nucleons}
The consideration of the elastic $\psi' N$ scattering amplitude parallels that
for the $J/\psi \, N$ process with the obvious replacement of $\alpha_\psi$ by
the chromo-polarizability $\alpha_{\psi'}$ of the $\psi'$. The latter parameter
is expected to be larger than the `reference' value of 2\,GeV$^{-3}$, so that
the numerical value of the `minimal' binding energy in nuclear matter -21\,MeV
is very likely to be overly conservative. The shift of the mass of the $\psi'$
resonance in nuclear matter can be important for consideration of the decay
$\psi \to D {\bar D}$ which may become possible due to a matter-induced shift in
the mass of the $D$
mesons\cite{Tsushima2,Sibirtsev4,Friman,Tsushima3,Mishra,Sibirtsev5}.

The main difference between the nuclear interactions of
slow $J/\psi$ and $\psi'$ is that for the latter there exist
subthreshold scattering processes: the charm-exchange process
$\psi' {+}N{\to}\Lambda_c{+}{\bar D}$, the charmonium transition
scattering $\psi'{+}N{\to}J/\psi{+}N$, and generally additional channels
where in the latter process instead of a single nucleon excited
states are being produced such as $N\pi$, $\Delta\pi$, etc. The
processes other than $\psi'{+}N{\to}J/\psi{+}N$ are beyond the scope of the
present discussion. It can only be noted here that due to the
discussed relation of the relevant gluonic matrix element to
the energy-momentum tensor in QCD, the processes with non-diagonal
transitions, such as $N{\to}N\pi$, should be suppressed with
respect to the diagonal one $N{\to}N$. One can also notice
that similar transitions from $\psi'$ to lower charmonium states
other than $J/\psi$ should also be suppressed in comparison with
$\psi'{\to}J/\psi$, since those other states cannot be produced in
the second order in the leading $E1$ term of the multipole expansion.

The process $\psi' + N \to J/\psi + N$ for a slow $\psi'$ involves a momentum
transfer to the nucleon $q^2 \approx -0.82\,{\rm GeV}^2$, so that the previous,
essentially static consideration is generally modified by an effect of an
unknown form factor $F(q^2)$. For this reason the presented here estimates are
somewhat approximate. The cross section for the discussed transitional process
is found, using the known value of $\alpha^{(12)}$ in complete analogy with
Eq.(\ref{xspsin}):
\be
\sigma(\psi' + N \to J/\psi + N) \approx 16\,{\rm mb}\, \left ( {1\, {\rm GeV}
\over p_i} \right ) \,  |F(q^2)|^2~,
\label{xspsipn}
\ee
where $p_i$ is the c.m. momentum of the initial particles. The inverse-velocity,
$1/p_i$, behavior
of the cross section is due the subthreshold kinematics of
the process. Assuming, conservatively, that the form factor
$|F(q^2)|$ suppresses the amplitude by not more than a factor
of two, one comes to the conclusion that the cross section of
the considered process can reach tens of millibarn at rather
moderately low values of the initial momentum $p_i$.

The discussed process gives rise to a decay rate of the $\psi'$ in a nuclear
medium, which can be evaluated\cite{sibmv} as
\be
\Gamma(\psi' \to J/\psi) \approx 70 \, {\rm MeV} \, \left ( {\rho_N \over 0.16
\, {\rm fm}^{-3}} \right ) \, |F(q^2)|^2~,
\label{ppg}
\ee
and is likely reaching tens of MeV at the nominal average nuclear
density.

\section{Charmonium above the $D {\bar D}$ threshold}
The charmonium resonances that are heavier than the $D^0 {\bar D}^0$ threshold
at 3.73\,GeV are kinematically allowed to decay into $D$ meson pairs and are
generally expected to be significantly broader that the states below the
threshold. The exception from such behavior being for the resonances that might
have mass still below the $D^0 {\bar D}^{*0}$ threshold at 3.87\,GeV and the
quantum numbers that forbid their decay into a pair of pseudoscalar mesons, such
as those with unnatural spin-parity, $P=(-1)^{J+1}$, or with negative CP parity.
Some potential models point at existence in this narrow mass range of such
`exceptional' resonances $^1D_2$ and $^3D_2$ both having unnatural parity.
However no such states have been observed thus far. Instead, an already
plentiful and growing suite of very interesting states is being observed at the
$D^0 {\bar D}^{*0}$ threshold and above, some of which almost definitely cannot
be explained as simple $c {\bar c}$ states, but rather should additionally
contain light quarks and/or gluons as dynamical constituents. Thus in this mass
range the dynamics of the heavy $c {\bar c}$ pair closely intermixes with the
dynamics of charmed meson pairs and with general nonperturbative dynamics in
QCD.

\subsection{$\psi(3770)$}
\subsubsection{\it General Properties}
The resonance $\psi(3770)$, or $\psi''$, with the quantum numbers
$J^{PC}=1^{--}$ can and does decay into $D {\bar D}$ meson pairs, which explains
its relatively large total decay width\cite{pdg} $\Gamma[\psi(3770)] = 25.2 \pm
1.8\,$MeV. The $e^+e^-$ decay width of the $\psi(3770)$,
$\Gamma_{ee}[\psi(3770)]=0.247^{+0.028}_{-0.025}\,$keV  is about ten times
smaller than that of the nearby $\psi'$. Thus this resonance is considered to be
dominantly a $^3D_1$ state of charmonium with a small admixture of $^3S_1$. The
latter admixture enhances the $e^+e^-$ decay rate, which otherwise would be very
small for a pure $^3D_1$ state. The amount of mixing can be estimated from the
value of $\Gamma_{ee}$. Using a simple model with a two-state $\psi' -
\psi(3770)$ mixing Rosner\cite{rosner,rosner2} estimates the mixing angle as
$(12\pm 2)^o$, which certainly agrees with the notion of the $^3D_1 - {^3S_1}$
mixing being an $O(v^2/c^2)$ effect and the estimate $v^2/c^2 \approx 0.2$. It
should be understood however that the particular estimated value of the mixing
provides only an approximate guidance, not only due to its theoretical model
dependence, but also because
the experimental data, especially for $\psi(3770)$, are still somewhat volatile.

In particular, the data are still not conclusive on the decay properties of
$\psi(3770)$, most notably on the fraction of the decay rate that is not
associated with the decay $\psi(3770) \to D {\bar D}$. Namely, the reported by
CLEO\cite{cleoppp} result for the total resonance production cross section at
the maximum of the $\psi(3770)$ peak (at $E_{c.m.}=3773\,$MeV), $\sigma [e^+e^-
\to \psi(3770)]= (6.38 \pm 0.08^{+0.41}_{-0.30})\,$nb leaves very little if any
room for non-$D {\bar D}$ decays, if combined with their latest
measurement\cite{cleoppp2} of the $D \bar D$ production at the same maximum:
$\sigma(e^+e^- \to D \bar D) = (6.57 \pm 0.04 \pm 0.10)\,$nb. On the other hand,
the BES measurement of the total cross section\cite{besppp2} gives $\sigma
[e^+e^- \to \psi(3770)]= (7.25 \pm 0.27 \pm 0.34)\,$nb, and their directly
reported result\cite{besppp} for the branching fractions: ${\cal B}[\psi(3770)
\to D^0 \bar D^0]=(46.7 \pm 4.7 \pm 2.3)\%$, ${\cal B}[\psi(3770) \to D^+
D^-]=(36.9 \pm 3.7 \pm 4.2)\%$ and ${\cal B}[\psi(3770) \to {\rm non}-D \bar
D]=(16.4 \pm 7.3 \pm 4.2)\%$ leaves an ample room for non-$D \bar D$ decays of
$\psi(3770)$ and their branching fraction for $\psi(3770) \to D \bar D$ is also
in agreement with the CLEO result for $\sigma(e^+e^- \to D \bar D)$.

A sizable fraction of non-$D \bar D$ decays of $\psi(3770)$ would present a
serious difficulty for considering it as a pure $c \bar c$ state. Indeed, its
total annihilation rate should be small in comparison with such rate for the
$\psi'$, similarly to its leptonic width $\Gamma_{ee}$. The radiative and
hadronic transitions to lower charmonium levels are expected to be small and in
fact are measured to be small: the hadronic transitions to $J\psi$ all together
contribute less than approximately 0.5\,\% of the total decay
rate\cite{cleop3hadr,besp3hadr}, while the total fraction due radiative decays
to $\gamma+ \chi_{cJ}$\cite{cleop3gamma} likely amounts to at most about 1\,\%.
Thus if the non-$D {\bar D}$ decay rate of the $\psi(3770)$ is measured to
exceed the small rate that can be accounted for, it would imply an enhanced
decay of this resonance into light hadrons. Such enhancement, if found, can be
attributed\cite{rosner05,mv05} to a presence in the wave function of
$\psi(3770)$ of a certain four-quark component: $c \bar c u \bar u$ and $c \bar
c d \bar d$, where the annihilation of the heavy quark pair is
enhanced\cite{ov76}. A mixture of the $\psi(3770)$ with four-quark states can be
viewed as a `re-annihilation'\cite{rosner05} of $D \bar D$ meson pairs, which
are strongly coupled to the resonance. Furthermore, within such mechanism one
can expect that in the four-quark component an enhanced violation of the
isotopic spin due to the fact that the mass difference between the $D^+D^-$ and
$D^0 \bar D^0$ thresholds, $\Delta \approx 9.6\,$MeV, is not much smaller than
the excitation energy of the $\psi(3770)$ resonance above these thresholds.

A presence of light quark-antiquark pairs in the wave function of $\psi(3770)$
should generally enhance\cite{mv05} both the $\pi \pi$ and $\eta$ transitions to
$J/\psi$ with a larger enhancement for the latter transition, which is otherwise
suppressed by the flavor SU(3) symmetry. Experimentally\cite{cleop3hadr} the
branching fractions are ${\cal B}[\psi(3770) \to \pi^+ \pi^- J/\psi]=(0.189 \pm
0.020 \pm 0.020)\%$ and  ${\cal B}[\psi(3770) \to \eta J/\psi]=(0.087 \pm 0.033
\pm 0.022)\%$, so that the ratio of the $\eta$ emission rate to that of the
$\pi^+ \pi^-$ pairs is approximately 0.5. Such ratio indicates a significant
relative enhancement of the $\eta$ transition, if compared with the similar
ratio, 0.1, for the $\psi'$ decays and given the fact that an increased phase
space favors the two-pion transitions more than the $\eta$ emission.

Furthermore, an isospin violation in the four-quark admixture in $\psi(3770)$
implies a presence of an isovector, $I=1$, component in its wave function. Such
component should then enhance the single-pion transition $\psi(3770) \to \pi^0
J/\psi$, and, through the $\psi' - \psi(3770)$ mixing, also provide an
additional contribution to the amplitude of the decay $\psi' \to  \pi^0 J/\psi$,
much needed given the previously discussed mismatch between the data and the
theory. Starting with the needed contribution to the latter decay one can
estimate\cite{mv05} the expected rate of the former transition as ${\cal
B}[\psi(3770) \to \pi^0 J/\psi] \sim 2 \times 10^{-4}$, which can be compared
with the current experimental limit\cite{cleop3hadr} ${\cal B}[\psi(3770) \to
\pi^0 J/\psi] < 2.8 \times 10^{-4}$ at 90\% C.L.

\subsubsection{\it Isospin Breaking in Production of $D \bar D$ pairs at
$\psi(3770)$}
The production of the $D \bar D$ meson pairs at and near the $\psi(3770)$
resonance in $e^+e^-$ annihilation is essentially completely dominated by the
electromagnetic current of the charmed quarks, which is a pure isoscalar.
However the yield of the two isotopic components in the final state, $D^+D^-$
and $D^0 \bar D^0$, is not the same. Rather the ratio is
measured\cite{cleoppp2,besppp} to be $\sigma(e^+e^- \to D^+ D^-)/\sigma(e^+e^-
\to D^0 \bar D^0)=0.79 \pm 0.01 \pm 0.01$ at the maximum of the $\psi(3770)$
peak. This is certainly not unexpected since, as previously mentioned, the mass
difference between the charged and neutral $D$ mesons is substantial at the
energy of the resonance, and also the Coulomb interaction between the produced
slow charged $D$ mesons modifies their production cross section. Usually the
effect of the mass difference in the cross section is estimated by the $P$ wave
kinematical factor $p^3$ with $p$ being the momentum of each meson in the c.m.
frame. The Coulomb effect in the limit of slow point-like particles produced by
a point source reduces to the well known factor $[1+ \pi \alpha/(2 v)]$, where $v$
is the c.m. velocity of either of the charged mesons ($v \approx 0.13$ for the
$D^+D^-$ pairs produced at the $\psi(3770)$ peak). A straightforward estimate of
the product of the kinematical and the Coulomb factors gives $(p_+/p_0)^3\,[1+ \pi \alpha/(2 v)] \approx 0.75$ in a reasonable agreement with the experimental
number for the charged-to-neutral yield ratio. However, it would be premature to
conclude that the issue of this ratio is solved. Indeed, a similar estimate does
not work at all for the $B$ meson pair production at a similar near-threshold
resonance $\Upsilon(4S)$, where the isotopic mass difference for the $B$ mesons
is practically nonexistent, and the Coulomb factor amounts to about 1.19, while
the most precise data give the yield ratio very close to one\cite{babarrcn}:
$1.006 \pm 0.036 \pm 0.031$. On the theoretical side, it is well
understood\cite{am,lepage,be} that the form factors in the production vertex and
in the Coulomb interaction between the charged mesons generally modify the
charged-to-neutral yield ratio. Another related effect\cite{mvrcn,5aut}, that
modifies both the kinematical and the Coulomb correction factors is the strong
interaction between the mesons, which certainly is relevant since there is a
resonance in the production channel.

Due to small velocity of the mesons near the resonance energy one can apply the
methods of the standard nonrelativistic quantum mechanics. With these methods
the strong interaction is assumed to be confined to a certain radius $r < a$,
and that in the isotopically symmetric case  the ($P$ wave)  states of meson
pairs with definite isospin, $I=0$ and $I=1$, are characterized at $r > a$ by
the scattering phases $\delta_0$ and $\delta_1$. Both phases behave as $p^3$
near the threshold, while the $I=0$ phase $\delta_0$ also makes a rapid
variation across the isoscalar $\psi(3770)$ resonance. The isospin violating
effects due to the mass difference and the Coulomb interaction (including the
form factor) can be treated as being due to a difference $\delta V(r)$ at
distances $r > a$ in the potential for the $D^+D^-$ and $D^0 \bar D^0$ channels.
Then if a source producing the meson pairs is localized entirely within the
region of strong interaction, and the production amplitude has the isotopic
composition $A = A_0 |I=0 \rangle + A_1 |I=1 \rangle$,  the charge-to-neutral
yield ratio $R^{c/n}$ is given to the first order in $\delta V$ by the
formula\cite{5aut}
\be
R^{c/n}=\left | {A_0+A_1 \over A_0 - A_1} \right |^2 \,
\left \{ 1 + { 1 \over v} \, {\rm Im}\left [ {{A_0 \, e^{2i \delta_1}- A_1 \,
e^{2i \delta_0}} \over {A_0 - A_1 }}  \,
 \int_a^\infty e^{2ipr} \,
\left ( 1+ {i \over p r} \right )^2 \, \, \delta V(r) \, dr \, \right ] \right
\}~.
\label{rcnres1}
\ee
In the case of an essentially pure isoscalar source, relevant to the charmed
meson pair production in $e^+e^-$ annihilation, this general expression reduces
to the following
\be
R^{c/n}=1+{ 1 \over v} \, {\rm Im}\left [ e^{2i \delta_1} \,
\int_a^\infty e^{2ipr} \, \left ( 1+ {i \over p r} \right )^2 \,
\delta V(r) \, dr \right ]~,
\label{rcnres0}
\ee
so that the strong interaction effect in the discussed corrections is determined
by the scattering phase in the isovector state $\delta_1$ and does not depend on
the resonant isoscalar phase $\delta_0$.

The dependence of the effect on the parameter $a$ is an inevitable consequence
of a $P$ wave dynamics. Only in the limit of vanishing phase $\delta_1$ this
parameter can be set equal to zero. It can be also mentioned that if one
considers the Coulomb interaction of the charged $D$ mesons as that of point
particles, the effect of this interaction as well as that of the mass difference
corresponds to the potential difference $\delta V= \Delta - \alpha/r$, the
result of the simplified approach is recovered after integration in
Eq.(\ref{rcnres0}) down to $a=0$: $R^{c/n}=1-3 \Delta/( 2 \, v p) + \pi
\alpha/(2 v)$. For a nonvanishing $\delta_1$ the correction depends in an
essential way on both $a$ and $\delta_1$\cite{5aut}. Due to the $p^3$ dependence
of this scattering phase one can expect a measurable variation of the ratio
$R^{c/n}$ with energy near the threshold. An experimental study of this
variation can thus provide an information on the strong interaction between
heavy mesons, which information would not be available by other means.

\subsection{X(3872)}
\subsubsection{\it General Properties}
In the summer of 2003 the Belle Collaboration announced\cite{bellex} an
observation  of a narrow resonance $X(3872)$ produced in the decays $B \to K \,
X$ and decaying as $X(3872) \to \pi^+ \pi^- \, J/\psi$. The statistical
significance of the new peak in the invariant mass of $\pi^+ \pi^-  J/\psi$ was
in excess of $10\sigma$. Shortly after the initial discovery the new resonance
was confirmed by observation with similar significance in the inclusive
production in $p \bar p$ collisions at the Tevatron\cite{cdfx,d0x} and by
another independent observation in the $B$ decays\cite{babarx}. The first
observed peculiar features of this resonance were its small width, $\Gamma_X <
2.3\,$MeV\cite{bellex} and the exceptional proximity of its mass to the
threshold of $D^0 {\bar D}^{*0}$: with the recent improvement in the precision
of the $D^0$ mass\cite{cleod0}, which
placed the $D^0 {\bar D}^{*0}$ threshold at $3871.81 \pm 0.36\,$MeV, the mass of
the $X(3872)$  corresponds to $M_X-M(D^0 {\bar D}^{*0})=-0.6 \pm 0.6\,$MeV.

The small width of $X(3872)$ implies that its decay into $D \bar D$ is forbidden
either by its unnatural spin-parity, or by negative CP parity, which, as
discussed, would be possible for certain states of charmonium. However such
states would undergo radiative transitions into $\gamma + \chi_{cJ}$, which
transitions were not observed in the experiment\cite{bellex}. An
analysis\cite{bellex2} of the angular correlations\cite{rosner:x} in the process
with the decay $X  \to \pi^+ \pi^-  J/\psi$ prefers the assignment
$J^{PC}=1^{++}$. A similar analysis by CDF\cite{cdfx2} allows either $1^{++}$ or
$2^{-+}$. The latter assignment however would greatly suppress the decay of $X$
to $D^0 \bar D^0 \pi^0$, which is very close to its threshold. It is generally
believed that the observation of this decay mode by Belle \cite{belledd} and
BaBar\cite{babardd} rules out the possibility of $J^{PC}=2^{-+}$.

The positive $C$ parity of $X(3872)$ is in fact directly mandated by the
observations\cite{bellexg,babarxg} of the decay $X \to \gamma \, J/\psi$. When
combined with the existence of the discovery mode $X \to \pi^+ \pi^- J/\psi$
this implies that the pions in the latter decay have to be in a C-odd state, and
thus the total isospin of the pion pair has to be equal to one! In particular no
emission of a $\pi^0 \pi^0$ pion pair can take place. Thus the $X$ resonance
definitely cannot be a pure $c \bar c$ system, but has to contain light quarks
in its wave function.

Furthermore, the Belle data\cite{bellexg} also indicate that the decay $X(3872)
\to \pi^+ \pi^- \pi^0 \, J/\psi$ has a rate approximately equal to that of the
decay into $\pi^+ \pi^- \, J/\psi$. By the G parity a system of three pions in a
state with a fixed C parity cannot have the same isospin as a system of two
pions. Specifically, at the negative C parity the only possible values of the
isospin are $I(2\pi)=1$, and $I(3\pi)=0$, or $2$. Therefore, not only the
isospin of $X$ is nontrivial, but it is not definite altogether. This is also
supported by the negative results of the search\cite{babarnoxm} for charged
states, which would be the isospin partners of $X$ if it had $I=1$ and the
isospin was a good quantum number.

\subsubsection{\it $X(3872)$ as a (Dominantly) Molecular State}
The unusual properties of the $X(3872)$ state gave rise to the
suggestion\cite{clopag,ess,nat,mvx} that the wave function of this state has a
significant `molecular' component made out of mesons rather than out of quarks.
In particular the quantum numbers $J^{PC}=1^{++}$ imply this component in $X$
contains the state $D^0 {\bar D}^{*0} + \bar D^{0}  D^{*0}$ in the $S$ wave,
which is quite natural, given the extreme proximity of the mass of $X$ to the
corresponding threshold. Furthermore, the threshold for the pairs of charged
mesons, $D^+  D^{*-}+D^- D^{*+}$ is heavier by $\Delta \approx 8\,$MeV, and this
mass gap is large in the scale of the possible binding energy $w$ for the
neutral mesons. For this reason the isospin is badly broken in $X(3872)$ and the
wave functions of the $D^0 {\bar D}^{*0} + \bar D^{0}  D^{*0}$ and  $D^+
D^{*-}+D^- D^{*+}$ components are significantly different, which then explains
the unusual isotopic decay properties of the $X$ resonance.

An existence of molecular states of loosely bound  heavy hadrons was argued on
general grounds long ago\cite{ov76} and a molecular interpretation was
considered\cite{drgg} for explaining the properties of the then already known
$\psi(4040)$ resonance. The argument for the existence of the bound states is
essentially quite straightforward: the strong force between hadrons containing
heavy and light quarks arising due to the interaction between the light
components does not depend on the mass of the heavy quark. Therefore at a
sufficiently large heavy quark mass there inevitably are bound states in the
channels where the strong interaction gives an attraction. However it had to be
tested experimentally, whether the charmed quark is `heavy enough' to form such
states  in some channels.

It should be understood however that there is no reason to expect that  only the
molecular component is present in the wave function and that it determines all
of the properties of the $X(3872)$ boson. Rather one should consider the wave
function in terms of a general Fock decomposition:
\be
\psi_X=a_0 \, \psi_0 + \sum_i a_i \, \psi_i~,
\label{fock}
\ee
where $\psi_0$ is the state of the neutral $D$ mesons $(D^0 {\bar
D}^{*0}+ {\bar D^0}  D^{*0})/\sqrt{2}$, while $\psi_i$ refer to `other'
hadronic states. Due to the extreme proximity of the mass of $X$ to the
$D^0 {\bar D}^{*0}$ threshold, the $\psi_0$ part should be dominant at
long distances. Indeed, assuming that the mass of $X$ is below the
threshold by the binding energy $w$: $m_{D^0}+m_{D^{*0}}-M_X=w$, the
spatial extent of the $\psi_0$ is determined as $(m_D \, w)^{-1/2}
\approx 5 \, {\rm fm} \, (1 \, {\rm MeV}/  w)^{1/2} $, and $\psi_0$ thus
describes the `peripheral' part of the wave function, in fact beyond the
range of strong interaction. On the other hand, the `other' states in
the sum in the Fock decomposition (\ref{fock}) are localized at shorter
distances and constitute the `core' of the $X(3872)$ wave function. In
other terms, one may think of this picture as that of a mixing in
$X(3872)$ of the molecular component $D^0 {\bar D}^{*0} + D^{*0} {\bar
D}^0$ with `other' states, such as e.g. a `pure' $c {\bar c}$
charmonium, which then has to be in a $^3P_1$ state, also favored by the
heavy quark spin selection rule\cite{mvssr}. The notion of $X$ being a
`molecular' system is helpful inasmuch as the probability weight $|a_0|^2$ of
the meson component $\psi_0$ makes a large portion of the total normalization.
In particular the model of Ref.\cite{esw} includes $S$ and $D$ wave states of
the
neutral and charged charmed meson pairs as well as the channels $\rho J/\psi$
and $\omega J/\psi$, and estimates the weight factor of the $S$ wave $(D^0 {\bar
D}^{*0}+ {\bar D^0}  D^{*0})$ component as 70-80\% at $w \approx 1 \, MeV$.

Since the internal composition of the $X(3872)$ can be quite different at
different distances, one or the other part of the Fock decomposition
(\ref{fock}) may be important in specific processes. It appears that the pionic
transitions from $X(3872)$ to $J/\psi$ are determined by a long distance
dynamics, where the $D^0 {\bar D}^{*0} + D^{*0} {\bar D}^0$ component dominates,
so that the isospin states are mixed, and the $\pi^+ \pi^-$ and $\pi^+ \pi^-
\pi^0$ transitions have approximately the same strength. The production of $X$
however is determined by short distances, and proceeds through the core
component, which is approximately an isospin singlet, as
evidenced\cite{babarxcn} by a comparable relative rate of the decays $B^+ \to X
\, K^+$ and $B^0 \to X \, K^0$, while an exclusive contribution of the molecular
$D^0 {\bar D}^{*0} + D^{*0} {\bar D}^0$ state would likely correspond to a
strong suppression\cite{bkus,suzuki} of the  $B^0 \to X \, K^0$ decay in
comparison with $B^+ \to X \, K^+$.

The mesons in the $D^0 {\bar D}^{*0} + D^{*0}
{\bar D}^0$ component move freely beyond the range of the strong interaction,
where their wave function in the coordinate space is given by
\be
\phi_n(r)=c \, {\exp (-\kappa_n \, r) \over r}~,
\label{phic}
\ee
where $\kappa_n$ is determined by the binding energy $w$ and the reduced mass
$m_r \approx 966\,$MeV in the $D^0 {\bar D}^{*0}$ system as $\kappa_n = \sqrt{2
\, m_r \, w}$. The normalization coefficient $c$ determines the statistical
weight of the $D^0 {\bar D}^{*0} + D^{*0}
{\bar D}^0$ component in $X(3872)$, and its definition is correlated with that
of the coefficient $a_0$ in the Fock decomposition (\ref{fock}). We resolve this
ambiguity in the definition by requiring that the coordinate wave function of
the neutral meson pair be normalized to one, so that the statistical weight of
the state $(D^0 {\bar D}^{*0} + D^{*0}
{\bar D}^0)/\sqrt{2}$ is given as $|a_0|^2$. If the wave function of the form
(\ref{phic}) is used down to $r=0$, this requirement corresponds to
$c=\sqrt{\kappa_n/(2\pi)}$.

The dominance of the $D^0 {\bar D}^{*0} + D^{*0}{\bar D}^0$ at long distances
translates into a substantial isospin violation in the processes determined by
the `peripheral' dynamics, examples of which are apparently the observed decays
$X \to \pi^+  \pi^- J/\psi$ and  $X \to \pi^+  \pi^- \pi^0 J/\psi$. It is quite
likely however that this isospin-breaking behavior is only a result of the
`accidentally' large mass difference $\Delta \approx 8 \,$MeV between $D^+
D^{*-}$ and $D^{*0} {\bar D}^{*0}$. Therefore
it is natural to expect that at shorter distances within the range of the strong
interaction the isospin symmetry is restored and at those distances the wave
function of $X(3872)$ is dominated by $I=0$. In this picture the wave function
of a $D^+ D^{*-}+D^-D^{*+}$ state within the region beyond the range of the
strong interaction can be found from Eq.(\ref{phic}) by requiring that at short
distances the pairs of charged and neutral mesons combine into an $I=0$
state\footnote{The effects of the Coulomb interaction between the charged mesons
are neglected here.}, so that the wave function of the charged meson pair has
the form
\be
\phi_c(r)= c \, {\exp (-{\kappa_c} \, r) \over r}~,
\label{phicc}
\ee
where ${\kappa_c} = \sqrt{2 m_r \, (\Delta+w)} \approx 125 \,$MeV. It should be
noticed that both the neutral (Eq.(\ref{phic})) and the charged
(Eq.(\ref{phicc})) meson wave function have the same normalization factor $c$
(determined by $\kappa_n$) and differ only in the exponential power.
This simple picture allows one to estimate the relative statistical weight of
the charged and neutral $D$ meson components in the $X(3872)$:
\be
\lambda \equiv {\left |\langle X | D^+ D^{*-}+D^-D^{*+}\rangle \right |^2 \over
\left | \langle X | D^0 {\bar D}^{*0} + D^{*0}{\bar D}^0 \rangle \right |^2} =
{\kappa_n \over {\kappa_c}}~.
\label{wrat}
\ee

Clearly, the wave functions in Eq.(\ref{phic}) and Eq.(\ref{phicc}) cannot be
applied at short distances in the region of strong interaction, where the mesons
overlap with each other and cannot be considered as individual particles. In
order to take into account this behavior an `ultraviolet' cutoff should be
introduced. One widely used method for introducing such cutoff is to consider
the meson wave functions only down to a finite distance $r_0$, at which distance
the boundary condition of the state being that with $I=0$ is imposed. An
alternative, somewhat more gradual cutoff, described by parameter $\Lambda$, can
be introduced\cite{suzuki} by subtracting from the wave functions (\ref{phic})
and (\ref{phicc}) an expression $c \, e^{-\Lambda r}/r$. It should be noticed,
that such regularization also results in a modification of the normalization
coefficient $c$, which for the gradual cutoff takes the form
\be
c=\sqrt{\kappa_n \over 2 \pi} \, {\sqrt{\Lambda \, (\Lambda + \kappa_n)} \over
\Lambda - \kappa_n}~.
\label{norm}
\ee
One can also readily see, that an introduction of any such cutoff eliminates
relatively more of the charged meson wave function than of the neutral one, thus
reducing the estimate of the relative statistical weight as compared to that in
Eq.(\ref{wrat}), so that Eq.(\ref{wrat}) gives in fact the upper bound for the
ratio.

\subsubsection{\it Peripheral Decays to $D^0 \bar D^0 \pi^0$ and $D \bar D
\gamma$}
The `molecular' component of the $X(3872)$ dominating at large distances, the
periphery, should give rise to decays to $D^0 \bar D^0 \pi^0$ and $D^0 \bar D^0
\gamma$\cite{mvx,mvx2}. The underlying processes in these decays are the decays
of the very weakly bound $D^{*0}$ meson $D^{*0} \to D^0 \pi^0$ and $D^{*0} \to
D^0 \gamma$, and the charge-conjugate decays of the $\bar D^{*0}$. These decays
are relatively well studied for the $D^*$ mesons, and the relevant rates can be
deduced from the data in the Tables\cite{pdg} as
\be
\Gamma_\pi \equiv \Gamma(D^{*0} \to D^0 \pi^0) = 43 \pm 10\,{\rm keV ~~~and~~~}
\Gamma_\gamma \equiv \Gamma(D^{*0} \to D^0 \gamma) = 26 \pm 6\,{\rm keV~.}
\label{gd0}
\ee

\begin{figure}[tb]
  \begin{center}
\begin{minipage}[t]{18 cm}
\epsfxsize=9cm
\epsfbox{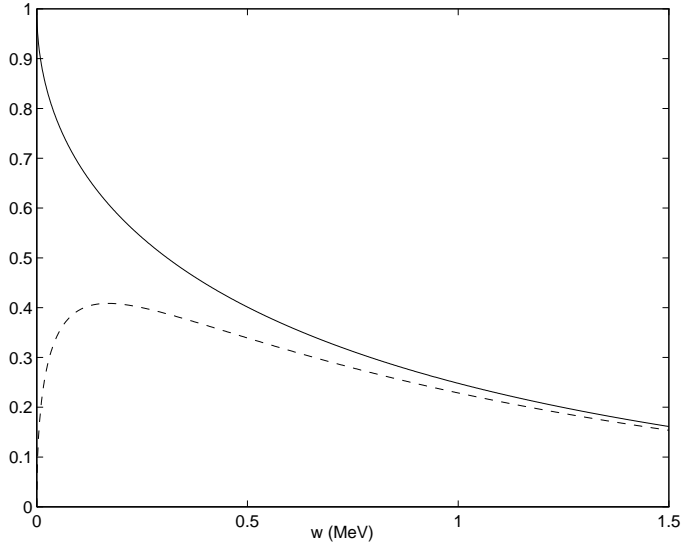}
\end{minipage}
\begin{minipage}[t]{16.5 cm}
\caption{The non-coherent contribution $A(w)$ (solid line) and the
interference term $B(w)$ (dashed) as defined in Eq.(\ref{gamt}),
calculated by a numerical integration in Eq.(\ref{dgpi}).\label{fig:abpi}}
\end{minipage}
\end{center}
\end{figure}

There is an interference between the amplitude of the decay of $D^{*0}$ and
$\bar D^{*0}$ in state with a fixed C parity of the initial meson pair. For the
C-even $X$ resonance the sign of the interference is positive for the decay into
$D^0 \bar D^0 \pi^0$ and is negative for the decay into $D^0 \bar D^0 \gamma$.
The differential over the Dalitz plot rate of the decay $X(3872) \to D^0 \bar
D^0 \pi^0$ can then be found as\cite{mvx}
\be
d \Gamma(X \to D^0 \bar D^0 \pi^0)=|a_0|^2 \, {\Gamma_\pi (\vec q_1 + \vec
q_2)^2 \over 12 \pi^2 \, p_0^3} \, \left | \phi(\vec q_1)+ \phi(\vec q_2) \right
|^2 \, d \vec q_1^2 d \vec q_2^2~,
\label{dgpi}
\ee
where, $p_0 = 43\,$MeV is the $\pi^0$ momentum is the pion momentum in the decay
of a free $D^{*0}$ meson, $\vec q_1$ and $\vec q_2$ are the momenta of the $D^0$
and $\bar D^0$ mesons in the rest frame of $X$, and $\phi(\vec q)$ is the
momentum-space wave function corresponding to that in Eq.(\ref{phic}):
\be
\phi(\vec q) = {4 \pi c \over \vec q^2 + \kappa_n^2}~.
\label{phimom}
\ee
The significance of the interference and the total rate of the decay depend on
the binding energy $w$:
\be
\Gamma(X \to D^0 \bar D^0 \pi^0) = |a_0|^2 \, \Gamma_\pi \left [A(w)
+  B(w) \right ]~,
\label{gamt}
\ee
where $A(w)$ describes the incoherent contribution of the decays of
individual $D^{*0}$ and $\bar D^{*0}$, and $B(w)$ describes the effect of the
interference between these two processes.
The result of a numerical calculation of the terms $A$ and $B$ with the
wave function from Eq.(\ref{phimom}) is shown in Fig.\ref{fig:abpi}. It is seen
from the
plot, that the interference between the two wave functions in
Eq.(\ref{dgpi}) significantly enhances the  decay from a C-even state.

\begin{figure}[tb]
  \begin{center}
\begin{minipage}[t]{18 cm}
\epsfxsize=9cm
    \epsfxsize=9cm
    \epsfbox{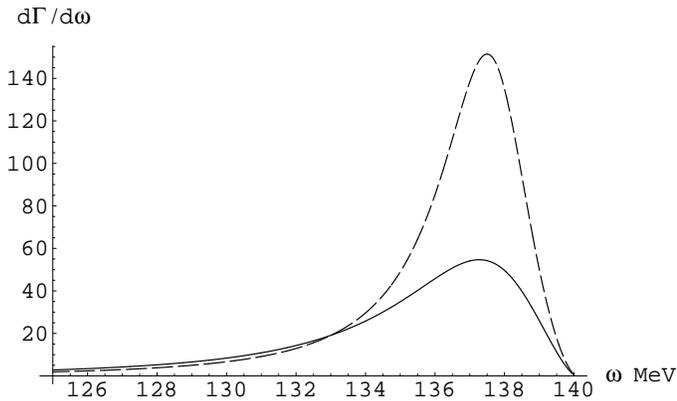}
\end{minipage}
\begin{minipage}[t]{16.5 cm} 
    \caption{The photon spectrum in the decay $X(3872) \to D^0 {\bar D}^0
\gamma$ for $\kappa_n = 44\,$MeV (solid) and for $\kappa_n=24\,$MeV (dashed),
both at $\Lambda \to \infty$. The vertical scale is in arbitrary units.
\label{gspectr}}
\end{minipage}
\end{center}
\end{figure}

The decay $X(3872) \to D \bar D \gamma$ can be considered in a similar way, and
the differential decay rate is found as
\be
{\rm d}\Gamma(X \to D {\bar D} \gamma)=\Gamma_\gamma \, {a_0^2 \over
2}
\, \left ( {\omega \over \omega_0} \right )^3 \, \left [ \phi \left
({{\vec k} \over 2} + {\vec p} \right )- \phi\left ({{\vec k} \over 2} -
{\vec p}\right ) \right ]^2 {{\rm d}^3 p \over (2 \pi)^3}~,
\label{dgamm0}
\ee
where $\omega_0=137\,$MeV stands for the photon energy in the corresponding
decay $D^{*0} \to D \gamma$, $\omega$ is the energy of the photon, and ${\vec
p}=(\vec p_{D}-\vec p_{\bar D})/2$ is the momentum of the $D$ meson in the c.m.
frame of the final $D {\bar D}$ meson pair. An integration with the wave
function with the cutoff $\Lambda$ at short distances gives the expression for
the total decay rate
\be
\Gamma(X \to D^0 {\bar D}^0 \gamma) =
\Gamma_\gamma \, a_0^2 \,
\left [ 1-
 {2 \kappa_n \over \omega_0} \, {\Lambda \, (\Lambda+\kappa_n) \over (\Lambda -
\kappa_n)^2} \left ( \arctan {\omega_0 \over 2 \kappa_n} + \arctan {\omega_0
\over 2 \Lambda} - 2 \arctan {\omega_0 \over \Lambda+\kappa_n} \right ) \right
]~.
\label{gtnr}
\ee
At $\omega_0=137\,$MeV and $\kappa_n \approx 44\,$MeV (corresponding to the
binding energy $w = 1\,$MeV) the numerical value of the expression in the square
braces, the interference factor, varies between 0.36 at $\Lambda \to \infty$ and
$0.61$ at $\Lambda=200\,$MeV. The spectrum of the photon energies is shown in
Fig.\ref{gspectr}. As can be expected on general grounds this spectrum peaks
near the energy corresponding to the decay $D^{*0} \to D^0 \gamma$ with the
spread induced by the slow `Fermi motion' of the initial meson in the loosely
bound state.

In the peripheral contribution the decay to $D^+ D^- \gamma$ is greatly
suppressed in comparison with $D^0 \bar D^0 \gamma$. The suppression results
from three contributing factors\cite{mvx2}: the small rate of the decay $D^{*+}
\to D^+ \gamma$, the relatively small statistical weight of the pair of charged
mesons in $X(3872)$ and the stronger negative interference for the charged
mesons located at shorter distances within the $X$ resonance wave function. One
can also argue\cite{mvx2} that the appearance of the pairs of charged mesons as
a result of rescattering $D^0 \bar D^0 \to D^+ D^-$ should be only a minor
effect.

A quite different final state composition should be expected for the $D \bar D
\gamma$ final state from the `core' component of the $X(3872)$. In particular
this component can give rise to decays through the intermediate $\psi(3770)$
resonance: $X(3872) \to \psi(3770) \gamma \to D \bar D \gamma$ in which case the
photon spectrum should have a peak at $\omega \approx 100\,$MeV and the isotopic
composition of the final $D \bar D$ state should be the same as in the decay of
$\psi(3770)$.

\subsubsection{\it $e^+e^- \to \gamma \, X(3872)$ as an Alternative Source of
$X(3872)$}

The resonance $X(3872)$ is observed experimentally only in the decays of $B$
mesons $B \to X \, K$\, and in inclusive production in proton - antiproton
collisions at the Tevatron. Both these types of processes present significant
challenges for precision measurements of the parameters of the resonance. In
particular, neither the total width of $X(3872)$ is yet resolved (the current
limit is $\Gamma_X < 2.3\,$MeV), nor its mass is known with a precision
sufficient to determine the mass gap $w$ from the $D^0 D^{*0}$ threshold. A
viable alternative source of the $X(3872)$ can be provided by the process
$e^+e^- \to \gamma X(3872)$\cite{dubmveex} at the c.m. energy within few MeV of
the $D^{*0} {\bar D}^{*0}$ threshold, where the kinematical simplicity of the
process would hopefully allow more detailed studies of $X(3872)$.

\begin{figure}[tb]
\begin{center}
\begin{minipage}[t]{18 cm}
\epsfxsize=17cm
\epsfbox{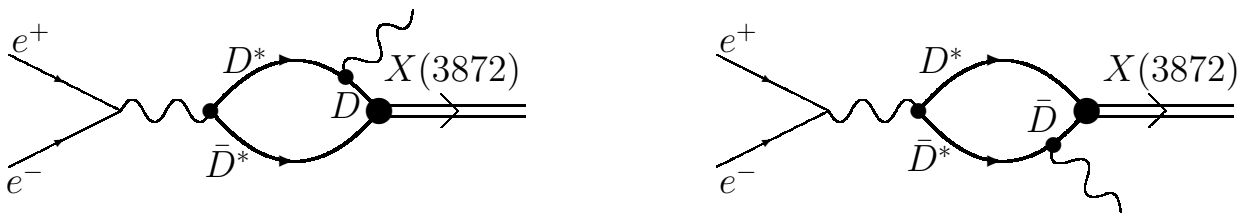}
\end{minipage}
\begin{minipage}[t]{16.5 cm}
\caption{The production process $e^+e^- \to D^{*0} {\bar D}^{*0} \to \gamma
X(3872)$. \label{eexgfig} }
\end{minipage}
\end{center}
\end{figure}
The cross section of the discussed process can be estimated by using the
unitarity relation and considering the process, shown in Fig.\ref{eexgfig}
$e^+e^- \to D^{*0} {\bar D}^{*0} \to \gamma X(3872)$, and also using the
amplitude of the known decay $D^{*0} \to \gamma \, D^0$. The amplitude for
production of the $D^* \bar D^*$ pair in $e^+e^-$ annihilation at a small energy
above the threshold $E=2M(D^{*0})+W$ can be generically written in the form
\be
A(e^+e^- \to D^{*0} {\bar D}^{*0})= A_0 \, ({\vec j} \cdot {\vec p}) \, ({\vec
a} \cdot {\vec b})^* + {3 \over 2 \, \sqrt{5}} \, A_2 \, j_i \, p_k \, \left [
a_i \, b_k + a_k \, b_i - {2 \over 3} \, \delta_{ik} \, ({\vec a} \cdot {\vec
b}) \right ]^*~,
\label{aee}
\ee
where ${\vec j}= ({\bar e} {\vec \gamma} e)$ stands for the current of the
incoming electron and positron, ${\vec p}$ is the momentum of one of the mesons
($D^{*0}$ for definiteness) in the c.m. frame, and $A_0$ and $A_2$ are the
factors corresponding to production of the vector meson pair in the states with
respectively the total spin $S=0$ and $S=2$. It can be also noted that the
amplitude in Eq.(\ref{aee}) describes the production of mesons in the $P$ wave.
Another kinematically possible amplitude, the $F$-wave, should be small near the
threshold, i.e. at a small $W$. Both $A_0$ and $A_2$ are generally functions of
the excitation energy $W$. Furthermore, their dependence on the energy near the
threshold is known to be nontrivial due to the $\psi(4040)$ resonance\cite{pdg},
with possible further complications in the immediate vicinity of the
threshold\cite{lang,cleoddfinal}. Neither the relative magnitude nor the relative phase
of the amplitudes $A_0$ and $A_2$ is presently known, but both of these can be
measured from angular correlations\cite{mvfpcp}. These amplitudes determine the
total cross section for production of $D^{*0} {\bar D}^{*0}$ in $e^+e^-$
annihilation:
\be
\sigma(e^+e^- \to D^{*0} {\bar D}^{*0}) = \int |A(e^+e^- \to D^{*0} {\bar
D}^{*0})|^2 \, 2\pi \, \delta \left ( W-{p^2 \over m}\right ) \, {d^3 p \over (2
\pi)^3} = C \, {m \, p^3 \over 2 \pi} \, \left ( |A_0|^2 + |A_2|^2 \right )~,
\label{sdd}
\ee
where $m = M(D^{*0})$, $p=|{\vec p}|$, and $C$ is an overall constant related to
the average value of the current $|{\vec j}|^2$. The specific value of the
latter constant is not essential for the discussed estimate, since it cancels
in the ratio of the cross sections. The latter ratio thus can be expressed in
terms of the amplitudes  $A_0$ and $A_2$ and of the parameters of the decay
$D^{*0} \to \gamma \, D^0$, the rate $\Gamma_\gamma$ and the photon energy
$\omega_0$,
\be
{\sigma_{\rm Abs} (e^+e^- \to  D^{*0} {\bar D}^{*0} \to \gamma  X) \over \sigma
(e^+e^- \to  D^{*0} {\bar D}^{*0})} = |a_0|^2 \, {\Gamma_\gamma \, m \, \omega
\,
\kappa_n \over 2 \omega_0^3 \, p} \, F^2 \, {  |  A_0- {A_2/ \sqrt{5}}  |^2 +
(9/20) \,  | A_2  |^2  \over |A_0|^2 + |A_2|^2 }~,
\label{sabsr}
\ee
where $\omega$ is the energy of the photon, and the form factor $F$ is defined
through the wave function $\phi(\vec q)$ (Eq.(\ref{phimom}) as
\be
F= {1 \over 2 } \, \int_{-1}^1 \, ({\vec p} \cdot {\vec k}) \,
\phi \left( {\vec p} - {{\vec k} \over 2} \right ) \, d \cos \theta~,
\label{f0}
\ee
with $\theta$ being the angle between the $D^*$ c.m. momentum ${\vec p}$ and the
photon momentum${\vec k}$. The explicit form of the form factor is
\be
F= { c \over p \, \omega} \left [ \left ( p^2 +{\omega^2 \over 4} + \kappa_n^2
\right) \, \ln { (p+\omega/2)^2+ \kappa_n^2 \over (p- \omega/2)^2+ \kappa_n^2} -
  \left ( p^2 +{\omega^2 \over 4} + \Lambda^2 \right) \, \ln { (p+\omega/2)^2+
\Lambda^2 \over (p- \omega/2)^2+ \Lambda^2} \right ]
\label{f1}
\ee
with the normalization coefficient $c$ given by Eq.(\ref{norm}).

\begin{figure}[tb]
  \begin{center}
\begin{minipage}[t]{18 cm}
\epsfxsize=9cm
    \epsfxsize=9cm
     \epsfbox{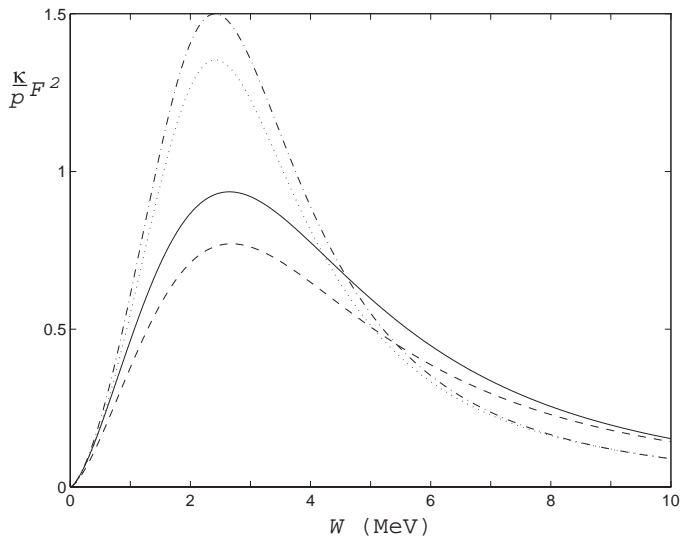}
\end{minipage}
\begin{minipage}[t]{16.5 cm} 
    \caption{The factor $\kappa_n \, F^2/p$  vs. the excitation energy $W$ above
the $D^{*0} {\bar D}^{*0}$ threshold  at
representative values of the binding energy $w$ in $X(3872)$ and the ultraviolet
cutoff parameter $\Lambda$: $w=1\,$MeV, $\Lambda=200\,$MeV (solid), $w=1\,$MeV,
$\Lambda=300\,$MeV (dashed), $w=0.3\,$MeV, $\Lambda=200\,$MeV (dashdot), and
$w=0.3\,$MeV, $\Lambda=300\,$MeV (dotted). \label{eexg}}
\end{minipage}
\end{center}
\end{figure}

The `absorptive' cross section $\sigma_{\rm Abs}$ in Eq.(\ref{sabsr}) is (most
likely) not the actual value of the cross section, since the amplitude of the
process $e^+e^- \to \gamma X$ can receive contribution from other mechanisms.
Nevertheless it is instructive to examine the numerical value and the behavior
with energy of this quantity as given by Eq.(\ref{sabsr}). The dependence on the
c.m. energy of the factor $(\kappa_n/p) \, F^2$ is shown in Fig.\ref{eexg} for
two representative values of the `molecular' binding energy in $X(3872)$,
$w=1\,$MeV ($\kappa_n \approx 44\,$MeV) and $w=0.3\,$MeV ($\kappa_n \approx
24\,$MeV). This factor peaks at the energy where $p \approx \omega/2 \approx
70\,$MeV. The appearance of such peak is easily understood qualitatively: at
${\vec p} \approx {\vec k}/2$ the $D^0$ meson emerging from the emission of the
photon in $D^{*0} \to D^0 \gamma$  moves slowly relative to the ${\bar D}^{*0}$
and forms a loosely bound state.\footnote{The same situation arises at ${\vec p}
\approx - {\vec k}/2$ for the ${\bar D}^0$ meson emerging from ${\bar D}^{*0}
\to {\bar D}^0 \gamma$.} The width of the peak is clearly determined by the
parameter $\kappa_n$.

As is seen from the plots of Fig.\ref{eexg} the numerical value of the factor
$(\kappa_n/p) \, F^2$ near its peak is of order one. Another factor in
Eq.(\ref{sabsr}), $\Gamma_0 \, m \, \omega /(2  \omega_0^3) \approx \Gamma_0 \,
m /(2  \omega_0^2) \approx 1.5 \times 10^{-3}$, sets the overall scale of the
discussed cross section. The factor in Eq.(\ref{sabsr}) depending on the
presently unknown ratio of the (generally complex) amplitudes $A_0/A_2$, takes
values between 0.34 (at $A_0/A_2 \approx 0.68$) and 1.31 (at $A_0/A_2 \approx
1.47$), and can thus be considered as being of order one. Finally, the
statistical weight factor $|a_0|^2$, as discussed, is likely to be a large
fraction of one. Summarizing these numerical estimates, the value of the ratio
in Eq.(\ref{sabsr}) at the peak can be estimated as being of order $10^{-3}$,
although the uncertainty is presently large.

In absolute terms, the measured\cite{lang,cleoddfinal} cross section $\sigma(e^+e^- \to
D^{*0} {\bar D}^{*0})$ at $E = 4015\,$MeV, i.e. at the energy above the $D^{*0}
{\bar D}^{*0}$ threshold $W \approx 1.6\,$MeV is about 0.15\,nb. This cross
section grows from the threshold as $p^3$. With this factor taken into account
the peak of the quantity $\sigma_{\rm Abs} (e^+e^-  \to \gamma  X)$ shifts to a
slightly higher value of $p$, $p \approx 100\,$MeV, corresponding  to $W \approx
5\,$MeV, where it should be numerically of the order of 1\,pb.

The considered mechanism of the process $e^+e^- \to \gamma X$ describes a `soft'
production of its peripheral $D^0 {\bar D}^{*0} + D^{*0} {\bar D}^0$ component
in radiative transitions from slow $D^{*0} {\bar D}^{*0}$ pairs. Other
intermediate states with charmed meson pairs, i.e. $D^+ D^{*-}+D^- D^{*+}$, $D
{\bar D}$ and $D {\bar D}^*$ (${\bar D} D^*$),   can potentially contribute to
the discussed process $e^+e^- \to \gamma X$. However, one can readily see that
in either of these processes the charmed meson emerging after the emission of
the photon is very far off the mass shell in the scale of $\kappa_n$. Thus
none of such processes can proceed due to the long-distance peripheral
component of the $X(3872)$ resonance, but rather is determined by the short-
distance dynamics of the `core' of $X$. For this reason such contributions, as
well as other possible mechanisms related to the `core' dynamics, should be
smooth functions of the c.m. energy on the scale of few MeV around the $D^{*0}
{\bar D}^{*0}$ threshold, where the considered absorptive amplitude experiences
a significant variation. Therefore even under the most conservative (and quite
unlikely) assumption that these mechanisms cancel the contribution of the latter
amplitude near its maximum, such cancellation cannot take place at all energies
in the considered range. Thus the cross section of the process $e^+e^- \to
\gamma X$ at an energy within few MeV of the $D^{*0} {\bar D}^{*0}$ threshold
has to be at least as large as the estimates for $\sigma_{\rm Abs}$ near its
maximum i.e. of the order of 1\,pb. The latter is a conservative estimate, since
one cannot exclude that the contribution of those `other' mechanisms exceeds the
calculated amplitude and that the actual cross section is larger than
$\sigma_{\rm Abs}$.

\subsubsection{\it One- and Two-Pion Transitions from $X(3872)$ to $\chi_{cJ}$}
Although the bulk of the data on $X(3872)$ indicate that it is very likely to be
related to dynamics of $D^0 \bar D^{*0}$ charmed meson pairs, a possibility is
still being considered\cite{cdfn,mecha} that the observed properties of
$X(3872)$ can be, to an extent, mimicked by a $2^3P_1$ state of charmonium, so
that any `molecular' admixture would be viewed as a secondary effect due to the
coupling to the $D {\bar D}^*$ states. In this picture the main available
indicator of a significant isospin violation in $X(3872)$, the approximately
equal rate of the decays $X(3872) \to \rho J/\psi \to \pi^+ \pi^- J/\psi$ and
$X(3872) \to \omega J/\psi \to \pi^+ \pi^- \pi^0 J/\psi$ is explained by the
kinematical suppression of the isospin-allowed transition $X(3872) \to \omega
J/\psi$.

The transitions from $X(3872)$ to the $\chi_{cJ}$
charmonium states with emission of one or two pions, which can be studied in
addition to the observed processes $X(3872) \to \pi^+ \pi^- J/\psi$ and $X(3872)
\to \pi^+ \pi^- \pi^0 J/\psi$, may prove to be instrumental in further
exploration of the $X$ resonance. Such  transitions  may be accessible
for experimental observation and may hold the clue to understanding the isotopic
structure of the $X(3872)$ and of the prominence of the four-quark component in
its internal dynamics. The characteristics of such transitions are
generally completely different between the possible charmonium and molecular
components of the resonance $X(3872)$. The rate of the one-pion transition
relative to the process with two pions is sensitive to the $I=1$ four-quark
component of the $X(3872)$, while the isoscalar four-quark component should give
rise to relative rates of two-pion transitions to the $\chi_{cJ}$ states with
different $J$, which are very likely at variance from those expected for
transitions between charmonium levels\cite{dubmvxpi}.

The pion transitions from a charmonium excited $^3P_1$ state are described by
the multipole expansion. The two-pion emission is not suppressed by the isotopic
symmetry and dominantly proceeds to the $\chi_{c1}$ state. Indeed, this is the
only final state in the transitions from $^3P_1$ where the pions can be emitted
in the $S$ wave, for which the pion pair creation by the gluonic fields is
enhanced by the conformal anomaly in QCD (Eq.(\ref{kapf})). A transition to the
$\chi_{c2}$ resonance requires a presence of a $D$ wave, proportional to a small
parameter $\kappa \approx 0.2$, and is also suppressed by smaller available
phase space. Numerically, the suppression turns out to be quite strong, about a
factor of $10^{-4}$\cite{dubmvxpi}. Moreover, due to the spin-parity properties
the $\chi_{c0}$ resonance cannot be produced in either $S$ or $D$ wave dipion
emission, so that such transition arises only starting with the fourth power of
pion momenta in the chiral expansion and should also be quite small.
Furthermore, if the quarkonium matrix element (the chromo-polarizability) for
the $2 {^3P_1} \to 1 {^3P_1}$ transition is evaluated from the known amplitude
of the transition $\psi' \to \pi \pi J/\psi$, the transition rate can be
estimated\cite{dubmvxpi} as $\Gamma(2 {^3P_1} \to \pi^+ \pi^- \, 1 {^3P_1})
\approx 1.5\,$keV, which is most likely only a tiny fraction of the total width
of $X(3872)$.

The isospin-violating single pion transitions in charmonium, $2 {^3P_1} \to
\pi^0 \, 1 {^3P_J}$ can also be described within the multipole expansion,
similarly to $\psi' \to \pi^0 \, J/\psi$. The ratio of the transition rates
$\Gamma_J$ to final states with different $J$ is then given by
\be
\Gamma_2\,:\Gamma_1\,:\Gamma_0= 3 p_{\pi\,(2)}^{\,3}\,: 5
p_{\pi\,(1)}^{\,3}\,:0\approx 1:\,2.70 :0\,,
\label{sprx}
\ee
where $p_{\pi\,(J)}$ stands for the pion momentum in the corresponding process.
One can notice that the $\chi_{c0}$ final state is not accessible in this
transition too. It can be noted however that in the case of pure charmonium the
single pion transition should be even weaker than the two-pion: in the ratio of
the rates the unknown quarkonium matrix element cancels, and one
finds\cite{dubmvxpi}
\be
{\Gamma \left( 2\, ^3 P_1\to \chi_{c1} \pi^0 \right)\over \Gamma
\left( 2\, ^3 P_1\to \chi_{c1} \pi^+ \pi^-\right)}\approx 0.04\,.
\ee

A somewhat different pattern of the transition rates can be expected if the
resonance $X(3872)$ is dominantly a molecular state or, generically, is a
four-quark state. Then the pion transitions to the charmonium states $\chi_{cJ}$
can be treated as a `shake off' of the light quarks. In particular, the spin
dependent `heavy-light' quark interaction is proportional to the inverse
power of the heavy quark mass $m_Q^{-1}$, so that any exchange of the
polarization between the light and heavy degrees of freedom is expected to be
suppressed. Neglecting such exchange one can arrive at the following estimate of
the relative rate of the single-pion transitions:
\be
\Gamma_0\,:\Gamma_1\,:\Gamma_2= 4 p_{\pi\,(0)}^{\,3}\,: 3
p_{\pi\,(1)}^{\,3}\,:5 p_{\pi\,(2)}^{\,3}\approx 2.88:0.97 :1\,,
\ee
so that unlike for a pure charmonium the rate of the transition to the lowest
$\chi_{cJ}$ state should be the largest. Furthermore, due to the apparent strong
isospin violation in the wave function of the $X(3872)$ resonance, the single
pion process should not be suppressed by the isospin, and in fact is likely to
dominate over the two-pion transitions. For the relative strength of the latter
decays to $\chi_{cJ}$ with different $J$, it can be mentioned that the
transition to $\chi_{c0}$ is still suppressed in the chiral expansion by the
spin-parity properties, while the $\chi_{c2}/\chi_{c1}$ ratio is uncertain with
the kinematics obviously favoring the $\chi_{c1}$ final state.

\subsubsection{\it $X(3872)$ as a Virtual State}
The expected decay $X(3872) \to D^0 \bar D^0 \pi^0$ was sought for and
observed\cite{belledd,babardd}, however the characteristics of the observed
process do not quite look like what one would expect for decay of a bound state.
Namely, the experimental study of the $B$ meson decays $B \to D^0 {\bar
D}^0 \pi^0 \, K$\,\cite{belledd,babardd} and $B \to D^0 {\bar D}^0 \gamma \,
K$\,\cite{babardd} revealed that the invariant mass recoiling against the Kaon
displays a significant enhancement with a maximum at approximately 3875\,MeV,
which is only about 3\,MeV above the $D^0 {\bar D}^{*0}$ threshold. The observed
events can all be in fact attributed to the process $B \to (D^0 {\bar D}^{*0} +
{\bar D} D^{*0}) \, K$ since no distinction between the $D^{*0}$ mesons and
their decay products was done. Moreover, the yield of the heavy meson pairs
within the above-threshold peak is about ten times larger than that of the
$\pi^+ \pi^- J/\psi$ and $\pi^+ \pi^- \pi^0 J/\psi$ channels at the peak of
$X(3872)$. It has been most recently argued\,\cite{hkkn} that a very plausible
explanation of the observed enhancement of the $D^0 {\bar D}^{*0}$ production
combined with the smaller observed $X(3872)$  peak in the $\pi^+ \pi^- J/\psi$
channel is that both these phenomena are due to a virtual state\,\cite{bugg,yuk}
in the $D^0 {\bar D}^{*0}$ channel. In this picture the observed peak in the
$\pi^+ \pi^- J/\psi$ and $\pi^+ \pi^- \pi^0 J/\psi$ mass spectra is in fact a
cusp with a sharp maximum at the $D^0 {\bar D}^{*0}$ threshold.

The isospin properties of such near-threshold virtual state can be
analyzed\cite{mvxthr} within the approximation of small interaction radius. Such
approach is similar to the `universal scattering length'
approximation\cite{bkus}, and differs in including the effect of the nearby
threshold for charged mesons $D^+ D^{*-}$. An interesting energy-dependent
behavior of the isotopic properties arises from the mere fact of the mass
splitting $\Delta$ between the two isospin-related and coupled $D {\bar D}^*$
channels. In particular it can be argued that the expected pattern of the
isospin breaking is consistent with the observed relative yield of $\pi^+ \pi^-
J/\psi$ and $\pi^+ \pi^- \pi^0 J/\psi$ at the peak which
experimentally\,\cite{bellex2} corresponds to ${\cal B}(X \to \pi^+ \pi^- \pi^0
J/\psi)/{\cal B}(X \to \pi^+ \pi^-  J/\psi) = 1.0 \pm 0.4 \pm 0.3$. Moreover,
the production amplitude for the $I=1$ state $\pi^+ \pi^- J/\psi$ in the
considered approximation necessarily has a zero between the $D^0 {\bar D}^{*0}$
and $D^+ {D}^{*-}$ thresholds, thus reducing the apparent width of the cusp and
putting it in line with the experimental limit\,\cite{bellex} $\Gamma <
2.3\,$MeV on the width of the peak in this particular channel.

The approximation of small interaction radius is applicable at small energy
$E=M(D {\bar D}^*) - M(D^0)- M(D^{*0})$ for a consideration of the strong
dynamics of two coupled channels $D^0 {\bar D}^{*0} + {\bar D} D^{*0}$
and $D^+ D^{*-}+D^- D^{*+}$, which for brevity can be called $n$ and $c$
channels. The energy range of interest for the discussion of the $X$ peak is
from few MeV below the $n$ threshold and up to the $c$ threshold, i.e. up to $E
\approx \Delta = M(D^+ D^{*-}) - M(D^{0} {\bar D}^{*0}) \approx 8.1\,$MeV. In
this range the  scale of the c.m. momentum (real and
virtual) in either channel is set by $\sqrt{2 m_r \Delta} \approx 127\,$MeV,
where $m_r \approx 970\,$MeV is the reduced mass for the meson pair. One can
apply in this region of soft momenta the standard picture of the strong-
interaction scattering (see e.g. in the textbook \cite{ll}), where the strong
interaction is localized at distances $r < r_0$ such that $r_0 \, \sqrt{2 m_r
\Delta}$ can be considered as a small parameter. Considering
for definiteness an energy value between the two thresholds, $0 < E < \Delta$,
one can write the corresponding wave functions (up to an overall normalization
constant) as
\be
\chi_n(r)=\sin (k_n r +\delta), ~~~~~\chi_c(r) = \xi \, \exp
(-\kappa_c r)~,
\label{chinc}
\ee
where $k_n = \sqrt{2 m_r E}$ and
$\kappa_c=\sqrt{2 m_r (\Delta - E)}$, $\delta$ is the elastic\footnote{In this
consideration the small inelasticity due to the $\pi^+ \pi^- J/\psi$ and $\pi^+
\pi^- \pi^0 J/\psi$ channels is neglected and will be included later. Also the
small width of the $D^*$ mesons is entirely neglected. The effects of the latter
width are considered in the most recent papers \cite{blu,blu2}.}
scattering phase in the $n$ channel and the constant $\xi$, generally energy-
dependent, describes the relative normalization and phase of the wave function
for the two channels.

The wave functions (\ref{chinc}) should be matched at $r \approx r_0$ to the
solution of the `inner' problem, i.e. that in the region of the strong
interaction. In the limit of small $r_0$ all the complexity of the `inner'
problem reduces to only two parameters. Namely, in the region of the strong
interaction the $n$ and $c$ channels are not independent and get mixed. Due to
the isotopic symmetry of the strong interaction the independent are the channels
with definite isospin, $I=0$ and $I=1$, corresponding to the functions
$\chi_0=\chi_n+  \chi_c$ and $\chi_1=\chi_n-  \chi_c$, and the matching
parameters are the logarithmic derivatives $-\kappa_0$ and $-\kappa_1$ of these
functions at $r=r_0$.  Using the assumption of small $r_0$ the matching
condition for the functions from Eq.(\ref{chinc}) can be shifted to $r =0$, so
that one can write the resulting matching equations as
\be
{  k_n \cos \delta - \xi \kappa_c \over \sin \delta + \xi} = - \kappa_0~,~~~~~
{  k_n \cos \delta + \xi \kappa_c \over \sin \delta - \xi} = - \kappa_1~.
\label{match}
\ee
These equations determine both the scattering phase $\delta$ and the constant
$\xi$ as
\be
\cot \delta = -{\kappa_{\rm eff} \over k_n}
\label{cotdel}
\ee
with
\be
\kappa_{\rm eff}={2 \kappa_0 \kappa_1 -\kappa_c \kappa_1 - \kappa_c \kappa_0
\over
\kappa_0 + \kappa_1 - 2 \kappa_c}~,
\label{kapeff}
\ee
and
\be
\xi={\kappa_0 - \kappa_1 \over 2 \kappa_c - \kappa_1- \kappa_0} \, \sin \delta~.
\label{xires}
\ee

The nonrelativistic scattering amplitude in the $n$ channel is therefore given
by\,\cite{ll}
\be
F=- {1 \over \kappa_{\rm eff} + i \, k_n}~,
\label{ampf}
\ee
and the scattering length $a$ is thus found from the $E =0$ limit of this
expression as
\be
a=\left. {1 \over \kappa_{\rm eff}} \right |_{E=0}= {\kappa_0+\kappa_1- 2 \,
\sqrt{2
m_r \Delta} \over 2 \kappa_0 \kappa_1 - (\kappa_0 + \kappa_1) \sqrt{2 m_r
\Delta}}~.
\label{scata}
\ee
The whole approach is applicable if the scattering length is
large in the scale of strong interaction. A large positive value of $a$ implies
an existence of a shallow bound state, while a large negative $a$ corresponds to
the situation with a virtual state\,\cite{ll}. According to the estimates of
Ref.\cite{hkkn} the required by the data scattering length in the problem
considered is $-(3 \div 4)\,$fm, corresponding to a negative and quite small
indeed
parameter $\kappa_{\rm eff}(E=0)\approx (50 \div 60)\,$MeV.

The physical picture, consistent with a small $\kappa_{\rm eff}$, and which
could be argued on general grounds\,\cite{ov76}, is that an attraction in the
$I=0$ channel is strong enough to provide a small value of $\kappa_0$, while the
interaction in the $I=1$ channel is either a weak attraction or, more likely, a
repulsion. In both cases the absolute value of $\kappa_1$ is large, i.e. of a
normal strong interaction scale, with the sign being respectively negative or
positive. Another, purely phenomenological, argument in favor of large
$|\kappa_1|$ is that no peculiar near-threshold behavior is observed in the
production of the $I=1$ charged states, e.g. $D^0 D^{*-}$. At large $|\kappa_1|$
the expression (\ref{kapeff}) simplifies and takes the approximate form
\be
\kappa_{\rm eff} \approx 2 \kappa_0 - \kappa_c~.
\label{kapefa}
\ee
Using this approximation, one can readily see that in order for
$\kappa_{\rm eff}(E=0)$ to be negative and small, the parameter $\kappa_0$ has
to
be
positive and quite small:
\be
\kappa_0 < \sqrt{m_r \Delta/2} \approx 63\,{\rm MeV}.
\label{ineq}
\ee

It is interesting to note that in the discussed picture the interaction in the
$I=0$ state is strong enough by itself to produce a shallow bound state in the
limit of exact isospin symmetry, i.e. at $\Delta \to 0$. In reality the isospin
breaking by the mass difference between the charged and neutral charmed mesons
turns out to be sufficiently significant to deform the bound state into a
virtual one, i.e. to shift the pole of the scattering amplitude from the first
sheet to the second sheet of the Riemann surface for the amplitude as a complex
function of the energy $E$.

The inelasticity in the $n$ and $c$ channels, related to decays to the observed
final states $\pi^+ \pi^- J/\psi$ ($\rho J/\psi$), $\pi^+ \pi^-
\pi^0 J/\psi$ ($\omega J/\psi$), $\gamma J/\psi$ and probably other, appears to
be reasonably
small, as one can infer from the observed\,\cite{belledd,babardd} dominance of
the $D^0 {\bar D}^{*0}$ production in the threshold region, and can be
parametrized by a small imaginary shift $i \, \gamma$ of the denominator of the
scattering amplitude in Eq.(\ref{ampf}):
\be
F=-{1 \over \kappa_{\rm eff}+ i \, k_n + i \, \gamma} \approx  -{1 \over  2
\kappa_0
- \kappa_c+ i \, k_n + i \, \gamma}~.
\label{ampfg}
\ee
If one further assumes\,\cite{mvssr,hkkn} that the `seed'
decay $B \to X K$ is a short-distance process, one would find that the yield in
each final
channel coupled to $X$ is proportional to that channel's contribution to the
unitary cut of the amplitude $F$. This implies in particular that
\be
{\cal B} [B \to (D^0 {\bar D}^{*0} + {\bar D} D^{*0}) \, K]\, :\, {\cal B} (B
\to \omega  J/\psi \, K)\, : \, {\cal B} (B \to \rho  J/\psi \, K) = \
k_n \, |F|^2 \, : \, \gamma_\omega \, |F|^2 \, : \, \gamma_\rho \, |F|^2~,
\label{ratio}
\ee
where the specific expression for $|F|^2$ depends on the value of the energy $E$
relative to the $n$ and $c$ thresholds, as given by Eq.(\ref{ampfg}) an its
analytical continuation across the thresholds. Besides the energy dependence of
the overall factor $|F|^2$, the heavy meson channel contains the phase space
factor $k_n$, while for the $\omega J/\psi$ and $\rho J/\psi$ yields an
additional dependence on the energy arises from the factors $\gamma_\omega$ and
$\gamma_\rho$.

A certain variation of the width parameter $\gamma_\omega$ for the $\pi^+ \pi^-
\pi^0 J/\psi$ channel in the discussed range of energy is of a well known
kinematical origin. Indeed, the central value of the mass of the $\omega$
resonance puts the threshold for the channel $\omega J/\psi$ at
$3878.5\,$MeV, which corresponds to $E \approx 6.7\,$MeV in our conventions,
i.e. squarely between the $n$ and $c$ thresholds. Any production of the $\pi^+
\pi^- \pi^0 J/\psi$ states at smaller invariant mass is a sub-threshold process,
possible due to the width $\Gamma_\omega$ of the $\omega$ resonance. In other
words, the energy dependence of the width factor $\gamma_\omega$ can be
estimated as
\be
\gamma_\omega=|A_\omega|^2 \, q^{(\omega)}_{\rm eff}~,
\label{gw}
\ee
where $A_\omega$ is the amplitude factor for the coupling to the $\omega J/\psi$
channel and $q^{(\omega)}_{\rm eff}$ is the effective momentum of $\omega$ at
the
invariant mass $M$ calculated as
\be
q^{(\omega)}_{\rm eff} (M) = \int_{m_0}^{M-m_{J/\psi}} |{\vec q}(m)| \,
{m_\omega \,
\Gamma_\omega \over (m^2-m_\omega^2)^2 + m_\omega^2 \, \Gamma_\omega^2} \, {d
m^2 \over \pi}~
\label{qeff}
\ee
with the c.m. momentum $|{\vec q}(m)|$ found in the standard way:
\be
|{\vec q}(m)|={ \sqrt{ [(M-m_{J/\psi})^2- m^2] \, [(M+m_{J/\psi})^2- m^2] }
\over 2 \, M}~.
\label{qofm}
\ee
The lower limit $m_0$ in the integral in Eq.(\ref{qeff}) can be chosen anywhere
sufficiently below $m_\omega - \Gamma_\omega$, since the Breit-Wigner curve in
the integrand rapidly falls off away from the resonance.

Numerically, the effective momentum $q^{(\omega)}_{\rm eff}$ can be estimated as
varying from approximately 20\,MeV to 50\,MeV between the $n$ and $c$
thresholds, i.e. when $E$ changes from $E=0$ to $E = \Delta$.
In the $\rho J/\psi$ channel the expected energy behavior of the yield is quite
different. If one writes the corresponding width factor $\gamma_\rho$ similarly
to Eq.(\ref{gw}) as
\be
\gamma_\rho= |A_\rho|^2 \, q^{(\rho)}_{\rm eff}~,
\label{gr}
\ee
the effective momentum $q^{(\rho)}_{\rm eff}$ can be estimated as varying only
slightly due to the large width of the $\rho$ resonance: $q^{(\rho)}_{\rm eff}
\approx (125 \div 135)\,$MeV as the energy changes between $E=0$ and $E=\Delta$.

The amplitudes $A_\omega$ and $A_\rho$ on the other hand should display a
noticeably different behavior in the energy range of interest, due to the rapid
(and different from each other) variation of the $I=0$ and $I=1$ scattering
amplitudes of the meson-meson scattering. Namely assuming that the amplitudes
for production of the $I=0$ and $I=1$ at short distances are described by
constant (in the considered energy range) factors $\Phi_\omega$ and $\Phi_\rho$
and that the effective strong production radius is $R$, one can
arrive\cite{mvxthr}, after taking into account the rescattering of mesons, at
the following estimate for the ratio of the amplitudes
\be
{A_\rho \over A_\omega} = {\kappa_0-\kappa_c \over \kappa_1-\kappa_c} \, (1
-\kappa_1 \, R) \, {\Phi_\rho \over \Phi_\omega}~.
\label{rorat}
\ee
Since the situation where the $X$ peak is a virtual state corresponds to a small
positive $\kappa_0$ satisfying the condition (\ref{ineq}), the amplitude
$A_\rho$ described by Eq.(\ref{rorat}) should necessarily change sign between
the $n$ threshold, where $\kappa_c=\sqrt{2 m_r \Delta}$, and the $c$ threshold,
where $\kappa_c=0$.

\begin{figure}[tb]
  \begin{center}
\begin{minipage}[t]{18 cm}
\epsfxsize=9cm
    \epsfxsize=9cm
     \epsfbox{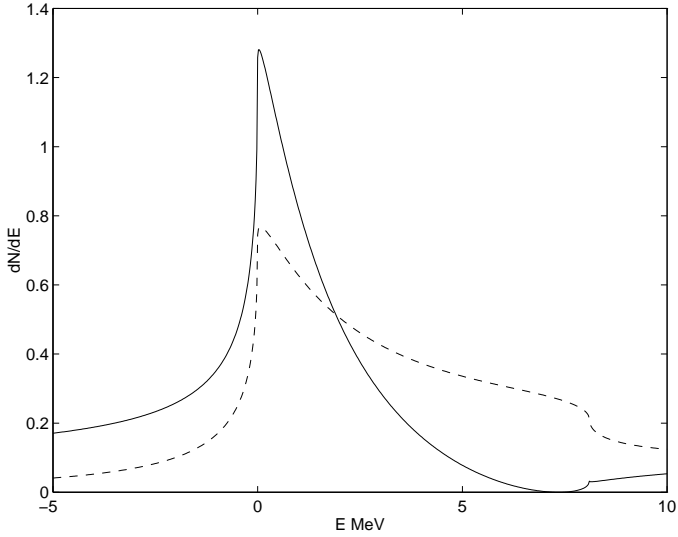}
\end{minipage}
\begin{minipage}[t]{16.5 cm} 
     \caption{The expected shape (in arbitrary units) of the virtual state peak
in the yield of $\pi^+ \pi^- J/\psi$ (solid) and $\pi^+ \pi^- \pi^0 J/\psi$
(dashed) channels. \label{xthresh}}
\end{minipage}
\end{center}
\end{figure}
The expected difference in the shape of the cusp in the $\rho J/\psi$ and
$\omega J/\psi$ channels is illustrated in Fig.\ref{xthresh}. In these plots the
parameters of the virtual state correspond to the scattering length $a=-(4 +
0.5\, i)\,$fm, which is close to the possible fit values of the scattering
length found in Ref.\cite{hkkn}. In the limit of large $\kappa_1$ this value of
$a$ translates into $\kappa_0 \approx 38\,$MeV and $\gamma \approx 6\,$MeV. One
can see from Fig.\ref{xthresh} that due to the discussed zero of the amplitude,
the peak in the $\pi^+ \pi^- J/\psi$ channel is expected to be quite narrow in
agreement with the experimental limit on the width of $X(3872)$. The plots in
Fig.\ref{xthresh} are normalized to the same total yield in each channel over
the shown energy range in order to approximate the experimentally observed
relative yield. Such normalization corresponds to setting $$\left | {\kappa_1
\over 1-\kappa_1 R} \, {\Phi_\omega \over \Phi_\rho} \right| \approx 175\,{\rm
MeV}~,$$ which value does not appear to be abnormal, even though at present we
have no means of independently estimating this quantity.

Summarizing the situation with the $X(3872)$ resonance it can be stated that as
of the time of this writing, full four years after its discovery and even though
this resonance is listed among established particles in the Tables\,\cite{pdg},
its real status is still lively debated in the literature. It is clear that this
peak is very closely related to the near-threshold dynamics of the $D \bar D^*$
meson pairs, but it is still unclear whether this is a shallow bound molecular
state, or a virtual state, or something else. It may well be that a further
pursuit of the puzzles arising in connection with the $X(3872)$ peak may provide
important clues to understanding multi-quark dynamics.

\subsection{\it Higher $J^{PC} = 1^{--}$ Resonances}
The existence of reasonably broad resonances above the open charm threshold has
been fully expected on the basis of potential models\cite{cornell}. What has not
been fully expected however are the peculiar properties of individual
resonances, different for different states, and never failing to bring
unexpected surprises.
\subsubsection{\it The Vicinity of $\psi(4040)$}
An unusual behavior of the cross section of $e^+e^-$ annihilation into charmed
meson pairs in the energy region, which is now associated with the resonance
$\psi(4040)$, was first pointed out in Ref.\cite{drgg}. Namely, it was noticed
that the production of the $D^* \bar D^*$ is greatly enhanced relative to $D
\bar D$ and $D \bar D^*$ in comparison with a simple spin-model
estimate\cite{drgg,cornell}. The cross section in each channel is proportional
to the $P$ wave factor $p^3$ with $p$ being the c.m. momentum in the
corresponding final state. The model predicted the ratio 1:3:7 of the extra
factors on top of the $p^3$ in the cross section for production of respectively
$D \bar D$, $D \bar D^* + \bar D D^*$ and $D^* \bar D^*$. The then existing data
however indicated that the cross section for production of the pairs of vector
mesons $D^* \bar D^*$ near their threshold was enhanced by a factor of tens to
hundreds in comparison with this estimate. This observation gave the reason for
the suggestion\cite{drgg} that $\psi(4040)$ is in fact a molecular state made of
the vector mesons. However a more conventional potential model\cite{cornell}
could well accommodate the $\psi(4040)$ peak as a dominantly $3^3S_1$ state, and
it was also argued\cite{yopr} that the suppression of the decay modes for this
resonance with one or two pseudoscalar mesons is due to an `accidental'
kinematical zero of the overlap integrals in a model of such decay.

\begin{figure}[tb]
  \begin{center}
\begin{minipage}[t]{18 cm}
\epsfxsize=9cm
    \epsfxsize=11cm
     \epsfbox{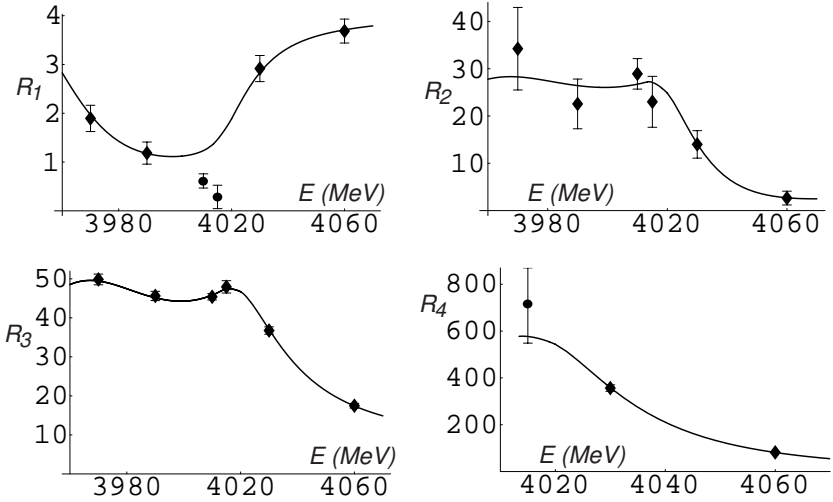}
\end{minipage}
\begin{minipage}[t]{16.5 cm} 
    \caption{The plots of the rate coefficients $R$ corresponding to the fit
with excluded data points at energy 4010 and 4015\,MeV for the $D {\bar D}$
channel and at 4015\,MeV for the $D^* {\bar D}^*$ channel. The excluded points
are shown by filled circles.\label{newres}}
\end{minipage}
\end{center}
\end{figure}

A significantly more detailed, than previously available, data on the production
of charmed meson pairs have been accumulated recently\cite{lang,cleoddfinal,belleisr} and
the new data also include the cross section for production of pairs of strange
charmed mesons $D_s \bar D_s$. The new data still indicate an unusually strong
production of the vector mesons, $D^* \bar D^*$, near their threshold and also
reveal intricate features of the behavior of the cross section in other channels
as the energy sweeps across the $D^* \bar D^*$ threshold(s) and the $\psi(4040)$
resonance, which definitely points at a strong coupling between the channels.
This behavior is more illustrative in terms of dimensionless rate coefficients
$R_i$ defined as follows\cite{dubmvnres}
\bea
\label{ris}
\sigma(e^+e^- \to D {\bar D}) = \sigma_0(s)\, 2 \, v_D^3 \, R_1~,&~&~
\sigma(e^+e^- \to D_s {\bar D}_s) = \sigma_0(s)\, v_{D_s}^3 \, R_2~,  \\
\nonumber
\sigma(e^+e^- \to D {\bar D}^*+D^* {\bar D}) = \sigma_0(s)\, 6 \, \left ({2 p}
\over \sqrt{s} \right )^3 \, R_3~,&~&~\sigma(e^+e^- \to D^* {\bar D}^*) =
\sigma_0(s)\, 7 \, (v_0^3+v_+^3) \, R_4~,
\eea
with $\sigma_0=\pi \, \alpha^2/(3 s)$. Here $v_D$, $v_{D_s}$, $v_0$ and $v_+$
stand for the c.m. velocities of each of the mesons in respectively the channels
$ D {\bar D}$, $D_s {\bar D}_s$, $D^{*0} {\bar D}^{*0}$ and $D^{*+} {\bar
D}^{*-}$, while for the channel $D {\bar D}^*+D^* {\bar D}$ with mesons of
unequal mass the velocity factor is replaced by $(2 p/\sqrt{s})$ with $p$ being
the c.m. momentum.  The values of $R_i$ calculated from a preliminary version of
the data\cite{lang} are shown as data points in Fig.\ref{newres}. The extra
factors 1, 3 and 7 in Eq.(\ref{ris}) for respectively the
pseudoscalar-pseudoscalar, pseudoscalar-vector and the vector-vector channels
correspond to the ratio of the corresponding production cross section in the
simplest model\cite{drgg,cornell} of independent quark spins, so that the
inequality between $R_1$, $R_3$ and $R_4$ also illustrates a conspicuous
deviation from this model. In particular the very large values of $R_4$ describe
the unusually strong enhancement of the vector-vector channel.

The curves in the plots of Fig.\ref{newres} correspond to a fit\cite{dubmvnres}
with one resonance and a non-resonant background with coupling among the
channels and with a proper treatment of the onset of the inelasticity at the two
thresholds for the vector meson pairs, $D^{*0} \bar D^{*0}$ at 4013\,MeV and
$D^{*+} D^{*-}$ at 4020\,MeV. A fit to the full set of data between 3970\,MeV
and 4060\,MeV turns out to be impossible with an acceptable $\chi^2$ (the best
fit corresponds to $\chi^2/NDF=17.6/8$. The offending are the data points at
4010 and 4015\,MeV for the $D \bar D$ channel and, to some extent, the data
point at 4015\,MeV for the $D^* \bar D^*$ production. If those points are
removed, a fit to the data can be found with $\chi^2/NDF=3.0/5$ and the
resulting central values of the resonance nominal mass, width and $\Gamma_{ee}$
are $M=4019\,$MeV, $\Gamma = 65\,$MeV and $\Gamma_{ee}=1.7\,$keV. A more
detailed analysis\cite{langth} of the final data including the radiative
corrections resulted in the values $M=4013 \pm 4\,$MeV, $\Gamma = 66 \pm 8\,$MeV
and $\Gamma_{ee}=1.9 \pm 0.7\,$keV.

The peculiar behavior of the $D \bar D$ points at 4010 - 4015\,MeV may indicate
a presence of another narrow resonance at this energy at or between the $D^{*0}
\bar D^{*0}$ and $D^{*+} D^{*-}$ thresholds, which has a quite small coupling to
$e^+e^-$, corresponding to $\Gamma_{ee}$ in the range of a few tenths of
eV\cite{dubmvnres}. Such resonance could then be explained as a $P$ wave $D^*
\bar D^*$ molecule which is expected\cite{ov76} to have the $e^+e^-$ width in
the same ballpark.

\subsubsection{$\psi(4170)$}
The main surprise of the peak $\psi(4170)$ (also labeled as $\psi(4160)$ in the Tables \cite{pdg}) is that it corresponds to the
absolute maximum of the cross section for production in $e^+e^-$ annihilation of
strange charmed meson pairs $D_s \bar D_s^* + \bar D_s D_s^*$. The cross section
at the maximum reaches\cite{lang,cleoddfinal} nearly 1\,nb, which is a very large value for an
exclusive state with hidden strangeness and charm. This property of the peak is
intensively used in the experimental studies of the $D_s$ mesons, but it has no
theoretical explanation.

\subsubsection{$Y(4260)$}
The total cross section for charm production in $e^+e^-$ annihilation has a well
known\cite{pdg} deep minimum around the c.m. energy 4260\,MeV. It is at about
the same energy where the BaBar experiment, while performing a survey of the
region around 4\,GeV in the radiation return process $e^+e^- \to \gamma \, X$,
found\cite{babar4260} a peak in the cross section for the {\it exclusive} final
state $\pi^+ \pi^- J/\psi$. The existence of the peak was further confirmed by
Belle\cite{belle4260} and by CLEO\cite{cleo4260} using the same radiative return
method and also by CLEO\cite{cleo4260a} by a direct scan of the $e^+ e^-$
annihilation in the peak region. The peak at the same invariant mass is also
indicated by the data\cite{babarxcn} on the decays $B \to \pi^+ \pi^- J/\psi K$.
The values for the mass and the width measured by independent experiments are in
a statistical agreement with each other, and the average values of these
parameters according to the updated Tables\cite{pdg} are $M(Y)=4264^{+10}_{-
12}\,$MeV and $\Gamma(Y)=83^{+20}_{-17}\,$MeV.

The only so far observed decays channels of $Y(4260)$ are $\pi^+ \pi^- J/\psi$,
$\pi^0 \pi^0 J/\psi$ (in a fair agreement with the isospin of $Y$ being equal to
zero) and $K^+ K^- J/\psi$. Notably, the data show no peak in any channels with
$D$ meson pairs. In particular the most stringent upper limit is
found\cite{babarydd} for the $D \bar D$ channel: $\Gamma(Y \to  D \bar
D)/\Gamma(Y \to \pi^+ \pi^- J/\psi) < 1.0$ at 90\% confidence level. Such
behavior is next to impossible to explain by considering the resonance as a
charmonium state even though the mass of $Y$ is close to the expectation for the
$4S$ state\cite{llanes}, since lower $J^{PC}=1^{--}$ resonances $\psi(3770)$,
$\psi(4040)$ decay practically exclusively to $D$ meson pairs. A whole spectrum
of interpretations of the state $Y(4260)$ has been suggested\cite{slzhu}
including $c \bar c + glue$ hybrids\cite{closepagehyb,koupene} and a tetraquark
$cs\bar c \bar s$ state assignment\cite{mrpp}, which does not look very
promising, given that the ratio of the decay rates\cite{cleo4260} $\Gamma(Y \to
K^+ K^- J/\psi)/\Gamma(Y \to \pi^+ \pi^- J/\psi) \approx 0.15$ does not show any
enhanced presence of strangeness within $Y(4260)$.

\subsubsection{\it $J^{PC}=1^{--}$ Peaks in $\pi^+ \pi^- \psi'$}
Subsequent studies of the radiative return events have lead to an observation by
BaBar\cite{babar432} of a `broad structure' in the final state $\pi^+ \pi^-
\psi'$. A single resonance fit to the data yielded the mass of $M= 4324 \pm
24\,$MeV and the width of $\Gamma = 172 \pm 33\,$MeV. A further investigation of
this final channel resulted in an observation by Belle\cite{belle432} of a peak
with the mass of $4361 \pm 9 \pm 9\,$MeV and the width of $74 \pm 15 \pm
10\,$MeV, possibly compatible with the structure observed by BaBar, and an {\it
additional} narrower peak at $4664 \pm 11 \pm 5\,$MeV with the width of $48 \pm
15 \pm 3\,$MeV.

\subsubsection{\it $Z(4430)$ and Remarks on New States}

Most recently the Belle experiment presented data\cite{bellez} on observation of
a peak $Z(4430)$ in the charged system $\pi^\pm \psi'$ emerging from the decays
$B \to K \pi^\pm \psi'$. The parameters of the peak are $M=4433 \pm 4 \pm
2\,$MeV, $\Gamma= 45^{+18}_{-15}\, ^{+30}_{-13}\,$MeV.  The statistical
significance of the observation corresponds to $6.5\sigma$, however in view of an
utmost importance of the observed peak an additional confirmation is eagerly awaited. Unlike the
previously discussed electrically neutral states, which all at least had a
chance of being a pure charmonium, this one is charged and has isospin $I=1$ and
clearly cannot be a pure $c \bar c$ but has to contain light quarks in addition
to the $c \bar c$ pair. Naturally, a variety of interpretations of $Z(4430)$ has
been suggested: a threshold peak\cite{rosnerz,buggz} or a
resonance\cite{mengchao} a loosely bound molecular state\cite{liuzhu} in the
$D^* \bar D_1 (2420)$ meson system, a radially excited tetraquark\cite{mpr}, a
QCD-string based model\cite{glp}, and a baryonium state\cite{qiao}. In either
case, whether or not the peak $Z(4430)$ is related to the $D^* \bar D_1 (2420)$
meson system, its `affinity' to the particular decay channel $\pi \psi'$, rather
than the multitude of available channels with $D$ mesons presents a very
intriguing riddle.

Clearly, a similar riddle also relates to the states $Y(4260)$, and the
`structures' in the $\pi^+ \pi^- \psi'$ channel at 4.32 and 4.66\,GeV. Perhaps,
the simplest explanation of such unusual prominence in the decays of each state
of a particular charmonium level accompanied by light hadrons would be a
picture, where a charmonium state, e.g. $J/\psi$, or $\psi'$ is `stuck' in a
light hadronic state. In a sense, such picture can be viewed as that of a bound
state of a relatively compact charmonium inside a light hadron having a larger
spatial size. This possibility in fact returns us to the previous discussion of
existence of bound states of $J/\psi$ and/or $\psi'$ in light nucleons. It may
well be that some of the recently found high mass states are in fact such bound
states in mesonic rather than baryonic matter. If this indeed is the case, one
can naturally expect an existence of similar baryonic states, e.g. bound states
of either $J/\psi$ or $\psi'$ `inside' a proton, or even `inside' a deuteron.
Needless to mention that an observation of such baryonic states would be of an
immense interest.

\section{Summary}

The potential models of interaction within charmonium generally agree with the
lower-mass part of the observed spectrum of the resonances, where two
long-missing states, the $^1P_1$ $h_c$ and the $2 {^1S_0}$ $\eta'_c$, have
eventually been located. In the mass region at and above the threshold for
charmed mesons their dynamics apparently plays an important role in determining
the spectrum of states. At lower masses some fine effects of the interaction
between the quark and the antiquark still remain unsolved, such as the behavior
of the spin-dependent forces, describing the splittings of the $\chi_{cJ}$
states, and the spin-spin interaction giving rise to the small splitting between
the c.o.g. of the $\chi_{cJ}$ resonances and the spin-singlet $h_c$.

Less model-dependent and based on the underlying theory of QCD spectral methods
for analyzing the charmonium states, either analytical, or numerical lattice
simulations, reproduce reasonably well the properties of the lowest state in
each channel. However, these methods are intrinsically less appropriate for
handling radially excited states, and they also eventually run into limitations
discussed in the Section 2.3.2.

The overall picture of the hidden charm decay through strong or electromagnetic
annihilation is in agreement with the data, still leaving us with a number of
puzzles in some particular cases, by solving which we might gain some new
insights into hadron dynamics. Among such puzzles are the larger than expected
total widths of $\eta_c$ and $\chi_{c0}$ and the non-similarity of some
exclusive decay channels for $J/\psi$ and $\psi'$ - a violation of the `12\%
rule'. These yet to be explained properties of the well known charmonium states
very likely indicate a presence of substantial nonperturbative effects in the
annihilation processes.

The radiative transitions between charmonium levels are in line with generic
considerations based on nonrelativistic quantum mechanics and also with the
results of specific potential models. A notable exception is the $M1$ transition
$J/\psi \to \gamma \eta_c$, which is significantly weaker than any theoretical
estimates. Coupled with the large total width of $\eta_c$ this may signal a
mixing of the $\eta_c$ with light $J^{PC}=0^{-+}$ degrees of freedom.

The hadronic transitions between charmonium resonances offer a window into the
interaction of the charmonium states with soft gluon fields, and also into the
details of the conversion of soft gluons to light mesons. The former interaction
is described by the multipole expansion in QCD, while the latter conversion is
tractable due to the chiral algebra and low-energy theorems in QCD based on the
conformal and axial anomalies. Such approach turned out to be successful in
describing the transitions $\psi' \to \pi \pi J/\psi$ and $\psi' \to \eta
J/\psi$ and in predicting finer details such as the small $D$ wave in the former
transition. Due to this success it can be believed that the recently established
discrepancy by a factor of about 1.5 between the observed and the theoretical
rate of the decay $\psi' \to \pi^0 J/\psi$ is not a result of a failure of the
multipole expansion and the low-energy theorems, but rather very likely signals
a presence of a small four-quark isovector component in the $\psi'$ resonance.

The knowledge of the strength of the interaction of charmonium with soft gluon
fields, the chromo-polarizability, can be applied to considering the behavior of
charmonium inside a hadronic media, such as the interaction of slow charmonium
with nucleons. The estimates of the chromo-polarizability  from the rate of the
decay $\psi' \to \pi \pi J/\psi$ indicate that slow charmonium interacts quite
strongly with light hadrons. This interaction may well result in existence of
bound states of charmonium in light nuclei or with ordinary hadronic resonances.

The unusual new states at and above charmed meson thresholds keep mushrooming in
the most recent experimental data suggesting that this mass region is a real
playground for much of the exotics that has been previously speculated about:
tetraquarks, hybrids, molecules, and mixtures of those. The peak at $X(3872)$
that started it all still offers intriguing puzzles, and further studies are
required in order to conclusively assess its internal structure and even its
very status as of a resonance. The newly found peaks in the $\pi \pi J/\psi$ and
$\pi \pi \psi'$ invariant mass spectra and especially the latest peak $Z(4430)$
in the $\pi^\pm \psi'$ channel lead us further down the path of acceptance of
multiquark hadronic states being as commonplace,  as are nuclei if viewed as
multiquark systems.

Many of the properties of charmonium, starting with a very small width of the
$J/\psi$ resonance, came as a great surprise when first observed. Understanding
these properties and in some cases making their description routine, has greatly
advanced the knowledge of the dynamics of quarks and gluons. Now, thirty three
years after the discovery of $J/\psi$, the studies of charmonium keep bringing
new surprises and puzzles, solving which will hopefully result in further
advances.

\section*{Acknowledgments}
This work is supported in part by the DOE grant DE-FG02-94ER40823. Preliminary
notes for the manuscript were done during a visit to the Institute for Nuclear
Physics of the Bonn University with support from the Alexander von Humboldt
Foundation. Several sections of this paper were written at the Aspen Center for
Physics.

\end{document}